\documentclass[jmp,
 amsmath,amssymb,
reprint, twocolumn 
]{revtex4-1}

\usepackage{graphicx}
\usepackage{dcolumn}
\usepackage{bm}
\usepackage{lipsum}
\usepackage[utf8]{inputenc}
\usepackage[T1]{fontenc}
\usepackage{mathptmx}
\usepackage{etoolbox}
\usepackage{color}

\newcommand{\sech}{\operatorname{sech}}

\newtheorem{prop}{Proposition}
\newtheorem{lemma}[prop]{Lemma}

\newcommand{\bea}{\begin{eqnarray}}
\newcommand{\eea}{\end{eqnarray}}
\newcommand{\be}{\begin{equation}}
\newcommand{\ee}{\end{equation}}
\def\nn{\nonumber}
\usepackage[LGR,T1]{fontenc}
  \newcommand{\textgreek}[1]{\begingroup\fontencoding{LGR}\selectfont#1\endgroup}
\newcommand{\DD}{\mathcal{D}}  

\makeatletter
\def\@email#1#2{%
 \endgroup
 \patchcmd{\titleblock@produce}
  {\frontmatter@RRAPformat}
  {\frontmatter@RRAPformat{\produce@RRAP{*#1\href{mailto:#2}{#2}}}\frontmatter@RRAPformat}
  {}{}
}%
\makeatother

\usepackage{hyperref}

\begin{document}


\title[Scalar, vector and tensor fields on $dS_3$ with  arbitrary sources]{Scalar, vector and tensor fields on $dS_3$ with  arbitrary sources:\vspace{2pt} \\
harmonic analysis and antipodal maps}
\author{G. Comp\`ere}
 \email{geoffrey.compere@ulb.be}
\author{S. Robert}%
 \email{sebastien.robert@ulb.be}
\affiliation{ 
Universit\'e Libre de Bruxelles, International Solvay Institutes and Brussels Laboratory of the Universe (BLU-ULB), C.P. 231, B-1050 Bruxelles, Belgium}%

\date{\today}

\begin{abstract}
The scalar, vector and tensor spherical harmonics on three-dimensional de Sitter spacetime are defined and analyzed. Each harmonic defines two sets of asymptotic data on the two-sphere in the asymptotic expansion close to both the past and the future of de Sitter spacetime. For each case, we make explicit the antipodal relationship of both sets of asymptotic data between past and future infinity, which can be non-local. A procedure is defined to extract these asymptotic data in the presence of sources. This provides for each class of propagating field on de Sitter the relationship between two independent sets of data defined on the sphere in the asymptotic future with the corresponding data defined in the asymptotic past. We also provide several theorems on the decomposition of vectors and tensors on de Sitter such as one proving that a large class of tensors obeying an inhomogeneous wave equation can be expressed locally in terms of a symmetric transverse traceless tensor.  These results are instrumental in the description of interacting four-dimensional asymptotically flat fields at spatial infinity. 
\end{abstract}

\maketitle

\tableofcontents

\section{Introduction}

Tensor harmonic decompositions have been a decisive mathematical tool to extract physics from several relevant partial differential equations. Tensor harmonic analysis around the Schwarzschild background performed in the seminal work of Regge-Wheeler and Zerilli \cite{PhysRev.108.1063,1970PhRvD...2.2141Z} is now considered as the foundation of black hole perturbation theory. As a second example, the study of perturbations of Friedmann-Lema\^itre-Robertson-Walker spacetimes \cite{Bardeen:1980kt,1984PThPS..78....1K}, which underlies  structure growth in cosmology \cite{Seljak:1996gy,Kamionkowski:1996zd}, is built upon harmonic analysis of $S^3$ \cite{Lifshitz:1945du,Lifshitz:1963ps,PhysRevD.8.4297,10.1063/1.1666537,10.1063/1.523778,Gerlach:1978gy,10.1143/PTP.68.310}. The motivation of the present work is the holographic description of 4-dimensional asymptotically flat spacetimes \cite{beig1982einstein,beig1984integration,deBoer:2003vf,Dappiaggi:2005ci,Mann:2005yr,Barnich:2009se,Bagchi:2010zz,He:2014laa,Kapec:2016jld,Cheung:2016iub,Pasterski:2016qvg,Pasterski:2017kqt,Donnay:2022wvx}:   field theories, including gravity, can be rewritten outside the lightcone of a given point in terms of infinite sets of 3-dimensional fields obeying a sourced wave equation of the form 
\begin{align}
(\square - \alpha ) \Psi_{i_1 \dots i_T} = S_{i_1...i_T}  \label{wave}
\end{align}
where $\Box=q^{ab}\mathcal{D}_b\mathcal{D}_a$, $q_{ab}$ is the unit metric on 3-dimensional de Sitter spacetime $dS_3$, $q^{ab}$ is its inverse metric, $\mathcal{D}_a$ is its covariant derivative with Christoffel connection, $\alpha \in \mathbb R$ and  $T=0,1,2$. The source $S_{i_1...i_T}$ is smooth on $dS_3$ and does not depend on $\Psi_{i_1 \dots i_T}$. 

A systematic study of scalar ($T=0$), vector ($T=1$) and tensor ($T=2$) harmonics obeying equation \eqref{wave} with $S_{i_1...i_T}=0$ is  currently lacking in the literature. In 1986, Higuchi \cite{Higuchi:1986wu} built all symmetric scalar, vector and tensor harmonics on $S^N$, $N \geq 2$, from associated Legendre functions, extending earlier analyses \cite{1984AnPhy.156..412C,1985JMP....26...65R}.  He then built all harmonics on $dS_{N}$ which are obtained from analytic continuation from harmonics of $S^N$. His work was reviewed and further expanded by Marolf and Morrison \cite{Marolf:2008hg}. Since $dS_N$ is non-compact, it admits a countably infinite set of harmonics, a subset of which were uncovered by the work of Higuchi.
Some properties of tensors on $dS_3$ were derived in the work of Ashtekar and Hansen \cite{Ashtekar:1978zz} and subsequent work at spatial infinity \cite{beig1982einstein,beig1984integration,Ashtekar:1991boundary,Herberthson:1992gcz,Perng_1999,Compere:2011db}. In 2017, Troessaert identified for two particular cases of scalar and tensor equations that two types of harmonics exist on $dS_3$: the $p$ type and $q$ type, related to Legendre $P$ and $Q$ functions \cite{Troessaert:2017jcm}. They are distinguished by their eigenvalue under the antipodal map (i.e. combined time reversal and parity flip) on $dS_3$, which is instrumental in formally proving  relationships for four-dimensional asymptotically flat spacetimes between past and future null infinity such as the ones conjectured in \cite{Strominger:2013jfa,Strominger:2013lka}. The properties of scalar and vector harmonics were more systematically derived by Troessaert and Henneaux \cite{Henneaux_2019,Henneaux:2019yqq}, Fuentealba and Henneaux \cite{Fuentealba:2024lll,Fuentealba:2025ekj} and Brice\~no, Gonz\'alez and P\'erez \cite{Briceno:2025ivl}. However, the antipodal relationship between the future and past of $dS_3$ of the asymptotic data of scalar and vector harmonics at subleading order was not addressed. In the literature, the effect of sources $S_{i_1 \dots i_T} \neq 0$ on the antipodal relationships, which are however relevant for interacting theories such as theories of interacting scalars, Yang-Mills theory or General Relativity, was also never  studied thoroughly.  

In this work, we systematically construct the scalar, vector and tensor harmonics on $dS_3$ with an emphasis on their complete antipodal relationships and the fate of these antipodal relationships in the presence of sources. We treat sequentially the scalar, vector and tensor cases. Inner products are studied, which allows us to derive orthogonality relations and conserved quantities. The asymptotic expansions of harmonics at large early and late times are derived, which allows us to obtain the antipodal relationships of their asymptotic data. In the tensorial case, we derive new lemmae for tensors on $dS_3$ beyond the known ones summarized in \cite{Compere:2011db}. The presentation will be self-contained. 

In what follows we will use the following properties of tensors on $dS_3$. Let $t$, $t_a$, $t_{ab} = t_{(ab)}$ be arbitrary tensors on $dS_3$, then
\begin{subequations}
\label{CommunationRel}
\bea
\left[ \DD_a,\square \right] t &=& -2 \DD_a t,\label{commuBoxPartialscal}\\
\left[ \DD_a,\DD_b \right]t_{c} &=& q_{ac}t_b - q_{bc}t_a,\\
\left[ \DD_a,\square \right]t_b &=& 2 q_{ab} \DD_c t^c - 4 \DD_{(a} t_{b)},\\
\left[ \DD^c,\DD_a \right] t_{cb} &=& 3 t_{ba} - q_{ab} t^c_{\; c},\\
\left[ \DD_a,\DD_b \right] t_{cd} &=& 2 q_{a(c} t_{d)b} - 2 q_{b(c} t_{d)a},\\
\left[ \DD_a,\square \right] t_{bc} &=& 4 q_{a(b}\DD^d t_{c)d} -6\DD_{(a}t_{bc)},\\
\left[ \DD^b,\square \right] t_{bc} &=& 4 \DD^d t_{cd} -2 \DD_c t^d_{\; d} .
\eea
\end{subequations}

\vspace{6pt}\noindent \textbf{Data availability.} For the ease of reproducibility and usage of our study, we provide an openly accessible  Mathematica notebook on the following \href{https://github.com/gcompere/scalar_vector_tensor_harmonics_on_dS3}{GitHub repository}, where all key formulae of this manuscript are derived. 
\vspace{6pt}

\section{Scalar harmonics}

We define scalar harmonics on $dS_3$ as the solutions to the following equation \cite{Higuchi:1986wu}:
\begin{align}
\Box \psi_n = -\left[(n+1)^2-1\right]\psi_n,
\label{generalscalarequation}
\end{align}
In this paper, we only consider $n\in\mathbb{Z}$ which corresponds to the class of equations that are satisfied by the fields appearing in the expansion of the massless scalar field at spatial infinity \cite{Henneaux:2018mgn}. The $n=-1$ case also appears in the Beig-Schmidt expansion \cite{beig1984integration}. 
We trivially have $\psi_{n} = \psi_{-2-n}$ and therefore assume $n \geq -1$ without loss of generality.

We denote as $(\tau,x^A) \equiv (\tau,\theta,\phi)$ the global coordinates of $dS_3$ where the metric reads as 
\begin{align}
q_{ab}dx^a dx^b = -d\tau^2+\cosh^2\tau (d\theta^2+\sin^2\theta d\phi^2).
\end{align}
We denote as $q$ the determinant of the metric and as 
\begin{align}
\oint_{S^2(\tau)} d\Omega\,  f &\equiv  \int_{S^2}d\theta d\phi \sqrt{-q}f \nn\\ 
&=\cosh^2\tau \int_{0}^\pi d\theta \int_0^{2\pi} d\phi \sin\theta  f
\end{align}
the integration measure of the function $f$ on $dS_3$ on a constant $\tau$ slice. We simply denote as $\oint_{S^2} d\Omega := \oint_{S^2(0)} d\Omega$ the unit integration measure over $S^2$. Indices $A,B,\dots$ refer to tensors on $S^2$. The metric on $S^2$ is $\gamma_{AB}$, the inverse metric $\gamma^{AB}$, the Christoffel connection is $\nabla_A$ and the volume form is $\epsilon^{AB}$ with orientation convention $\epsilon^{\theta\phi}=+\csc\theta$. The measure on $dS_3$ and $S^2$ are related as $\epsilon^{\tau \theta\phi}=\sech^2\tau\epsilon^{\theta\phi}$. 

\subsection{Construction and definition}

We decompose the general solution of Eq. \eqref{generalscalarequation} as a sum of modes  
\begin{align}
\psi_{n\ell m}(\tau,x^A):= Y_{\ell m}(\theta,\phi)\psi_{n \ell }(\tau), \quad 0\le \vert m \vert \le \ell \label{firstscalaransatz},
\end{align}
where $ Y_{\ell m}(\theta,\phi)$ are the canonically normalized spherical harmonics.

It is necessary to distinguish 2 cases : $\ell<(n+1)$ and $\ell\ge(n+1)$. In the first case, the solution to Eq. \eqref{generalscalarequation} can be derived with the ansatz
\begin{align}
    \psi_{n \ell}(\tau) = \frac{1}{(\cosh\tau)^{\frac{1}{2}}}\hat{f}_{\ell n}(i \sinh{\tau}) .
\end{align}

We get the associated Legendre equation
\begin{align}
    (x^2-1)&\frac{d^2 \hat{f}_{\ell n}(x)}{dx^2}\!+2x \frac{d \hat{f}_{\ell n}(x)}{dx} \nonumber \\ 
    &-\!\left((n+\frac{1}{2})(n+\frac{3}{2}) + \frac{(\ell +\frac{1}{2})^2}{x^2-1}\!\right)  \hat{f}_{\ell n}(x)=0
\label{LegendreEquation}
\end{align}
with $x=i \sinh{\tau} $. The two families of independent solutions are the associated Legendre functions of the first kind $P^{\ell+\frac{1}{2}}_{n+\frac{1}{2}}(i \sinh{\tau})$ and the associated Legendre functions of second kind $Q^{\ell+\frac{1}{2}}_{n+\frac{1}{2}}(i \sinh{\tau})$. These solutions are the ones obtained by Higuchi in \cite{Higuchi:1986wu} from the analytic continuation of the harmonics of $S^3$. The main property that distinguishes the 2 families is their transformation under time inversion:
\begin{subequations}
\begin{align}
    P^{\ell+\frac{1}{2}}_{n+\frac{1}{2}}(i \sinh{(-\tau)}) &= \overline{P^{\ell+\frac{1}{2}}_{n+\frac{1}{2}}(i \sinh{(\tau)})} \nonumber \\
    &=(-1)^{n-\ell+1} P^{\ell+\frac{1}{2}}_{n+\frac{1}{2}}(i \sinh{\tau}) ,\\ 
    Q^{\ell+\frac{1}{2}}_{n+\frac{1}{2}}(i \sinh{(-\tau)}) &= \overline{Q^{\ell+\frac{1}{2}}_{n+\frac{1}{2}}(i \sinh{(\tau)})} \nonumber \\ 
    &=(-1)^{n-\ell} Q^{\ell+\frac{1}{2}}_{n+\frac{1}{2}}(i \sinh{\tau}).
\end{align}\label{parityllen+1}
\end{subequations}
These identities imply that the real solutions are obtained by multiplying $P^{\ell+\frac{1}{2}}_{n+\frac{1}{2}}(i \sinh{(\tau)})$ by $i^{n+\ell+1}$ and $Q^{\ell+\frac{1}{2}}_{n+\frac{1}{2}}(i \sinh{(\tau)})$ by $i^{n+\ell}$. Here overbars denote complex conjugation.

In the second case ($\ell\ge(n+1)$), the solutions to Eq. \eqref{generalscalarequation} can be found with the ansatz
\begin{align}
    \psi_{n \ell}(\tau) = \frac{1}{\cosh\tau}f_{\ell n}(\tanh\tau) 
\label{ansatzpsilge2}.
\end{align}

This gives the following associated Legendre equation
\begin{align}
    (x^2-1)&\frac{d^2 f_{ \ell n}(x)}{dx^2}+2x \frac{d f_{\ell n}(x)}{dx} \nonumber \\
    &-\left(\ell (\ell+1) + \frac{(n+1)^2}{x^2-1}\right) f_{\ell n}(x)=0
\label{LegendreEquationbis}
\end{align}
with $x=\tanh \tau$.

The two families of independent solutions are the associated Legendre functions of the first kind $P^{n+1}_\ell(\tanh \tau)$ and of second kind $Q^{n+1}_\ell(\tanh \tau)$. The two families also transform differently under time inversion:
\begin{subequations}
\begin{align}
    P^{n+1}_\ell(\tanh (-\tau)) &= (-1)^{n-\ell+1} P^{n+1}_\ell(\tanh \tau), \\
    Q^{n+1}_\ell(\tanh (- \tau)) &= (-1)^{n-\ell} Q^{n+1}_\ell(\tanh \tau).
\end{align}\label{paritylgen+1}
\end{subequations}
These solutions were the ones considered in \cite{Compere:2011db}. 

Assembling the solutions following the mode decomposition \eqref{firstscalaransatz}, we obtain the 2 families of harmonics on $dS_3$ defined $\forall \ell \ge \vert m \vert \ge 0$ and $\forall n \ge -1$:
\begin{align}  
&\psi_{n\ell m}^{(p)}(\tau,x^A) :=\psi_{n\ell }^{(p)}(\tau) Y_{\ell m}(\theta,\phi) \nonumber\\
&=\left\{
    \begin{array}{ll}
       i^{n+\ell+1}\frac{N^{(P)}_{n\ell}}{(\cosh\tau)^\frac{1}{2}} Y_{\ell m}(\theta,\phi) P^{\ell +\frac{1}{2}}_{n+\frac{1}{2}}(i \sinh{\tau}) & \mbox{for }\ell <(n+1) ;\\&\\
       \frac{N^{(P)}_{n\ell}}{\cosh\tau }Y_{\ell m}(\theta,\phi) P^{n+1}_\ell (\tanh\tau)  & \mbox{for } \ell \ge(n+1),
    \end{array}
\right.\label{psi(p)nlm}
\intertext{and}
&\psi_{n\ell m}^{(q)}(\tau,x^A) := \psi_{n\ell }^{(q)}(\tau) Y_{\ell m}(\theta,\phi) \nonumber \\
&=\left\{
    \begin{array}{ll}
       i^{n+\ell}\frac{N^{(Q)}_{n\ell}}{(\cosh\tau)^\frac{1}{2}} Y_{\ell m}(\theta,\phi) Q^{\ell +\frac{1}{2}}_{n+\frac{1}{2}}(i \sinh{\tau}) & \mbox{for } \ell <(n+1); \\&\\
       \frac{N^{(Q)}_{n\ell}}{\cosh\tau }Y_{\ell m}(\theta,\phi) Q^{n+1}_\ell (\tanh\tau)  & \mbox{for } \ell \ge(n+1),
    \end{array}
    \right.\label{psi(q)nlm}
\end{align}
where $N^{(P)}_{n\ell}$, $N^{(Q)}_{n\ell}$ are normalization coefficients defined as follows:
\begin{align}  
    &N^{(P)}_{n\ell} =\left\{
    \begin{array}{ll}
      \sqrt{2\pi} (-1)^{n+1} \frac{(n-\ell)!}{(n+\ell+1)!}& \mbox{for }\ell <(n+1) ;\\&\\
       \frac{(\ell-n-1)!}{(n+\ell+1)!}  & \mbox{for } \ell \ge(n+1),
    \end{array}
\right.\label{Npsi(p)nlm}
\intertext{and}
&N^{(Q)}_{n\ell}  =\left\{
    \begin{array}{ll}
       (-1)^\ell \sqrt{\frac{1}{2\pi}} &\qquad\qquad \mbox{for } \ell <(n+1); \\&\\
       1  &\qquad\qquad \mbox{for } \ell \ge(n+1).
    \end{array}
    \right.\label{Npsi(q)nlm}
\end{align}

In summary, the general solution to \eqref{generalscalarequation} can be decomposed as 
\begin{align}\label{solscal}
    \psi_n=\sum_{\ell,m} \left(A_{n\ell m} \psi_{n\ell}^{(p)}(\tau) Y_{\ell m}(\theta,\phi)+B_{n\ell m} \psi_{n\ell}^{(q)}(\tau) Y_{\ell m}(\theta,\phi) \right) ,
\end{align}
where the sum is over $\ell \geq 0$, $-\ell \leq m \leq \ell$ and $A_{n\ell m}$, $B_{n\ell m}$ are constants. 

\subsection{Properties}
We recall that spherical harmonics transform under the parity flip operator $\Upsilon$ as
\begin{align}
    \Upsilon^* [Y_{\ell m}(\theta,\phi)] = (-1)^\ell  Y_{\ell m}(\theta,\phi),\label{SphericalHarmonicsTransformations}
\end{align}
We define the antipodal map $\Upsilon_{\mathcal{H}}$ as the combined parity flip and time reversal which sends $(\tau,\theta,\phi)$ to $(-\tau,\pi-\theta,\phi+\pi)$. Using Eqs. \eqref{parityllen+1} and \eqref{paritylgen+1}, we deduce the eigenvalues under the antipodal map that characterize these two  families of harmonics:
\begin{align}
      \Upsilon_{\mathcal{H}}^*[\psi_{n\ell m}^{(p)}(\tau,x^A)] &= (-1)^{n+1} \psi_{n\ell m}^{(p)}(\tau,x^A),\\
      \Upsilon_{\mathcal{H}}^*[\psi_{n\ell m}^{(q)}(\tau,x^A)] &= (-1)^{n} \psi_{n\ell m}^{(q)}(\tau,x^A).
\label{scalarparity}
\end{align}
We will refer to the property of $p$ parity and $q$ parity on the hyperboloid  for $\psi_{n\ell m}^{(p)}(\tau,x^A)$ and  $\psi_{n\ell m}^{(q)}(\tau,x^A)$, respectively.

In $dS_3$, we can define the usual Klein-Gordon (KG) inner product. Let $\phi$, $\varphi$ be two complex scalars on $dS_3$, the KG inner product is defined as 
\begin{align}
    (\phi,\varphi)_{KG} :=  \oint_{S^2(\tau)} d\Omega \, n^a \left(\phi\mathcal{D}_a \overline{\varphi} - \overline{\varphi}\mathcal{D}_a \phi   \right)
\end{align}
where $n^a\partial_a = \partial_\tau$ is the future-oriented unit normal to the sphere within $dS_3$. This inner product is conserved when $\phi$ and $\varphi$ both satisfy Eq. \eqref{generalscalarequation} for the same $n$. The scalar KG inner product appears naturally when defining supertranslation charges for asymptotically flat spacetimes, see e.g. \cite{Compere:2011ve},  and asymptotic charges for the massless scalar field at spatial infinity \cite{Fuentealba:2024lll}. 

The harmonics have been normalized with respect to this KG inner product as follows
\begin{align}
    (\psi_{n\ell m}^{(c)},\psi_{n\ell' m'}^{(d)})_{KG}=\delta_{\ell \ell'} \delta_{m,m'} \epsilon_{cd}
\end{align}
with $c,d \in {p,q}$, $\epsilon_{pp}=\epsilon_{qq}=0$,  $\epsilon_{pq}=-\epsilon_{qp}=1$. Although the KG inner product is not positive definite, it can be adapted to define a Hilbert space (see e.g. \cite{wald1995quantumfieldtheorycurved}).  
After a change of basis 
\begin{align}
    \psi_{n\ell m}^{\pm} := \frac{1}{\sqrt{2}} \left(\psi_{n\ell m}^{(p)}\pm i  \psi_{n\ell m}^{(q)}\right),
\end{align}
we recover the positive energy states ($\psi_{n\ell m}^{+}$) and negative energy states ($\psi_{n\ell m}^{-}$) which are standard in the context of quantization in curved spacetime (see e.g. \cite{Bousso_2002,Wald:1995yp}). Throughout this paper, we will stick to the $\psi_{n\ell m}^{(c)}$, $c=p,q$ basis.

\subsection{Inhomogeneous solutions}

Let us consider a source $s(\tau,x^A)$ that is smooth on $dS_3$. Then, the general solution to the inhomogeneous equation 
\begin{align}
    \Big(\Box + n(n+2)\Big)\psi_{n} = s(\tau,x^A)\label{InhomognenousScaleq}
\end{align}
is given as follows: 
\begin{align}
  \psi_{n} &= \sum_{\ell,m} \left(A_{n\ell m }+A^{(S)}_{n\ell m}(\tau; \bar\tau)\right) \psi^{(p)}_{n\ell m}(\tau,x^A) \nonumber \\ 
   &+\left(B_{n\ell m }+B^{(S)}_{n\ell m}(\tau; \bar\tau)\right) \psi^{(q)}_{n\ell m}(\tau,x^A)\label{ScalNHSol}
\end{align}
with $A_{n\ell m }$ and $B_{n\ell m }$ constants and 
\begin{align}\label{ASscalar}
    A^{(S)}_{n\ell m}(\tau ; \bar\tau)&=  \int_{\bar\tau}^\tau d\tau'  \oint_{S^2(\tau')}  d\Omega \,s(\tau',x^A)\overline{\psi^{(q)}_{n\ell m}(\tau',x^A)} , \\
\label{BSscalar}
    B^{(S)}_{n\ell m}(\tau; \bar\tau)&= -\int_{\bar\tau}^\tau  d\tau' \oint_{S^2(\tau')} d\Omega \,s(\tau',x^A)\overline{\psi^{(p)}_{n\ell m}(\tau',x^A)} 
\end{align}
where $\bar\tau$ is the time corresponding to a Cauchy surface. 
The coefficients can be extracted with the KG inner product : 
\begin{align}
     (\psi_{n},\psi^{(q)}_{n\ell m})_{KG}(\tau) &=A_{n\ell m }+A^{(S)}_{n\ell m}(\tau ; \bar\tau),\\
     (\psi_{n},\psi^{(p)}_{n\ell m})_{KG}(\tau)&=-B_{n\ell m }-B^{(S)}_{n\ell m}(\tau; \bar\tau).
\end{align}
By construction, the inhomogeneous coefficients are the sum of two terms corresponding to the primitives of the integrals \eqref{ASscalar} and \eqref{BSscalar} evaluated at $\tau$ and $\bar\tau$, i.e.  $A^{(S)}_{n\ell m}(\tau ; \bar\tau) = A^{(S)}_{n\ell m}(\tau)-A^{(S)}_{n\ell m}(\bar\tau)$ and $B^{(S)}_{n\ell m}(\tau ; \bar\tau) = B^{(S)}_{n\ell m}(\tau)-B^{(S)}_{n\ell m}(\bar\tau)$. In the case where $s(\tau,x^A)$ has $p$ parity, then we have $A^{(S)}_{n \ell m}(-\tau) = A^{(S)}_{n \ell m}(\tau) $ and $B^{(S)}_{n \ell m}(-\tau) = -B^{(S)}_{n \ell m}(\tau) $ \footnote{More precisely, since a primitive is defined only up to an additive constant, one may choose it such that $B^{(S)}_{n \ell m}(-\tau) = -B^{(S)}_{n \ell m}(\tau) $ whenever $s(\tau,x^A)$ has $p$ parity.}. Similarly, in the case where $s(\tau,x^A)$ has $q$ parity, we have $A^{(S)}_{n \ell m}(-\tau) = -A^{(S)}_{n \ell m}(\tau) $ and $B^{(S)}_{n \ell m}(-\tau) = B^{(S)}_{n \ell m}(\tau) $. In both cases, the terms proportional to the coefficients $A^{(S)}_{n \ell m}(\tau)$ and $B^{(S)}_{n \ell m}(\tau)$ in \eqref{ScalNHSol} therefore have the same parity as $s(\tau,x^A)$. This is a consequence of the fact that the operator $\square$ is invariant under $\Upsilon_{\mathcal H}^*$. 

The KG inner product $(\psi_{n},\psi_n^H)_{KG}$ with $\psi_n^H(\tau,x^A)$ a homogeneous solution (to Eq. \eqref{generalscalarequation}) is conserved if $s(\tau,x^A)=0$.  



\subsection{Asymptotic behavior and antipodal matching}
\label{sec:asymptInh}

We want to establish the asymptotics of a general scalar satisfying Eq. \eqref{InhomognenousScaleq} and identify the antipodal relations between $\tau=\infty$ and $\tau=-\infty$. Throughout this paper, we will encounter particular cases for distinguished values $\bar n$ of $n$ in what follows. We will use the convention that for any term multiplying $1-\delta_{n,\bar n}$ (which is 0 when $n=\bar n$ and 1 otherwise), it will be identically zero even if the term multiplying $1-\delta_{n,\bar n}$ is ill-defined or infinite for $n=\bar n$. Also, we use $1/(n!)\vert_{n=-1}:=1/\Gamma[0]=0$. 

\noindent \vspace{-20pt} 
\begin{center}\textit{
 Homogeneous case   }
\end{center}
 \vspace{3pt} 

Let us start by studying the asymptotic behavior of the harmonics. For $\ell <(n+1)$, the associated Legendre functions that appear in the solutions \eqref{psi(p)nlm}-\eqref{psi(q)nlm} are defined as (see e.g. \cite{Mathematica})
\begin{subequations}
\begin{align}
    P^{\ell+1/2}_{n+1/2}(z) := &\frac{1}{\Gamma(\frac{1}{2}-\ell)} \frac{(1+z)^{\frac{1}{2}(\ell+1/2)}}{(1-z)^{\frac{1}{2}(\ell+1/2)}} \nonumber \\
    &\times {}_2F_1 \left(-n-\frac{1}{2},n+\frac{3}{2};\frac{1}{2}-\ell;\frac{1-z}{2}\right),\\
    Q_{n+1/2}^{\ell+1/2}(z) := & (-1)^{\ell+1} \frac{\pi}{2} (n-\ell+1)_{2\ell +1}P_{n+1/2}^{-\ell-1/2}(z),
\end{align}
\end{subequations}
where $ {}_2F_1 \left(a,b;c ; w\right)$ is the Gauss hypergeometric function and $z=i\sinh\tau$.

Taking the limit $\tau\to\pm\infty$, we find the following asymptotic behavior:
\begin{align}
 \psi_{n\ell}^{(p)}(\tau\to\pm\infty) &=(\pm 1)^{n+\ell+1} (-1)^{\ell} \sqrt{\pi}  \frac{2^{-2n} \Gamma (2+2n)}{\Gamma \left(n+\frac{3}{2}\right) \Gamma (n+\ell+2)n!}\nn\\
 &\hspace{-1cm} e^{n\vert\tau\vert}\left(\sum_{r=0}^{n}\;\sum_{k=0}^{r} \frac{(-1)^{r}(n-k)!\;r! (-\ell)_{k}(\ell+1)_{k}}{(k!)^2(r-k)!} e^{-2r\vert\tau\vert}\right.\nonumber\\
 &\hspace{-1cm} \left.+(-1)^{\ell+n+1}2\frac{\Gamma(n+\ell+2)\Gamma(n-\ell+1)}{(n+1)!}e^{-2(n+1)\vert\tau\vert}\right) \nn\\&\hspace{-1cm} +\mathcal{O}\left(e^{-(n+4)\vert\tau\vert}\right),\label{PsinlAsympt1}\\
 \psi_{n\ell}^{(q)}(\tau\to\pm\infty)&=(\pm 1)^{n+\ell} (-1)^{n+1} \sqrt{\pi}\frac{2^{-2(n+1)} \Gamma(2+2n)}{\Gamma(n+\frac{3}{2})\Gamma(n-\ell+1)n!} \nn\\&\hspace{-1cm} e^{n\vert\tau\vert}\sum_{r=0}^{n}\;\sum_{k=0}^{r} \frac{(-1)^{r}(n-k)!\;r! (-\ell)_{k}(\ell+1)_{k}}{(k!)^2(r-k)!} e^{-2r\vert\tau\vert} \nonumber\\ 
 &\hspace{-1cm}+\mathcal{O}\left(e^{-(n+4)\vert\tau\vert}\right).\label{PsinlAsympt2}
 \end{align}
In this case, $\psi_{n\ell}^{(p)}$ and $\psi_{n\ell}^{(q)}$ behave as $e^{n\vert\tau\vert}$ at leading order, which diverge for $n>0$ and converge for $n=0$. Up to a global factor, the asymptotic behavior of $\psi_{n\ell m}^{(p)}$ and $\psi_{n\ell m}^{(q)}$ are identical up to order $e^{-(n+2)\vert\tau\vert}$ where they start to differ.
 
Regarding $\ell\ge(n+1)$ harmonics, we use the following definitions (see e.g. \cite{Mathematica}) :
\begin{align}
P_{\ell}^{n+1}(z)=&2^{-(n+1)}(1-z^2)^{(n+1)/2} \nonumber \\
&\times \sum_{k=0}^{\ell-n-1} \frac{(-\ell)_{n+1+k}(\ell+1)_{n+1+k}}{k!(n+1+k)!} \left(\frac{1-z}{2}\right)^{k};\\
Q_\ell^{n+1}(z)=&\frac{1}{2} P_\ell^{n+1}(z) \ln\left(\frac{1+z}{1-z}\right)\nn\\& +\frac{1}{2} (1-\delta_{n,-1}) \sum_{k=0}^{n} \frac{(n+1)!}{k!(n+1-k)}  \nn\\& \times P_\ell^k(z) (1-z^2)^{\frac{n+1-k}{2}}[(z-1)^{k-n-1}-(z+1)^{k-n-1}]\nn\\&- (-1)^{n+1} (1-z^2)^{(n+1)/2}\sum_{r=n+1}^{\ell} \frac{(\ell+r)!}{(r-n-1)! }\nonumber \\ 
& \times \frac{\left(\psi(\ell+1)-\psi(r+1)\right)}{2^r (\ell-r)!r!}(z-1)^{r-n-1} ,
\end{align}
where $\psi(x)$ is the digamma function and $(x)_n$ are the Pochhammer symbols. 
Asymptotically, we obtain the following behaviors: 
\begin{align}
 \psi_{n\ell}^{(p)}(\tau\to\pm\infty) &= (\pm 1)^{(n+\ell+1)}\frac{(-1)^{n+1} 2}{(n+1)!}e^{-(n+2)\vert\tau\vert} \nonumber \\ 
 &+\mathcal O\left(e^{-(n+4)\vert\tau\vert}\right) ;\label{asymptotscalPlgn}\\
 \nonumber \psi_{n\ell}^{(q)}(\tau\to\pm\infty)&=(\pm 1)^{(n+\ell)} (1-\delta_{n,-1})e^{n\vert\tau\vert} \nonumber\\
 &\hspace{-1.5cm}\times \sum_{r=0}^{n}\;\sum_{k=0}^{r} \frac{(-1)^{n+r+1}(n-k)!\;r! (-\ell)_{k}(\ell+1)_{k}}{(k!)^2(r-k)!} e^{-2r\vert\tau\vert}  \nn\\
 & \hspace{-1.5cm}+2 (\pm 1)^{(n+\ell)} e^{-(n+2)\vert\tau\vert} \frac{(-\ell)_{n+1}(\ell+1)_{n+1}}{(n+1)!} \nonumber \\
 & \hspace{-1.5cm}\times \Big(\vert\tau\vert+\frac{1}{2} H_{n+1}-  H_{\ell}\Big)\left\{1+\mathcal O\left(e^{-2\vert\tau\vert}\right)\right\}, \label{asymptotscalQlgn}
\end{align}
where $H_{n}$ is the $n$-th harmonic number. 
In this case, $\psi_{n\ell}^{(p)} \propto e^{-(n+2)\vert\tau\vert}$, $\psi_{n\ge0,\ell}^{(q)} \propto e^{n \vert\tau\vert}$ and $\psi_{-1\ell}^{(q)} \propto \vert\tau\vert e^{- \vert\tau\vert}$ at leading order. Therefore, $\psi_{n\ell}^{(p)}$ converges for all $n \geq -1$ while $\psi_{n\ell}^{(q)}$ diverges for $n\ge1$ and converges for $n=-1,0$.

Let us define the operator $\Delta_{k,0}$ for $k \in \mathbb N$, 
\begin{align}\label{scalDelta}
\Delta_{k,0} :=    \prod_{q=0}^{k-1}(\nabla^2+q(q+1)), \;\;  k \in \mathbb N_0,\quad  \Delta_{0,0} := 1. 
\end{align}
When acting on a harmonic $Y_{\ell m}$, it obeys 
\begin{align}
\Delta_{k,0} Y_{\ell m} &= \prod_{q=0}^{k-1}(-\ell(\ell+1)+q(q+1))Y_{\ell m} = (-\ell)_{k}(\ell+1)_{k} Y_{\ell m} 
\end{align}
which vanishes for all $\ell\le k-1$. 

Combining Eqs. \eqref{PsinlAsympt1}, \eqref{PsinlAsympt2}, \eqref{asymptotscalPlgn}, \eqref{asymptotscalQlgn} and using the decomposition \eqref{solscal}, we find that generic functions with $p$ parity $\psi_n^P(\tau,x^A)$ and generic functions with $q$ parity $\psi_n^Q(\tau,x^A)$, at fixed $n$, behave as 
\begin{align}
    &\psi_n^P(\tau\to\pm\infty,x^A) = &\nn\\&=e^{n\vert\tau\vert} \sum_{r=0}^{n}\;\sum_{k=0}^{r} \frac{(-1)^{r}(n-k)!\;r!\Delta_{k,0} O^{(p)n}[\psi^{(P)\pm}_n(x^A)]}{(k!)^2(r-k)!} e^{-2r\vert\tau\vert}  \nn\\&+\psi^{(P)\pm}_n(x^A)e^{-(n+2)\vert\tau\vert}+\mathcal{O}\left(e^{-(n+4)\vert\tau\vert}\right),\label{PsinPasympt}\\
 \nonumber&\psi_{n}^{Q}(\tau\to\pm\infty,x^A) &\\&=(1-\delta_{n,-1})e^{n\vert\tau\vert} \sum_{r=0}^{n}\;\sum_{k=0}^{r} \frac{(-1)^{r}(n-k)!\;r! \Delta_{k,0} \psi^{(Q)\pm}_n(x^A)}{(k!)^2(r-k)!} e^{-2r\vert\tau\vert}  \nn\\&+e^{-(n+2)\vert\tau\vert}\Big(2\vert\tau\vert (-1)^{n+1}  \frac{ \Delta_{n+1,0}}{(n+1)!} \psi^{(Q)\pm}_n(x^A) \nn\\&+ O^{(q)n}[\psi^{(Q)\pm}_n(x^A)]\Big)\left\{1+\mathcal O\left(e^{-2\vert\tau\vert}\right)\right\},\label{PsinQasympt} 
\end{align}
where $\psi^{(P)\pm}_n(x^A)$, $\psi^{(Q)\pm}_n(x^A)$ are free functions on the $S^2$. We defined the non-local operators $O^{(p)n}$ and $O^{(q)n}$ as follows
\begin{align}
    O^{(p)n}[f(x^A)] := & \sum_{\ell=0}^{+\infty}\sum_{m=-\ell}^{+\ell}(-1)^{\ell+n+1}\frac{(n+1)!}{2\Gamma(n+\ell+2)\Gamma(n-\ell+1)} \nonumber \\ 
    &\times  Y_{\ell m}(x^A)\oint_{S^2}d\Omega' \overline{Y_{\ell m}(x^{A'})}f(x^{A'}),
    \end{align}
    \begin{align}
    O^{(q)n}[f(x^A)] := &\sum_{\ell=0}^{+\infty}\sum_{m=-\ell}^{+\ell} (-1)^{n+1}\frac{\Delta_{n+1,0}}{(n+1)!}(H_{n+1} - 2 H_{\ell} )\nonumber \\ 
    &\times Y_{\ell m}(x^A)\int_{S^2}d\Omega' \overline{Y_{\ell m}(x^{A'})}f(x^{A'}).
\end{align}
For any function $f(x^A)$, $O^{(p)n}[f(x^A)]$ annihilates harmonics $\ell\ge n+1$ and $O^{(q)n}[f(x^A)]$ annihilates harmonics $\ell\le n$. Hence, $O^{(p)n}O^{(q)n}=O^{(q)n}O^{(p)n}=0$.

The free data on the sphere $\psi^{(P)\pm}_n(x^A)$ (resp.  $\psi^{(Q)\pm}_n(x^A)$) are related due to the well-defined parity on $dS_3$ of $\psi_n^P$ (resp. $\psi_n^Q$). We find the following map 
\begin{align}\label{Matchingscalarend0}
    \Upsilon^*\psi^{(P)+}_n(x^A) &=(-1)^{n+1}\psi^{(P)-} _n(x^A),\\ 
    \Upsilon^*\psi^{(Q)+}_n(x^A) &=(-1)^{n}\psi^{(Q)-} _n(x^A). \label{Matchingscalarend1}
\end{align}

From Eqs. \eqref{PsinPasympt} and \eqref{PsinQasympt}, we find that the general solution $\psi_n(\tau,x^A)$ to Eq. \eqref{generalscalarequation} behaves as 
\begin{align}
    &\psi_n(\tau,x^A) = e^{n\vert\tau\vert} \sum_{r=0}^n N^{(r,0)}_{n,0}[\psi^{(\mathbb L)\pm}_{n}(x^A)]e^{-2r\vert\tau\vert} + e^{-(n+2)\vert\tau\vert} \nonumber \\ 
&    \times \left(\psi^{(\mathbb S)\pm}_{n}(x^A)+\vert\tau\vert N^{(n+1,1)}_{n,0}[\psi^{(\mathbb L)\pm}_{n}(x^A)]\right)+o(e^{-(n+2)\vert\tau\vert})\label{PsiHomBehavior}
\end{align}
where, for $0\le r\le n$, we defined the operators 
\begin{align}
    N^{(r,0)}_{n,0}[\cdot] &:=  (1-\delta_{n,-1}) \sum_{k=0}^{r} \frac{(-1)^{r}(n-k)!\;r! \Delta_{k,0} [\cdot]}{(k!)^2(r-k)!},\label{PsiHomBehaviorfdef0}\\
    N^{(n+1,1)}_{n,0}[\cdot]&:= 2(-1)^{n+1} \frac{\Delta_{n+1,0}[\cdot]}{(n+1)!}.\label{PsiHomBehaviorfdef}
\end{align}
$\psi^{(\mathbb L)+}_n(x^A)$ and $\psi^{(\mathbb S)+}_{n}(x^A)$ are arbitrary functions on the $S^2$ and are related to $\psi^{(\mathbb L)-}_n(x^A)$, and $\psi^{(\mathbb S)-}_{n}(x^A)$ through the following antipodal matching conditions
\begin{subequations} \label{Matchingscalarend}
\begin{align}
    &\Upsilon^*N^{(n+1,1)}_{n,0}[\psi^{(\mathbb L)+}_{n}(x^A)]=(-1)^{n}N^{(n+1,1)}_{n,0}[\psi^{(\mathbb L)-}_{n}(x^A)];\label{Matchingscalarbegin}\\&
    \Upsilon^*\psi^{(\mathbb L)+}_n(x^A)-\Upsilon^*O^{(p)n}[\psi^{(\mathbb S)+}_{n}(x^A)]=(-1)^{n}\big(\psi^{(\mathbb L)-}_n(x^A) \nonumber \\
    &-O^{(p)n}[\psi^{(\mathbb S)-}_{n}(x^A)]\big);\\
    &
    \Upsilon^*\psi^{(\mathbb S)+}_{n}(x^A)-\Upsilon^*O^{(q)n}[\psi^{(\mathbb L)+}_{n}(x^A)]=(-1)^{n+1}\big(\psi^{(\mathbb S)-}_{n}(x^A) \nonumber \\
    &-O^{(q)n}[\psi^{(\mathbb L)-}_{n}(x^A)]\big). 
\end{align}
\end{subequations}
Since the operators $O^{(p)n} O^{(q)n}$ and $O^{(q)n} O^{(p)n}$ are identically zero, these antipodal relationships imply 
\begin{align}
   &
    \Upsilon^*O^{(q)n}[\psi^{(\mathbb L)+}_n(x^A)]=(-1)^{n}O^{(q)n}[\psi^{(\mathbb L)-}_n(x^A)];\label{rest}\\ 
    & \Upsilon^*O^{(p)n}[\psi^{(\mathbb S)+}_{n}(x^A)]=(-1)^{n+1}O^{(p)n}[\psi^{(\mathbb S)-}_{n}(x^A)].
\end{align}
The antipodal relations \eqref{Matchingscalarend} are obtained by removing either the $q$-parity or the $p$-parity contributions from the asymptotic data of the general solution \eqref{solscal} and subsequently using the relations \eqref{Matchingscalarend0}-\eqref{Matchingscalarend1}. 

Without imposing any assumptions on the solution $\psi_n(\tau,x^a)$, the antipodal matching conditions for the asymptotic data $\psi^{(\mathbb L)\pm}_{n}(x^A)$ and $\psi^{(\mathbb S)\pm}_{n}(x^A)$ are non-local on the sphere, whereas those for the leading behavior proportional to $\tau$ (i.e. $N^{(n+1,1)}_{n,0}[\psi^{(\mathbb L)\pm}_{n}(x^A)]$) are local. If one assumes the absence of $\ell<n+1$ harmonics, then the leading behavior as $\tau\to\pm\infty$ is driven by the $q$-parity harmonics, which leads to local antipodal matching conditions for $\psi^{(\mathbb L)\pm}_{n}(x^A)$. Conversely, if there is no $\ell\ge n+1$ harmonics, the antipodal matching conditions for the subleading asymptotic data $\psi^{(\mathbb S)\pm}_{n}(x^A)$ become local. The relationships \eqref{Matchingscalarbegin} and \eqref{rest} are equivalent for $n\ge0$ in the sense that they both provide the antipodal maps of all $\ell \geq n+1$ harmonics of $\psi^{(\mathbb L)\pm}_{n}$ \footnote{For $n=-1$, these relationships are equivalent only for $\ell\ge1$ harmonics since $O^{(q)n=-1}$ vanishes when applied on a constant.}. This can be seen by comparing the relations \eqref{PsiHomBehaviorfdef} and \eqref{rest}, which both contain operators that annihilate all $\ell < n+1$ harmonics of $\psi^{(\mathbb L)\pm}_{n}$. Eq. \eqref{Matchingscalarbegin} is simpler because it is formulated locally in terms of the asymptotic data. 

\noindent \vspace{0pt} 
\begin{center}\textit{
Inhomogeneous case   }
\end{center}
 \vspace{6pt}

The asymptotics of a general inhomogeneous solution to Eq. \eqref{InhomognenousScaleq} can be computed with the following procedure. Let $F(\tau;\bar\tau)$ be a function of the form 
\begin{align}\label{defF0}
   F(\tau;\bar\tau) = \int_{\bar\tau} ^\tau f(\tau')d\tau'
\end{align}
where the function $f(\tau)$ is known to behave as $f(\tau\to\pm\infty)= g^\pm(\tau)+o(e^{-k\vert\tau\vert})$ for $k\in\mathbb{N}$. Let us define  $G^\pm(\tau)$ and $F(\tau)$ the primitives of $g^\pm$ and $f$. We fix the ambiguity in the integration constants of $G^\pm(\tau)$ by  setting the finite part of $G^\pm(\tau)$ as $\tau \to \pm \infty$ to zero. We have 
\begin{align}\label{defF}
 F(\tau;\bar\tau)
 &=F(\tau)-F(\bar\tau)
 \nonumber\\
 &=G^\pm(\tau)+C^\pm-F(\bar\tau)
 +o(e^{-k|\tau|}),
 \quad \tau\to\pm\infty,
\end{align}
where
\begin{align}
 C^\pm:=
 \lim_{\tau\to\pm\infty}
 \left[F(\tau)-G^\pm(\tau)\right]= \text{FP}  \lim_{\tau\to\pm\infty} F(\tau).
\end{align}
It follows that
\begin{align}\label{Cjump}
 C^+-C^-
 &=
 \lim_{\lambda^+,\lambda^-\to\infty}\Big[\int_{-\lambda^-}^{\lambda^+}f(\tau)d\tau - G^+(\lambda^+)+G^{-}(-\lambda^-)\Big]
 \nonumber\\
 &:=
 \operatorname{FP}
 \int_{-\infty}^{+\infty}f(\tau)\,d\tau .
\end{align}
The finite-part (FP) prescription is needed
when the ordinary improper integral of $f$ does not converge.
Since $ A_{\ell m}^{(S)n}(\tau;\bar\tau)$ and $ B_{\ell m}^{(S)n}(\tau;\bar\tau)$ defined in Eqs. \eqref{ASscalar}-\eqref{BSscalar} take the form of $F(\tau;\bar\tau)$ as defined in Eq. \eqref{defF}, we can apply this procedure to find their asymptotics from the asymptotics of the source and then combine it with the asymptotics of the harmonics in Eq.\eqref{ScalNHSol}. 

As an illustrative example, if 
\begin{align}
    s(\tau\to\pm\infty,x^A)= s^{(0)\pm}(x^A) e^{-(n+2)\vert\tau\vert}+o(e^{-(n+3)\vert\tau\vert}),\label{sourceExample}
\end{align}
with $n > 0$ and which does not contain $\ell<(n+1)$ harmonics, then using the asymptotics \eqref{asymptotscalPlgn} and \eqref{asymptotscalQlgn} and  integrating, we find  
\begin{align}
    A_{\ell m}^{(S)n}(\tau\to\pm\infty;\bar\tau) &=(\pm1)^{n+\ell+1}\frac{\vert\tau\vert }{4}(-1)^{n+1} n! \oint_{S^2} d\Omega \, s^{(0)\pm}(x^A) \nonumber \\
    & \overline{Y_{\ell m}(x^{A})}+A^\pm_{n\ell m}(\bar \tau) + o(e^{-\vert\tau\vert})\\
    B_{\ell m}^{(S)n}(\tau\to\pm\infty;\bar\tau) &= \frac{(\pm1)^{n+\ell}(-1)^{n+1}}{2(n+1)!(2n+2)}e^{-(2n+2)\vert\tau\vert} \oint_{S^2}d\Omega   \nonumber \\ 
    &s^{(0)\pm}(x^A) \overline{Y_{\ell m}(x^{A})}+ B^\pm_{n\ell m}(\bar \tau) + o(e^{-(2n+3)\vert\tau\vert})
\end{align}
where $A^\pm_{n\ell m}(\bar \tau)$ and $B^\pm_{n\ell m}(\bar \tau)$ are integration constants. They can be written as :
\begin{align}
    &A^+_{n\ell m}(\bar \tau)=\frac{1}{2}\text{FP}\int_{-\infty}^{+\infty} d\tau\int_{S^2(\tau)} d\Omega\, s(\tau,x^A) \overline{\psi^{(q)}_{n\ell m}(\tau,x^A)}\nn\\&\qquad\qquad\quad +\bar A_{n\ell m}(\bar \tau)\\
    &A^-_{n\ell m}(\bar \tau)=-\frac{1}{2}\text{FP}\int_{-\infty}^{+\infty} d\tau\int_{S^2(\tau)} d\Omega\, s(\tau,x^A) \overline{\psi^{(q)}_{n\ell m}(\tau,x^A)}\nn\\&\qquad\qquad\quad +\bar A_{n\ell m}(\bar \tau)\\
    &B^+_{n\ell m}(\bar \tau)=-\frac{1}{2}\int_{-\infty}^{+\infty} d\tau\int_{S^2(\tau)} d\Omega\, s(\tau,x^A) \overline{\psi^{(p)}_{n\ell m}(\tau,x^A)}\nn\\&\qquad\qquad\quad +\bar B_{n\ell m}(\bar \tau)\\&B^-_{n\ell m}(\bar \tau)=\frac{1}{2}\int_{-\infty}^{+\infty} d\tau\int_{S^2(\tau)} d\Omega\, s(\tau,x^A) \overline{\psi^{(p)}_{n\ell m}(\tau,x^A)}\nn\\&\qquad\qquad\quad +\bar B_{n\ell m}(\bar \tau).
\end{align}
where $ \bar A_{n\ell m}(\bar \tau)$ and $\bar B_{n\ell m}(\bar \tau)$ are constants.

Let us consider only convergent solutions to Eq.\eqref{InhomognenousScaleq} as $\tau \to \pm\infty$ with a source \eqref{sourceExample}. The asymptotic behavior of these solutions will behave as 
\begin{align}
  &\psi_{n}(\tau\to\pm\infty,x^A) =\psi^{(P)\pm}_n(x^A)e^{-(n+2)\vert \tau \vert} \nn\\
    &+\frac{1}{2(n+1)}\Big[(\vert\tau\vert+\frac{1}{(2n+2)})s^{(0)\pm}(x^A)+\mathcal{S}^{(Q)\pm}_n(x^A)\Big] e^{-(n+2)\vert\tau\vert} \nn\\
    & +o(e^{-(n+2)\vert\tau\vert}),
\end{align}
where $\psi^{(P)\pm}_n(x^A)$ obeys the antipodal map \eqref{Matchingscalarend0} and $\mathcal{S}^{(Q)\pm}(x^A)$ is defined as 
\begin{align}
   \mathcal{S}^{(Q)\pm}_n(x^A):=& \frac{2 (-1)^{n+1}}{n!}\sum_{\ell m}(\pm1)^{\ell+n}Y_{\ell m}(x^A)\nn\\&\times \text{FP}\int_{-\infty}^{+\infty} d\tau\int_{S^2(\tau)} d\Omega'\,s(\tau,x^{A'})\overline{\psi^{(q)}_{n\ell m}(\tau,x^{A'})}.
\end{align}

In this example, we can identify the field $\psi^{(P)\pm}_n(x^A)$ from the asymptotic behavior of $\psi_n$ by first identifying $s^{(0)\pm}(x^A)$ in the asymptotics of $s(\tau,x^A)$, then subtracting $[1/(4(n+1)^2)s^{(0)\pm}(x^A)+1/(2n+2)\mathcal{S}^{(Q)\pm}_n(x^A)]e^{-(n+2)\vert \tau\vert}$ from the solution at order $e^{-(n+2)\vert \tau\vert}$ in the limit $\tau \to \pm\infty$. Also, $\mathcal{S}^{(Q)\pm}_n(x^A)$ satisfy the $q$-parity antipodal matching condition : 
\begin{align}
    \Upsilon^*\mathcal{S}^{(Q)+}_n(x^A)=(-1)^n\mathcal{S}^{(Q)-}_n(x^A) 
\end{align}

In general, for an arbitrary source, we can compute $A^{(S)n}_{\ell m}$ and $B^{(S)n}_{\ell m}$ and therefore identify the associated behavior as $\tau \to \pm \infty$. We can then subtract these contributions in order to identify functions $\hat\psi^{(\mathbb L)\pm}_{n}$ and $\hat\psi^{(\mathbb S)\pm}_{n}$, which satisfy the antipodal relationships \eqref{Matchingscalarend}. This identification is possible because we are left with a ``subtracted function'' $\hat\psi_n(\tau,x^A)$ that asymptotically behaves as a homogeneous solution, see Eqs. \eqref{PsiHomBehavior}-\eqref{PsiHomBehaviorfdef}. The antipodal relationships then imply the existence of conserved charges across spatial infinity as follows. We have 
\begin{subequations}\label{Q0}
\begin{align}
\mathcal Q_T[\hat \psi_n^{(P)}]& := \oint_{S^2}d\Omega \hat \psi_n^{(P)+}(x^A) \overline{T(x^A)}\nonumber \\&=(-1)^{n+1}\oint_{S^2}d\Omega \hat \psi_n^{(P)-}(x^A)\Upsilon^*\overline{T(x^A)} ,\label{ScalConservLawsP}\\
   \mathcal Q_T[\hat \psi_n^{(Q)}]&:= \oint_{S^2}d\Omega \hat \psi_n^{(Q)+}(x^A)\overline{T(x^A)} \nonumber \\&=(-1)^{n}\oint_{S^2}d\Omega  \hat \psi_n^{(Q)-}(x^A)\Upsilon^*\overline{T(x^A)},\label{ScalConservLawsQ}
\end{align}    
\end{subequations}
where $T(x^A)$ is an arbitrary function on the sphere and $\hat\psi_n^{(P)\pm}(x^A)$, $\hat\psi_n^{(Q)\pm}(x^A)$ are respectively the $p$-parity and the $q$-parity part in the asymptotics of $\hat\psi_n(\tau,x^A)$, i.e. 
\begin{align}\label{PsitildenP}
    \hat \psi_n^{(P)\pm}(x^A) :=  \hat\psi^{(\mathbb S)\pm}_{n}(x^A)-O^{(q)n}[\hat\psi^{(\mathbb L)\pm}_{n}(x^A)],\\
    \hat \psi_n^{(Q)\pm}(x^A) :=  \hat\psi^{(\mathbb L)\pm}_{n}(x^A)-O^{(p)n}[\hat\psi^{(\mathbb S)\pm}_{n}(x^A)].\label{PsitildenQ}
\end{align}
One could also use $\Delta_{n+1,0}\hat \psi^{(\mathbb L)\pm}_{n}(x^A)$ to define a conserved charge but this charge will be redundant with $\mathcal Q_T[\hat \psi_n^{(Q)}]$ and vanish for all $\ell\le n$ harmonics. 

Alternatively, one can obtain these conservation laws from the KG inner product. 
For generic $p$ parity function $\psi_n^{P}(\tau,x^A)$ and $q$ parity function $\psi_n^{Q}(\tau,x^A)$ that satisfy Eq. \eqref{generalscalarequation}, we have
\begin{align}
   &\lim_{\tau\to+\infty}(\hat\psi_n(\tau,x^A),\psi_n^{Q}(\tau,x^A))_{KG}\nn\\&=\frac{(n+1)!}{2}\oint_{S^2}d\Omega \hat \psi_n^{(P)+}(x^A)\overline{\psi_n^{(Q)+}(x^A)}  \\&=-\frac{(n+1)!}{2}\oint_{S^2}d\Omega \hat\psi_n^{(P)-}(x^A)\overline{\psi_n^{(Q)-}(x^A)}   \\&=\lim_{\tau\to-\infty}(\hat\psi_n(\tau,x^A),\psi_n^{Q}(\tau,x^A))_{KG}
\end{align}
and 
\begin{align}
   &\lim_{\tau\to+\infty}(\hat\psi_n(\tau,x^A),\psi_n^{P}(\tau,x^A))_{KG}\nn\\&=-\frac{(n+1)!}{2}\oint_{S^2}d\Omega \hat \psi_n^{(Q)+}(x^A)\overline{\psi_n^{(P)+}(x^A)}  \\&=\frac{(n+1)!}{2}\oint_{S^2}d\Omega   \hat\psi_n^{(Q)-}(x^A) \overline{\psi_n^{(P)-}(x^A)}\\&=\lim_{\tau\to-\infty}(\hat\psi_n(\tau,x^A),\psi_n^{P}(\tau,x^A))_{KG}.
\end{align}
where $\psi_n^{(P)\pm}(x^A)$ and  $\psi_n^{(Q)\pm}(x^A)$ are the functions appearing in Eqs. \eqref{PsinPasympt}-\eqref{PsinQasympt} and $\hat\psi_n^{(P)\pm}(x^A)$, $\hat\psi_n^{(Q)\pm}(x^A)$ correspond to Eqs. \eqref{PsitildenP}-\eqref{PsitildenQ}. We used the fact that the asymptotics \eqref{PsinPasympt} and \eqref{PsinQasympt} only start to differ at order $e^{-(n+2)\vert \tau \vert}$ and the identity
\begin{align}
    \oint_{S^2}d\Omega O^{(q)n}[f(x^A)] O^{(p)n}[g(x^A)]=0
\end{align}
for any functions $f(x^A)$ and $g(x^A)$. 

Using the antipodal matching conditions given in Eqs. \eqref{Matchingscalarend0}-\eqref{Matchingscalarend1}, we recover the conservation laws \eqref{Q0} with $T(x^A)$ being proportional to $\psi_n^{(P)+}(x^A)$ or $\psi_n^{(Q)+}(x^A)$.  
\section{Vector harmonics}

Let us now derive properties of vector harmonics in $dS_3$, which appear notably as solutions to the conformal Killing equation on $dS_3$. The vector harmonics in $dS_3$ are closely related to the vector harmonics of $S^3$, see e.g. \cite{Higuchi:1986wu,Marolf_2009,Higuchi_2003}.

An arbitrary vector on dS$_3$ can be expressed as the sum of a transverse vector (i.e. divergence-free vector) and a longitudinal vector (i.e. which is the gradient of a scalar). We will treat them separately in what follows. In this section, we consider vector fields $V^{a}$ which obey
\begin{align}
    (\Box - \alpha)V^{a} = S^a\label{generalVectorEqu}
\end{align}
with $\alpha\in \mathbb{R}$ and $S^a$ an arbitrary vector on $dS_3$.

Similarly to the scalar case, we define an inner product between vector fields as follows. Let $V_1^a$ and $V_2^a$ be two complex vectors on $dS_3$, we define the vector KG inner product 
\begin{align}
(V_1,V_2)_{KG}^V:= \oint_{S^2(\tau)} d\Omega\, n^a  \left(V_1^{b} \mathcal{D}_a \overline{V_{2\;b}} - \overline{V_{2\;b}}\mathcal{D}_a V_1^{b} \right).
\end{align}
For $V_1^a$, $V_2^a$ obeying Eq. \eqref{generalVectorEqu} with $S^a=0$ and the same value of $\alpha$, the vector KG inner product is conserved after using the equations of motion.

\subsection{Longitudinal vector harmonics}
Longitudinal vector harmonics are built from taking the gradient of scalar harmonics
\begin{align}
    V_a^{(L,c)\:n\ell m} := \partial_a \psi_{n\ell m}^{(c)},\qquad c=p,q. \label{longitudinalVectHarm}
\end{align}
They obey the following equation
\begin{align}
     \Box V_a^{(L,c)\:n\ell m} 
  & = - [(n+1)^2-3]V_a^{(L,c)\:n\ell m} \label{eqL}
\end{align}
where $c=p,q$. We used the commutator \eqref{commuBoxPartialscal} and Eq. \eqref{generalscalarequation} to derive the equality.  Under the action of the hyperboloidal antipodal map they transform as 
\begin{align}
      \Upsilon_{\mathcal{H}}^*[V_a^{(L,p)\:n\ell m}] &= (-1)^{n+1} V_a^{(L,p)\:n\ell m},\\ \Upsilon_{\mathcal{H}}^*[V_a^{(L,q)\:n\ell m}] &= (-1)^{n} V_a^{(L,q)\:n\ell m}.
\end{align}
 
For $n=0$, $\partial_a \psi_{000}^{(q)} = 0$ and therefore we are missing a solution to Eq. \eqref{eqL}. The $q$ parity solution with $\ell=0$, $n=0$ can be constructed from an inhomogeneous solution to the equation $\Box\psi_0=\psi_{000}^{(q)}$. Solving this equation using the method described previously, we define
\begin{align}
    M(\tau,x^A) &:= \Big(\frac{1}{8}(\tau+\cosh\tau\sinh\tau)\psi_{000}^{(p)}+\frac{1}{2}\sinh^2\tau\psi_{000}^{(q)}\Big)\nn\\&=\frac{1}{8 \sqrt{\pi}}\tau\tanh\tau \label{psi0}
\end{align}
This scalar defines the $n=0$, $\ell=0$ $q$-parity vector  as
\begin{align}
    V^{(L,q)000}_a := \partial_a M(\tau,x^A).\label{longitudinalVectHarmn0}
\end{align}
We can compute the divergence and the curl of longitudinal harmonics 
\begin{align}
    \mathcal{D}^a V_{a}^{(L,c)n\ell m} &= -\Big(n-\frac{1}{2}\delta_{n,0}\delta_{cq}\delta_{\ell,0}\Big)(n+2) \psi_{n\ell m}^{(c)}, \\ 
    \text{Curl}(V^{(L,c)n\ell m}_a)\ &= 0\label{CurlDphizero}
\end{align}
where we define $\text{Curl}(V_a):=\epsilon_{abc}\mathcal{D}^b V^c$.
The longitudinal harmonics also satisfy orthogonality properties under the vector KG inner product : 
\begin{align}
    (V^{(L,c)n\ell m},V^{(L,d)n\ell' m'})^V_{KG} = (n(n+2)-\delta_{n,0}\delta_{\ell,0})\delta_{\ell \ell'}\delta_{m m'}\epsilon_{cd}.
\end{align}
For $n=0$, the vector KG inner product is trivial except for $\ell=0$ and the longitudinal vectors are also divergence-free for $\ell\ge1$ \cite{Marolf_2009}. 

As mentioned previously, an arbitrary vector on $dS_3$ can be decomposed into a transverse part and a longitudinal part. Since Curl($\mathcal{D}_a\phi$)=0 for any scalar $\phi$ and using Lemma \ref{lemma:V2} (see next Section \ref{sec:TransverseVectors}), we deduce that, for any vector $V_a$ in $dS_3$, Curl($V_a$)=0 if and only if $V_a$ is longitudinal. Hence, the solution to the set of equations
\begin{equation}
   (\Box+[(n+1)^2-3]) V_a= 0, \quad  \text{Curl}(V_a)=0,
\end{equation}
can be decomposed as follows 
\begin{equation}
    V_a = \sum_{\ell m} a_{\ell m} V^{(L,p)n\ell m} + b_{\ell m} V^{(L,q)n\ell m} ,
\end{equation}
where $a_{\ell m}, b_{\ell m}$ are constants.
\subsection{Transverse vector harmonics}\label{sec:TransverseVectors}

In the literature \cite{Higuchi:1986wu}, transverse vector harmonics on $dS_3$ are defined from 
\begin{align}
    \quad \Box V^{(T)n}_a  = -\left[ (n+1)^2-2\right] V^{(T)n}_a,\qquad \mathcal{D}^a V_{a}^{(T)n} = 0.
\label{generalEquationvectorsHarms}
\end{align}

Given the wave equation \eqref{generalEquationvectorsHarms} we can assume $n \geq -1$ without loss of generality. We are mostly interested in $n \in \mathbb Z$   for physical applications. Note that integer values of $n$ in Eq. \eqref{eqL} do not correspond to integer values of $n$ in Eq. \eqref{generalEquationvectorsHarms}.

In order to find the general solution to Eq. \eqref{generalEquationvectorsHarms} we start with the decomposition in spherical harmonics 
\begin{align}
   V_a^{\ell m}=  \left(
\begin{array}{c}
f_1(\tau) Y_{\ell m}(\theta,\phi) \\\\
f_2(\tau)\mathring{V}^{(E)\ell m}_\theta(\theta, \phi)+ f_3(\tau) \mathring{V}^{(B)\ell m}_\theta(\theta, \phi) \\\\
f_2(\tau)\mathring{V}^{(E)\ell m}_\phi(\theta, \phi) + f_3(\tau) \mathring{V}^{(B)\ell m}_\phi(\theta, \phi) \\
\end{array}
\right)\label{generaldivFreevector}
\end{align} 
where $\mathring{V}^{(E)\ell m}_A(\theta, \phi)$ and $\mathring{V}^{(B)\ell m}_A(\theta, \phi)$ are the normalized electric parity (E) and magnetic parity (B) vector harmonics on the sphere, defined for $\ell\ge1$: 
\begin{align}
    \mathring{V}^{(E)\ell m}_A(\theta, \phi) &:=  \frac{1}{\sqrt{\ell(\ell+1)}}\partial_A Y_{\ell m}(\theta,\phi),\\
    \mathring{V}^{(B)\ell m}_A(\theta, \phi) &:=  \frac{1}{\sqrt{\ell(\ell+1)}}\epsilon_{AB}\partial^B Y_{\ell m}(\theta,\phi). 
\end{align}
For $\ell=0$, we take $\mathring{V}^{(E)00}_A$ and $\mathring{V}^{(B)00}_A$ to vanish in Eq. \eqref{generaldivFreevector}. 
The eigenvalues of $\mathring{V}^{(E)\ell m}_A$ and $\mathring{V}^{(B)\ell m}_A$ under the action of parity $\Upsilon(\theta,\phi)=(\pi-\theta,\phi+\pi)$ is $(-1)^{\ell }$ and $(-1)^{\ell+1}$, respectively. The normalization condition is enforced using the scalar product as 
\begin{align}
\oint_{S^2}d\Omega \mathring{V}^{(E)\ell m}_A \overline{\mathring{V}^{(E)\ell' m'\;A}}=\oint_{S^2}d\Omega  \mathring{V}^{(B)\ell m}_A \overline{\mathring{V}^{(B)\ell' m'\;A}}=\delta_{\ell\ell'}\delta_{mm'}.     
\end{align}

Let us now impose that the divergence vanishes,   $\mathcal{D}^a V_{a}^{(T)\:\ell m} = 0$. In the case $\ell =0$, only one harmonic exists. This harmonic is also longitudinal and reads 
\begin{align}\label{eq:75}
    V_a^{(E,p)\:-1\,00} := \frac{1}{2\sqrt{\pi}} \mathcal{D}_a \tanh(\tau)=\frac{1}{2}\mathcal{D}_a \psi_{000}^{(p)}(\tau).
\end{align}
It obeys Eq. \eqref{generalEquationvectorsHarms} with $n=-1$.
This demonstrates the following lemma:
\begin{lemma}\label{lemma:V1}
Let $V_a$ be a vector field on $dS_3$ satisfying
\begin{align}
\left(\Box-\alpha\right)V_a=0,
\qquad
\mathcal{D}_a V^a=0.
\end{align}
If $\alpha\neq 2$, then $V_a$ has no $\ell=0$ harmonic.
\end{lemma}
This lemma was implicitly used in the analysis of \cite{Fuentealba:2022xsz,Fuentealba:2025ekj}. 

For $\ell \geq 1$, the divergence-free condition $\mathcal{D}^a V_{a}=0$ implies 
\begin{align}
   f_2(\tau) = \frac{-1}{\sqrt{\ell(\ell+1)}} \cosh^2\tau \left(\partial_\tau + 2 \tanh\tau \right)f_1(\tau). 
\end{align}
Since $f_1$ and $f_3$ are arbitrary, there are 2 sets of harmonics. We choose a basis close to the one defined on $S^3$ \cite{Marolf_2009} by setting 
\begin{align}
    &f_1(\tau) = \sqrt{\ell(\ell+1)}\sech\tau f(\tau) ,  \\
   &f_3(\tau) = \cosh\tau \,\tilde f(\tau)
\end{align}
which implies $f_2(\tau) =- \cosh\tau \left(\partial_\tau +  \tanh\tau \right)f(\tau)$. 

The two families of functions $f$ and $\tilde f$ are distinguished by the parity over the sphere of the resulting solutions. We define the electric parity (E) vector harmonics for $\ell \geq 1$ as 
\begin{align}
   V_a^{(E)\:\ell m} (f)=  \left(
\begin{array}{c}
\sqrt{\ell(\ell+1)} \sech\tau f(\tau)Y_{\ell m}(\theta,\phi) \\\\
-\cosh\tau \left(\partial_\tau + \tanh\tau\right)f(\tau)\mathring{V}^{(E)\ell m}_\theta(\theta, \phi)\\\\
-\cosh\tau\left(\partial_\tau + \tanh\tau\right)f(\tau)\mathring{V}^{(E)\ell m}_\phi(\theta, \phi) \\
\end{array}
\right),\label{VE}
\end{align} 
and the magnetic parity (B) harmonics for $\ell \geq 1$ as
\begin{align}
   V_a^{(B)\:\ell m}(\tilde{f})=  \left(
\begin{array}{c}
0 \\\\
\cosh\tau\tilde{f}(\tau) \mathring{V}^{(B)\ell m}_\theta(\theta, \phi)\\\\
\cosh\tau\tilde{f}(\tau)\mathring{V}^{(B)\ell m}_\phi(\theta, \phi) \\
\end{array}
\right).\label{VB}
\end{align}
Therefore, a generic divergence-free vector with $\ell\ge1$ harmonics can be written as 
\begin{align}
   V_a^{(T)}(\tau,x^A)=  \left(
\begin{array}{c}
-\nabla^B N^{(E)}_B \sech\tau \\\\
 -\cosh\tau \left(\partial_\tau + \tanh\tau\right)N^{(E)}_\theta +  \cosh\tau N^{(B)}_\theta\\\\
 -\cosh\tau \left(\partial_\tau + \tanh\tau\right)N^{(E)}_\phi +  \cosh\tau N^{(B)}_\phi \\
\end{array}
\right)\label{VDFGen}
\end{align} 
with 
\begin{align}
    N^{(E)}_A(\tau,x^A) = \nabla_A f(\tau,x^A),\quad  N^{(B)}_A(\tau,x^A)=\epsilon_{AB}\nabla^B\tilde f(\tau,x^A).\label{NAVDFGen}
\end{align}

Imposing that the transverse vector obeys Eq. \eqref{generalEquationvectorsHarms} exactly amounts to requiring that $f(\tau,x^A)$ and $\tilde f(\tau,x^A)$ obey the scalar equation \eqref{generalscalarequation}. Hence, both $f(\tau)$ and $\tilde{f}(\tau)$ are linear combinations of $\psi_{n\ell}^{(p)}(\tau)$ and $\psi_{n\ell}^{(q)}(\tau)$, see definitions \eqref{psi(p)nlm} - \eqref{psi(q)nlm}. We shall therefore distinguish the $p$ versus $q$ character of a vector harmonic in addition to its electric or magnetic character. In summary, there are 4 types of $\ell \geq 1$ transverse vector harmonics. For $n\ge 0$, we define 
\begin{align}
    V_a^{(E,p)\;n\ell m} & := \frac{1}{n+1}V_a^{(E)\:\ell m}(f=\psi_{n\ell}^{(p)}),\\ \label{TransverseVectorHarmp}V_a^{(B,p)\;n\ell m} & := V_a^{(B)\:\ell m}(\tilde{f}=\psi_{n\ell}^{(p)})
\end{align}
and
\begin{align}
    V_a^{(E,q)\;n\ell m} & := -\frac{1}{n+1} V_a^{(E)\:\ell m}(f=\psi_{n\ell}^{(q)}),\\  V_a^{(B,q)\;n\ell m} & := V_a^{(B)\:\ell m}(\tilde{f}=\psi_{n\ell}^{(q)})\label{TransverseVectorHarmq} . 
\end{align}

For $n=-1$, the $\ell\ge1$ transverse vector harmonics correspond to longitudinal harmonics:   
\begin{align}
    V_a^{(E)\:\ell m}(f=\psi_{-1\ell}^{(p)}) &= \sqrt{\ell(\ell+1)}V^{(L,p)0\ell m},\label{TransverseVectLongVect1} \\  V_a^{(E)\:\ell m}(f=\psi_{-1\ell}^{(q)}) &=\frac{1}{\sqrt{\ell(\ell+1)}} V^{(L,q)0\ell m}.\label{TransverseVectLongVect2}
\end{align}
Furthermore, starting from Eqs. \eqref{VE}-\eqref{VB} and imposing the condition $\text{Curl}(V_a)=0$, we find that the only solutions are a linear combination of 
\begin{equation}
    f(\tau)=\psi^{(c)}_{-1\ell}(\tau), \qquad \tilde f(\tau)=0,
\end{equation}
for $c=p,q$, $\ell\ge1$. This fact, combined with Eqs. \eqref{eq:75}, \eqref{TransverseVectLongVect1} and \eqref{TransverseVectLongVect2} leads to the following lemma 
\begin{lemma}\label{lemma:V2}
Any curl-free and divergence-free vector $V_a$ on $dS_3$ obeys
\begin{align}
   \left(\Box - 2\right)V_a = 0
\end{align}
and it can be expressed in terms of a scalar $\phi$ obeying $\Box\phi=0$ as 
\begin{equation}
    V_a=\mathcal{D}_a\phi.
\end{equation}
The scalar $\phi$ is ambiguous under a shift by a constant. 
\end{lemma}
 
For $n=-1$, we define 
 \begin{align}
    V_a^{(E,p)\;-1\ell m} & := \frac{1}{\sqrt{\ell(\ell+1)}}V_a^{(E)\:\ell m}(f=\psi_{-1\ell}^{(p)}), \\ 
    V_a^{(B,p)\;-1\ell m} & := V_a^{(B)\:\ell m}(\tilde{f}=\psi_{-1\ell}^{(p)})
\end{align}
and
\begin{align}
    V_a^{(E,q)\;-1\ell m} &:= \frac{1}{\sqrt{\ell(\ell+1)}}V_a^{(E)\:\ell m}(f=\psi_{-1\ell}^{(q)}),\\ 
    V_a^{(B,q)\;-1\ell m} &:= V_a^{(B)\:\ell m}(\tilde{f}=\psi_{-1\ell}^{(q)}). 
\end{align}

In summary, the general solution to Eq. \eqref{generalEquationvectorsHarms} can be decomposed as 
\begin{align}
    V^{(T)n}_a = &\delta_{n,-1} a_{0} V_a^{(E,p)\:-1\,00}\nn \\&+ \sum_{\ell \ge 1,m} \bigg(a_{\ell m}^{(E,p)}  V^{(E,p)n\ell m} +a_{\ell m}^{(B,p)}  V^{(B,p)n\ell m}\nn \\ 
    &+a_{\ell m}^{(E,q)}  V^{(E,q)n\ell m} +a_{\ell m}^{(B,q)}  V^{(B,q)n\ell m} \bigg),\label{expVT}
\end{align}
where $a_{0}$, $a_{\ell m}^{(E,p)}$, $a_{\ell m}^{(E,q)}$, $a_{\ell m}^{(B,p)}$, $a_{\ell m}^{(B,q)}$ are constants. 

The transformation laws under parity on the hyperboloid of the transverse harmonics are 
\begin{align}
      \Upsilon_{\mathcal{H}}^*[V_a^{(E,p)\;n\ell m}] &= (-1)^{n} V_a^{(E,p)\;n\ell m},\\  \Upsilon_{\mathcal{H}}^*[V_a^{(E,q)\;n\ell m}] &= (-1)^{n+1} V_a^{(E,q)\;n\ell m},\\
      \Upsilon_{\mathcal{H}}^*[V_a^{(B,p)\;n\ell m}] &= (-1)^{n} V_a^{(B,p)\;n\ell m},\\ \Upsilon_{\mathcal{H}}^*[V_a^{(B,q)\;n\ell m}] &= (-1)^{n+1} V_a^{(B,q)\;n\ell m},
\end{align}
and they obey the following duality properties
\begin{align}\label{curlpV}
    &\epsilon_a^{\; bc}\mathcal{D}_b V_c^{(E,p)\;n\ell m} = (n+1) V_a^{(B,p)\;n\ell m},\\ 
&    \epsilon_a^{\; bc}\mathcal{D}_b V_c^{(B,p)\;n\ell m} =-(n+1+\delta_{n,-1}\sqrt{\ell(\ell+1)})  V_a^{(E,p)\;n\ell m}, 
\\&\epsilon_a^{\; bc}\mathcal{D}_b V_c^{(E,q)\;n\ell m} = -(n+1) V_a^{(B,q)\;n\ell m},\\ 
& \epsilon_a^{\; bc}\mathcal{D}_b V_c^{(B,q)\;n\ell m} =(n+1-\delta_{n,-1}\sqrt{\ell(\ell+1)})  V_a^{(E,q)\;n\ell m}.
\end{align}
The KG inner product of the transverse vector harmonics is 
\begin{align}
   (V^{(D,c)n\ell m},V^{(D',d)n\ell' m'})_{KG}^V =\delta_{\ell\ell'} \delta_{mm'} \delta_{DD'}\epsilon_{cd}(1-\delta_{n,-1}\delta_{DE}).\label{KGproductTransV}
\end{align}
As we can see, it is zero for $n=-1$, $D=E$, which corresponds to the case where transverse vectors are also longitudinal.

Let us now construct an alternative inner product for transverse vector which will be useful later. 
Let $V_a^{(T)}$ be a transverse vector with $V_a^{(E)}$ its E parity piece and $V_a^{(B)}$ its B parity piece. We can define the two scalars  
\begin{align} 
 \psi^{E}_V &:=  \cosh\tau \, n^a V^{(T)}_a =\cosh\tau \, n^a V_a^{(E)} \nn\\&= -\nabla^2 f(\tau,x^A)+C\sech\tau ,\label{PsiEVect}\\
 \psi^{B}_V &:=  \cosh\tau \, n^a \text{Curl}(V^{(T)}_a) = \cosh\tau \, n^a \text{Curl}(V^{(B)}_a) = \nabla^2 \tilde{f}(\tau,x^A)\label{PsiBVect}
\end{align}
where $f(\tau,x^A)$ and $\tilde{f}(\tau,x^A)$ are the functions appearing in Eqs. \eqref{VDFGen} and \eqref{NAVDFGen} and $C$ is a constant corresponding to the $\ell=0$ piece \eqref{eq:75}.

If $V_a$ satisfies an equation of the form \eqref{generalVectorEqu}, then the associated scalars satisfy 
\begin{align}
    (\Box -\alpha+1) \psi^D_V =\psi^D_S,\qquad D=E,B\label{scalarTransvVectEqu}
\end{align}
after using the property $[\epsilon^{abc}\mathcal{D}_b,\Box]V_c=0$ and where $\psi^D_S$ are the associated scalars \eqref{PsiEVect}-\eqref{PsiBVect} to the source $S_a$ in \eqref{generalVectorEqu}. In particular, the scalars $\psi^{E(c)}_{n\ell m}:=\cosh\tau n^a V^{(E,c)n\ell m}_a $ and $\psi^{B(c)}_{n\ell m}:=\cosh\tau \, n^a \text{Curl}(V^{(B,c)n\ell m}_a) $ are proportional to the scalar harmonics $\psi^{(c)}_{n\ell m}$. 

The KG products of the scalars associated with the E/B vector harmonics defined for $\ell \geq 1$ are
\begin{align}
    &(\psi^{E(c)}_{n\ell m},\psi^{E(c')}_{n\ell' m'})_{KG} \nonumber\\ 
    &\hspace{0.5cm}=-\frac{\ell(\ell+1)}{(n+1)^2} (V^{(E,c)n\ell m},V^{(E,c')n\ell' m'})_{KG}^V,\quad n\ge0,\label{KGVEVE}\\
    &(\psi^{B(c)}_{n\ell m},\psi^{B(c')}_{n\ell' m'})_{KG} \nonumber\\ 
    &\hspace{0.5cm}=\ell(\ell+1)(V^{(B,c)n\ell m},V^{(B,c')n\ell' m'})_{KG}^V ,\label{KGVBVB}\\
 &(\psi^{B(c)}_{n\ell m},\psi^{E(c')}_{n\ell' m'})_{KG}\nonumber\\ 
    &\hspace{0.5cm}=\frac{\ell(\ell+1)\delta_{\ell\ell'}\delta_{mm'} ((1-\delta_{c,c'})(1-\delta_{n,-1})+\delta_{n,-1}\,\epsilon_{c,c'})}{n+1-\delta_{n,-1}\sqrt{\ell(\ell+1)}} ,\\
    &(\psi^{E(c)}_{-1\ell m},\psi^{E(c')}_{-1\ell' m'})_{KG}=\delta_{\ell\ell'} \delta_{mm'} \epsilon_{cc'},
\end{align} 
with $c=p,q$. The vector KG inner products in Eqs. \eqref{KGVEVE}-\eqref{KGVBVB} are given by Eq. \eqref{KGproductTransV}. 

For $n\ge0$, there is a bijection between the phase space of a vector satisfying Eq. \eqref{generalEquationvectorsHarms} and two scalars with no $\ell=0$ harmonics satisfying Eq. \eqref{generalscalarequation}. The vector KG inner product and the scalar inner product with scalars built from vectors are equivalent. They can be used to extract the coefficients in the decomposition \eqref{expVT}.

For $n= -1$, there is instead a bijection between the phase space of a vector satisfying Eq. \eqref{generalEquationvectorsHarms} and two scalars with no $\ell=0$ harmonics satisfying Eq. \eqref{generalscalarequation}, accompanied with one real or complex number (corresponding to the single $\ell=0$ harmonic). For the B parity vectors, one can use either the scalar KG inner product with the scalars $\psi^B_V$ or the vector KG inner product as an inner product. For the E modes, the vector KG inner product vanishes but the inner product between the scalars built from vectors can be defined instead for $\ell\ge1$. For $\ell=0$, the vector $V^{(E,p)-100}\propto V^{(L,p)000}$ has a non vanishing vector KG inner product with $V^{(L,q)000}_a$ which is equivalent to the non vanishing scalar inner product between $\psi^{E}_{V^{(E,p)-100}}$ and $\psi_{-100}^{(q)}$.


The vector KG inner product and the scalar inner product with scalars built from vectors are two ways of defining conserved quantities with transverse vectors that satisfy a homogeneous equation of the form \eqref{generalVectorEqu} with $S^a=0$. There is also a third way to construct conserved quantities. For any transverse vector, we have that  
\begin{equation} \label{electricCharge}
    Q_V:= \oint_{S^2(\tau)} d\Omega\, n^a V_a
\end{equation}
is conserved. Because of orthogonality of harmonics on the sphere, this quantity is non zero only for transverse vector containing $\ell=0$ harmonics. In particular, we have 
\begin{equation}
    Q_{V^{(E,p)-1,0,0}}=\oint_{S^2(\tau)} d\Omega\, n^a V^{(E,p)-1,0,0}_a =  2 \sqrt{\pi}.
\end{equation}
Since it extracts the $\ell=0$ harmonics of transverse vectors, this quantity is equivalent to $(\psi^{E}_V,\psi_{-100}^{(q)})_{KG}$ or $(V^a,V^{(L,q)000})_{KG}^V$. After using Eq. \eqref{eq:75}, this quantity can also be alternatively defined (up to an overall factor) as the scalar KG product between $\psi^{(p)}_{000}$ and the constant $\psi^{(q)}_{000}$. 

More generally, any longitudinal vector
$  V_a=\mathcal{D}_a\phi$,
 with $\phi$ an arbitrary scalar, satisfies 
\begin{equation}
    Q_V=-(\phi,1)_{KG}\propto(\phi,\psi^{(q)}_{000})_{KG},
\end{equation}
The charge \eqref{electricCharge} is precisely the structure used to define the electric charge at spatial infinity, see e.g. \cite{Perng_1999}.

\subsection{Killing and conformal Killing vectors of dS$_3$}
A special case of the vector harmonics are the Killing and conformal Killing vectors of dS$_3$ which are well studied in the literature, see e.g. \cite{Compere:2011db,Compere:2023qoa}. Conformal Killing vectors obey 
\begin{align}
    \mathcal{D}_a \hat{\chi}_b + \mathcal{D}_b \hat{\chi}_a = \frac{2}{3} \mathcal{D}_c \hat{\chi}^c q_{ab}, 
\label{ConformalKillingEquation}
\end{align}
while Killing vectors $\chi_a$ obey 
\begin{align}
    \mathcal{D}_a \chi_b + \mathcal{D}_b \chi_a = 0 \label{KillingEquation}.
\end{align}
Equivalently, Killing vectors are the divergence-free conformal Killing vectors. They form the algebra $so(3,1)$ under the Lie bracket, which is isomorphic to the Lorentz algebra. Taking the divergence of the Killing equation we find
\begin{align}
    0=\Box \chi_b + \mathcal{D}^a\mathcal{D}_b \chi_a = (\Box + 2 ) \chi_b.
\end{align}
Killing vectors therefore obey Eq. \eqref{generalEquationvectorsHarms} with $n=1$. Regarding conformal Killing vectors, they can be split into a transverse part and a longitudinal part. The transverse piece corresponds to a Killing vector and the longitudinal piece can be written as $\hat{\chi}_a=\mathcal{D}_aH$, where H is a scalar satisfying 
\begin{equation}\label{HlogEquation}
    \mathcal{D}_a\mathcal{D}_bH+q_{ab}H =0.
\end{equation}
We will denote as $\hat{\chi}_a$ the proper conformal Killing vectors which are longitudinal conformal Killing vectors. 

The Killing equation implies that only $\ell=1$ harmonics and $q$ harmonics appear. Hence, the 6 Killing vectors of $dS_3$ are the $V_a^{(E,q)\;11 m}$ and $V_a^{(B,q)\;11 m}$, $m=-1,0,1$ harmonics. The conformal Killing equation for proper conformal Killing vectors is equivalent to Eq. \eqref{HlogEquation} \footnote{More precisely, the conformal Killing equation is equivalent to $ \mathcal{D}_a\mathcal{D}_bH+q_{ab}H =c q_{ab}$, where $c$ is a constant. The general solution is $H=c+\psi_1^{(q)}$ with $\psi_1^{(q)}$ a linear combination of $n=1$, $\ell =0,1$ $q$-parity scalar harmonics. Since $H$ is defined up to an additive constant through the relation $\hat\chi^a=\mathcal{D}^aH$, we may set $c=0$.}. The solutions are therefore the $q$-parity scalar harmonics, explicitly given by Eq. \eqref{psi(q)nlm} with $n=1$ and $\ell =0,1$. Hence, there are 4 proper conformal Killing vectors of $dS_3$, corresponding to the longitudinal harmonics $V_a^{(L,q)\;10 0}$ and $V_a^{(L,q)\;11 m}$, with $m=-1,0,1$. 

\subsection{Inhomogeneous solutions}

Let $ V^{n}_a$ be a vector in $dS_3$ satisfying the equation
\begin{align}
    \quad \Box V^{n}_a +\left[ (n+1)^2-2\right] V^{n}_a &= S^{n}_a(\tau,x^A), \label{NHGeneralVectEqu0}\\  \mathcal{D}^aV^n_a &= S^n(\tau,x^A), \label{NHGeneralVectEqu}
\end{align}
with $S^{n}_a(\tau,x^A)$ and $S^n(\tau,x^A)$ being arbitrary sources. Taking the divergence of Eq. \eqref{NHGeneralVectEqu0}, $S^{n}_a(\tau,x^A)$ must satisfy
\begin{align}
  \mathcal{D}^aS^n_a= (\Box +(n+1)^2)S^n(\tau,x^A).
\end{align}
These equations appear notably in the description of the electromagnetic field at spatial infinity \cite{Fuentealba:2025ekj}. 

The generic solution to Eqs. \eqref{NHGeneralVectEqu0}-\eqref{NHGeneralVectEqu} can be built as follows. For $n\ge0$, we define the following vector 
\begin{align}
    W^{(T)n}_a &:=  -\frac{1}{2}\text{Curl}(\text{Curl}(V_a^n)) =  V^{n}_a +\frac{1}{2} \left(\mathcal{D}_a \mathcal{D}_b V^{n\;b}-\Box V_a^n\right) \nonumber\\ 
    &=\frac{(n+1)^2}{2}V^{n}_a +\frac{1}{2} \left(\mathcal{D}_a S^n-S_a^{n}\right),\label{DecompostionVect}
\end{align}
By construction, the vector $W^{(T)n}_a$ is transverse and the vector $V^{n}_a$ can be constructed once $W^{(T)n}_a$ is known. For $n=-1$, Eqs. \eqref{NHGeneralVectEqu0}-\eqref{NHGeneralVectEqu} are invariant under the shift $V^{n=-1}_a\to V^{n=-1}_a+\mathcal{D}_a\phi$ with $\phi$ a scalar obeying $\Box\phi=0$ and the construction \eqref{DecompostionVect} does not allow to reconstruct $V^{n=-1}_a$. Therefore, we define $W^{(T)n=-1}_a$ as being the transverse part in the decomposition
\begin{equation}
    V^{n=-1}_a = W^{(T)n=-1}_a + W^{(L)}_a(\phi^{(S)}_0),\label{Vnm1Decomp}
\end{equation}
where the longitudinal solution $W^{(L)}_a(\phi^{(S)}_0):=\mathcal{D}_a\phi^{(S)}_0
$ must be found by solving
\begin{align}
    \Box\phi^{(S)}_0 = S^{n=-1}(\tau,x^A). \label{BoxPhiS}
\end{align}

In summary, for all cases $n+1\in\mathbb N$, we defined a transverse vector $W^{(T)n}_a$ built from a vector $V^{n}_a$ satisfying Eqs. \eqref{NHGeneralVectEqu0}-\eqref{NHGeneralVectEqu}. Combining those definitions with the fact that $[\text{Curl},\Box]V_a=0$ for any vector $V_a$ in $dS_3$ and using the commutator \eqref{commuBoxPartialscal}, we find that $W^{(T)n}_a$ must satisfy the equations
\begin{align}
    \quad \Box W^{(T)n}_a +\left[ (n+1)^2-2\right] W^{(T)n}_a &= S^{(T)n}_a(\tau,x^A),\label{GeneralNHDFVectEqu0} \\ 
    \mathcal{D}^a W_{a}^{(T)n} &= 0\label{GeneralNHDFVectEqu}
\end{align}    
with
\begin{align}
    S^{(T)n}_a(\tau,x^A):= \left\{
    \begin{array}{ll}
       S^{n}_a +\frac{1}{2} \left(\mathcal{D}_a \mathcal{D}_b S^{n\;b}-\Box S_a^{n}\right) & \mbox{for } n\ge0; \\&\\
       S^{n=-1}_a - \mathcal{D}_aS^{n=-1}  & \mbox{for } n=-1,
    \end{array}
    \right.
\end{align}
 is a transverse vector on $dS_3$.

This set of equations can be converted into two scalars equations (see Eqs. \eqref{PsiEVect}-\eqref{scalarTransvVectEqu}). We define 
\begin{align}\label{eq133}
    \omega^E_n :=  \cosh\tau \, n^a W^{(T)n}_a,\qquad \omega^B_n:=\cosh\tau \, n^a \text{Curl}(W^{(T)n}_a),
\end{align}
and
\begin{align}\label{sigmaE}
    \sigma^E_n := \cosh\tau \, n^a S^{(T)n}_a,\qquad  \sigma^B_n:=\cosh\tau \, n^a \text{Curl}(S^{(T)n}_a),
\end{align}
which leads to the equations 
\begin{align}\label{sigmaB}
    \quad (\Box+n(n+2)) \omega^E_n &=  \sigma^E_n,\\ 
    (\Box +n(n+2) )\omega^B_n &=  \sigma^B_n . \label{GeneralVectScalEqu}
\end{align}
We recover the inhomogeneous equation \eqref{InhomognenousScaleq} we studied in the scalar case. The solutions are 
\begin{align}
    \omega^E_{n} &= \sum_{\ell,m} \left(A^{(E)}_{n\ell m }+A^{(S,E)}_{n\ell m}(\tau;\bar\tau)\right) \psi^{(p)}_{n\ell m}(\tau,x^A) \nonumber \\ 
    &+\left(B^{(E)}_{n\ell m }+B^{(S,E)}_{n\ell m}(\tau;\bar\tau)\right) \psi^{(q)}_{n\ell m}(\tau,x^A);\label{psiENHSolV}\\
    \omega^B_{n} &= \sum_{\ell,m} \left(A^{(B)}_{n\ell m }+A^{(S,B)}_{n\ell m}(\tau;\bar\tau)\right) \psi^{(p)}_{n\ell m}(\tau,x^A) \nonumber \\ 
    &+\left(B^{(B)}_{n\ell m }+B^{(S,B)}_{n\ell m}(\tau;\bar\tau)\right) \psi^{(q)}_{n\ell m}(\tau,x^A)\label{psiBNHSolV}
\end{align}
with $A^{(D)}_{n\ell m }$ and $B^{(D)}_{n\ell m }$, $D=E,B$, constants and
\begin{subequations}\label{ASE}
\begin{align}
    A^{(S,E)}_{n\ell m}(\tau;\bar\tau)&= \int_{\bar\tau}^\tau d\tau' \oint_{S^2(\tau')} d\Omega \, \sigma^E_n(\tau',x^A)\overline{\psi^{(q)}_{n\ell m}(\tau',x^A)},\\
    A^{(S,B)}_{n\ell m}(\tau;\bar\tau)&=\int_{\bar\tau}^\tau d\tau' \oint_{S^2(\tau')} d\Omega \,\sigma^B_n(\tau',x^A)\overline{\psi^{(q)}_{n\ell m}(\tau',x^A)},\\
    B^{(S,E)}_{n\ell m}(\tau;\bar\tau)&= -\int_{\bar\tau}^\tau d\tau' \oint_{S^2(\tau')} d\Omega \,\sigma^E_n(\tau',x^A)\overline{\psi^{(p)}_{n\ell m}(\tau',x^A)},\\
    B^{(S,B)}_{n\ell m}(\tau;\bar\tau)&=- \int_{\bar\tau}^\tau d\tau' \oint_{S^2(\tau')} d\Omega \,\sigma^B_n(\tau',x^A)\overline{\psi^{(p)}_{n\ell m}(\tau',x^A)}.
\end{align}
\end{subequations}
Using the KG inner product on the scalars, we can single out the coefficients 
\begin{subequations}
\begin{align}
     (\omega^E_n,\psi^{(q)}_{n\ell m})_{KG} &=A^{(E)}_{n\ell m }+A^{(S,E)}_{n\ell m}(\tau;\bar\tau),\\ 
     (\omega^E_n,\psi^{(p)}_{n\ell m})_{KG} &=-B^{(E)}_{n\ell m }-B^{(S,E)}_{n\ell m}(\tau;\bar\tau),\\
     (\omega^B_n,\psi^{(q)}_{n\ell m})_{KG} &=A^{(B)}_{n\ell m }+A^{(S,B)}_{n\ell m}(\tau;\bar\tau),\\ 
     (\omega^B_n,\psi^{(p)}_{n\ell m})_{KG} &=-B^{(B)}_{n\ell m }-B^{(S,B)}_{n\ell m}(\tau;\bar\tau). 
\end{align}
\end{subequations}
Finally, the inhomogeneous solution to Eqs. \eqref{GeneralNHDFVectEqu0}-\eqref{GeneralNHDFVectEqu} can be rewritten as 
\begin{align}
   & W^{(T)n}_a = \delta_{n,-1} A^{(E)}_{-100} V_a^{(E,p)\:-1\,00}\nn\\&\;\; +\sum_{\ell=1}^{+\infty}\sum_{m=-\ell}^\ell \bigg\{ -\frac{(A^{(B)}_{n\ell m}+ A^{(S,B)}_{n\ell m}(\tau;\bar\tau) )}{\sqrt{\ell(\ell+1)}}V^{(B,p)n\ell m}_a \nn\\&\;\;- \frac{(B^{(B)}_{n\ell m}+ B^{(S,B)}_{n\ell m}(\tau;\bar\tau) )}{\sqrt{\ell(\ell+1)}} V^{(B,q)n\ell m}_a \nn\\&\;\;- \epsilon_{a}^{\,bc}\mathcal{D}_b \left(\frac{(A^{(E)}_{n\ell m}+ A^{(S,E)}_{n\ell m}(\tau;\bar\tau) )}{\sqrt{\ell(\ell+1)}} V^{(B,p)n\ell m}_c\right)\nonumber \\  
    &\;\;- \epsilon_{a}^{\,bc}\mathcal{D}_b \left(\frac{(B^{(E)}_{n\ell m}+ B^{(S,E)}_{n\ell m}(\tau;\bar\tau) )}{\sqrt{\ell(\ell+1)}} V^{(B,q)n\ell m}_c\right) \bigg\} .\label{VTNHdecomp}
\end{align}
Using, Eq. \eqref{DecompostionVect} for $n\ge0$ or Eqs. \eqref{Vnm1Decomp}-\eqref{BoxPhiS} for $n=-1$, the generic solution to Eqs. \eqref{NHGeneralVectEqu0}-\eqref{NHGeneralVectEqu} can be obtained from the inhomogeneous solution \eqref{VTNHdecomp}. 

\subsection{Asymptotic behavior and antipodal matching for transverse vector harmonics}

We want to establish the asymptotics of a general transverse (i.e. divergence-free) vector as $\tau \to \pm \infty$ and identify the antipodal relations between $\tau=\infty$ and $\tau=-\infty$. We will not explicitly write the asymptotic behavior and antipodal matching conditions for the longitudinal vectors as they can be straightforwardly obtained from the scalar case.

Let us recall that a general transverse vector can be written as 
\begin{align}
    V_\tau^{(T)} &= -\nabla^B N^{(E)}_B(\tau,x^A) \sech\tau+C\;\sech^2\tau,\label{generalVectTtau} \\
    V_A^{(T)} &= -\cosh\tau \left(\partial_\tau + \tanh\tau\right)N^{(E)}_A(\tau,x^A) +  \cosh\tau N^{(B)}_A(\tau,x^A),\label{generalVectTA}
\end{align}
with 
\begin{align}\label{NAf}
    N^{(E)}_A(\tau,x^A) = \nabla_A f(\tau,x^A),\quad N^{(B)}_A(\tau,x^A)=\epsilon_{AB}\nabla^B\tilde f(\tau,x^A).
\end{align}
and $C$ a constant.
Solutions to Eq. \eqref{generalEquationvectorsHarms} are determined by the constant C and the functions $f$ and $\tilde{f}$ that obey Eq.  \eqref{generalscalarequation}. Similarly to Eq. \eqref{scalDelta}, let us define the operator $\Delta_{k,s}$ for $k,s \in \mathbb N$, 
\begin{align}\label{GenDelta}
\Delta_{k,s} :=    \prod_{q=0}^{k-1}(\nabla^2+q(q+1)-s^2), \;\;  k \in \mathbb N_0,\quad  \Delta_{0,s} := 1. 
\end{align}
The operator $\Delta_{k,1}$ annihilates all $\ell\le k-1$ vector harmonics. We will also use this operator in the tensorial case $s=2$. We note the identity
\begin{equation}
    \sum_{k=0}^n \frac{\Delta_{k,s}}{k!(k+1)!}=\frac{1}{n!(n+1)!}\Delta_{1,n+1,s}\label{DeltaIdentity}
\end{equation}
where 
\begin{equation}
 \Delta_{a,k,s}  :=    \prod_{q=a}^{k-1}(\nabla^2+q(q+1)-s^2)  =  \frac{\Delta_{k,s}}{\Delta_{a,s}}
\end{equation}
denotes the operator $\Delta_{k,s}$ defined as \eqref{GenDelta} but starting at $q=a$, $a\in\mathbb N$ and $a\le k$, instead of $q=0$. This operator is local. 

Using Eqs. \eqref{PsiHomBehavior}--\eqref{PsiHomBehaviorfdef} and \eqref{NAf}, we find that vectors $N^{(E)n}_A$ and $N^{(B)n}_A$ for the eigenvalue $n$ in Eq. \eqref{generalEquationvectorsHarms} admit the following asymptotic behavior 
\begin{align}
    N^{(E)n}_A(&\tau,x^A) \nn\\&= e^{n\vert\tau\vert}\sum_{r=0}^n N_{n,1}^{(r,0)}[\Gamma_{A}^{(\mathbb L,E)n\pm}(x^A)]e^{-2r\vert\tau\vert} \nonumber \\&+ e^{-(n+2)\vert\tau\vert}\left(\Gamma_{A}^{(\mathbb S,E)n\pm}(x^A)+\vert\tau\vert N_{n,1}^{(n+1,1)}[\Gamma_{A}^{(\mathbb L,E)n\pm}(x^A)]\right) \nonumber\\ 
    &+o(e^{-(n+2)\vert\tau\vert}),\\ N^{(B)n}_A(&\tau,x^A) \nn\\&= e^{n\vert\tau\vert}\sum_{r=0}^n N_{n,1}^{(r,0)}[\Gamma_{A}^{(\mathbb L,B)n\pm}(x^A)]e^{-2r\vert\tau\vert} \nonumber\\&+ e^{-(n+2)\vert\tau\vert}\left( \Gamma_{A}^{(\mathbb S,B)n\pm}(x^A)+\vert\tau\vert N_{n,1}^{(n+1,1)}[ \Gamma_{A}^{(\mathbb L,B)n\pm}(x^A)]\right) \nonumber \\ 
    &+o(e^{-(n+2)\vert\tau\vert}),
\end{align}
where we defined, for $r,n,s \in \mathbb N$ and $0\le r\le n$, the operators ,
\begin{align}
    N_{n,s}^{(r,0)}[\cdot]&:= (1-\delta_{n,-1}) \sum_{k=0}^{r} \frac{(-1)^{r}(n-k)!\;r! \Delta_{k,s}[\cdot]}{(k!)^2(r-k)!},\label{NnsOp1}\\
    N_{n,s}^{(n+1,1)}[\cdot]&:= \frac{ 2(-1)^{n+1}}{(n+1)!}\Delta_{n+1,s}[\cdot].\label{NnsOp2}
\end{align}
These operators are a generalization to an arbitrary s of the definitions \eqref{PsiHomBehaviorfdef0}-\eqref{PsiHomBehaviorfdef}.
The 8 vectors $\Gamma_{A}^{(\mathbb L,E)n\pm}(x^A)$, $\Gamma_{A}^{(\mathbb S,E)n\pm}(x^A)$, $\Gamma_{A}^{(\mathbb L,B)n\pm}(x^A)$ and $\Gamma_{A}^{(\mathbb S,B)n\pm}(x^A)$ are defined in terms of 8 functions $\psi^{(\mathbb L)\pm}_{n}(x^A)$, $\psi^{(\mathbb S)\pm}_{n}(x^A)$, $\tilde \psi^{(\mathbb L)\pm}_{n}(x^A)$ and $\tilde \psi^{(\mathbb S)\pm}_{n}(x^A)$ as  
\begin{align}
   \Gamma_{A}^{(\mathbb L,E)n\pm}(x^A)&:=\nabla_A \psi^{(\mathbb L)\pm}_{n}(x^A),\\
    \Gamma_{A}^{(\mathbb L,B)n\pm}(x^A)&:=\epsilon_{AB}\nabla^B \tilde\psi^{(\mathbb L)\pm}_{n}(x^A),\\
    \Gamma_{A}^{(\mathbb S,E)n\pm}(x^A)&:=\nabla_A \psi^{(\mathbb S)\pm}_{n}(x^A),\\
    \Gamma_{A}^{(\mathbb S,B)n\pm}(x^A)&:=\epsilon_{AB}\nabla^B \tilde\psi^{(\mathbb S)\pm}_{n}(x^A),
\end{align}
which obey the antipodal matching conditions \eqref{Matchingscalarend}. 

Then, from the relation \eqref{generalVectTtau}, we find that the $\tau$ component of $V^{(T)n}$ behaves as
\begin{align}
&V^{(T)n}_\tau(\tau\to\pm\infty,x^A) =-2(1-\delta_{n,-1}) e^{(n-1)\vert\tau\vert} \nn\\& \hspace{0.2cm}
\times \sum_{r=0}^{n}\;\sum_{k=0}^{r} \frac{(-1)^{r}(n-k)!\;(r+1)! \Delta_{k,0}}{k!(k+1)!(r-k)!}\nabla^{A}\Gamma_{A}^{(\mathbb L,E)n\pm}(x^A) e^{-2r\vert\tau\vert}  \nn\\ \hspace{-1.5cm}&-2e^{-(n+3)\vert\tau\vert}\Big[\frac{(-1)^{n+1}}{n!}\Delta_{1,n+1,0}\nabla^{A}\Gamma_{A}^{(\mathbb L,E)n\pm}(x^A)+\nabla^{A}\Gamma_{A}^{(\mathbb S,E)n\pm}(x^A)\nn \\ \hspace{-1.5cm}&-2\delta_{n,-1} C+\vert\tau\vert\frac{2(-1)^{n+1}}{(n+1)!} \Delta_{n+1,0}\nabla^{A}\Gamma_{A}^{(\mathbb L,E)n\pm}(x^A)\Big] \nn\\&+ o( e^{-(n+3)\vert\tau\vert})
\end{align}
after using the identity \eqref{DeltaIdentity}.

Regarding the angular component of $V^{(T)n}_a$ in Eq.\eqref{generalVectTA}, we distinguish the E parity piece and the B parity piece: 
\begin{align}
    & V_A^{(T)n(E)}(\tau\to\pm\infty,x^A) \nn\\&= -\frac{(\pm1)}{2}(1-\delta_{n,-1})e^{(n+1)\vert\tau\vert}\Big[(n+1)!\Gamma_{A}^{(\mathbb L,E)n\pm}(x^A)\nn\\&+(1-\delta_{n,0})\sum_{r=1}^{n}(n+1-2r)e^{-2r\vert\tau\vert}\sum_{k=1}^r \frac{k(-1)^{r}(n-k)! (r-1)!}{(k!)^2(r-k)!} \nonumber\\ 
    &\times \Delta_{k,1}\Gamma_{A}^{(\mathbb L,E)n\pm}(x^A)\Big]-
    \frac{(\pm1)}{2}e^{-(n+1)\vert\tau\vert}\Big[-2\vert\tau\vert (-1)^{n+1} \nonumber \\ 
    &\times \frac{ \Delta_{n+1,1}}{n!}\Gamma_{A}^{(\mathbb L,E)n\pm}(x^A)
    -(n+1)\Gamma_{A}^{(\mathbb S,E)n\pm}(x^A) \nn\\&+2\delta_{n,-1}\Gamma_{A}^{(\mathbb L,E)-1\pm}(x^A) \nn\\&+(1-\delta_{n,-1})(n+1)! \sum_{k=0}^{n} \frac{(-1)^{n+1}\;  \Delta_{k,1}\Gamma_{A}^{(\mathbb L,E)n\pm}(x^A)}{(k!)^2}\nn\\&+(1-\delta_{n,-1})2(-1)^{n+1} \frac{ \Delta_{n+1,1}}{(n+1)!}\Gamma_{A}^{(\mathbb L,E)n\pm}(x^A)\Big]\nn\\&+o(e^{-(n+1)\vert\tau\vert})
\intertext{and}
    &V_A^{(T)n(B)}(\tau\to\pm\infty,x^A) \nn\\&= \frac{1}{2}e^{(n+1)\vert\tau\vert}(1-\delta_{n,-1})\Big[n! \Gamma_{A}^{(\mathbb L,B)n\pm}(x^A)\nn\\&+(1-\delta_{n,0})\sum_{r=1}^{n}e^{-2r\vert\tau\vert}\sum_{k=1}^r \frac{k(-1)^{r}(n-k)! (r-1)!}{(k!)^2(r-k)!} \nonumber \\ 
    &\times \Delta_{k,1}\Gamma_{A}^{(\mathbb L,B)n\pm}(x^A)\Big]+
    \frac{1}{2}e^{-(n+1)\vert\tau\vert}\Big[2\vert\tau\vert (-1)^{n+1} \nn\\ 
    &\times \frac{\Delta_{n+1,1}}{(n+1)!}\Gamma_{A}^{(\mathbb L,B)n\pm}(x^A)
    +\Gamma_{A}^{(\mathbb S,B)n\pm}(x^A)-(1-\delta_{n,-1})\times \nn\\& \sum_{k=0}^{n} \frac{(-1)^{n+1}\;n!  \Delta_{k,1}\Gamma_{A}^{(\mathbb L,B)n\pm}(x^A)}{(k!)^2}\Big]+o(e^{-(n+1)\vert\tau\vert}).
\end{align}
We can summarize the generic asymptotic behavior of a transverse vector as follows. We define
\begin{align}
    K_A^{(\mathbb L)n\pm}(x^A) &:= (\pm1)(n+1) \Gamma_{A}^{(\mathbb L,E)n\pm}(x^A)-\Gamma_{A}^{(\mathbb L,B)n\pm}(x^A),\label{KAVect}\\
    K_A^{(\mathbb S)n\pm}(x^A) &:= (\pm1)(n+1) \Gamma_{A}^{(\mathbb S,E)n\pm}(x^A)+\Gamma_{A}^{(\mathbb S,B)n\pm}(x^A) \nn \\ 
    &\mp 2(n+1)!(-1)^{n+1}\Gamma_{A}^{(\mathbb L,E)n\pm}(x^A) ,\\
    K_A^{(\mathbb L,\tau)n\pm}(x^A) &:=  \frac{2(-1)^{n+1} }{(n+1)!}\nn\\ 
    &\times\Big(\Delta_{n+1,1}\delta_A^B-2 \Delta_{1,n+1,1}\nabla_A\nabla^B\Big)K_B^{(\mathbb L)n\pm}(x^A)\label{KAtauVectRel}\\
    k^{(\mathbb S)n\pm}(x^A)&:=(\pm 1)\nabla^A\Gamma_{A}^{(\mathbb S,E)n\pm}(x^A)-2(\pm 1)\delta_{n,-1} C.\label{hnVect}
\end{align}

For $n\ge0$, there is the relation 
\begin{align}
    &k^{(\mathbb S)n\pm}(x^A) \nn\\&\quad= \frac{1}{n+1}\Big(\nabla^AK_A^{(\mathbb S)n\pm}(x^A)+2n!(-1)^{n+1}\nabla^AK_A^{(\mathbb L)n\pm}(x^A)\Big).\label{hnVectRel}
\end{align}
Also, for any E parity vector $V^{(E)}_A$, we have 
\begin{equation}
    \nabla_A\nabla^B V^{(E)}_B =\Delta_{1,1} V^{(E)}_A
\end{equation}
and for any B parity vector $V^{(B)}_A$:
\begin{equation}
    \nabla^A V^{(B)}_A = 0. 
\end{equation}

The functions $K_A^{(\mathbb L)n\pm}(x^A)$, $K_A^{(\mathbb S)n\pm}(x^A)$, $K_A^{(\mathbb L,\tau)n\pm}(x^A)$ and $k^{(\mathbb S)n\pm}(x^A)$ can be identified from the asymptotic behavior of the transverse vectors contrary to the $\Gamma_{A}^{(\mathbb L,E)n\pm}$, $\Gamma_{A}^{(\mathbb L,B)n\pm}$, $\Gamma_{A}^{(\mathbb S,E)n\pm}$, $\Gamma_{A}^{(\mathbb S,B)n\pm}$ functions which require an electric-magnetic split. A special case is $n=-1$, for which $K_A^{(\mathbb L)-1\pm}(x^A)$ and $K_A^{(\mathbb L,\tau)-1\pm}(x^A)$ only have a magnetic parity on the sphere. 

This leads to
\begin{align}
&V^{(T)n}_\tau(\tau\to\pm\infty,x^A)=-\frac{2(\pm1)(1-\delta_{n,-1})}{(n+1)} e^{(n-1)\vert\tau\vert} \times  \nn\\ 
&\sum_{r=0}^{n}\;\sum_{k=0}^{r}\frac{(-1)^{r}(n-k)!\;(r+1)! \Delta_{k,0}}{k!(k+1)!(r-k)!}\nabla^{A}K_A^{(\mathbb L)n\pm}(x^A) e^{-2r\vert\tau\vert}  \nn\\
&-2(\pm1)e^{-(n+3)\vert\tau\vert}\Big[\frac{(-1)^{n+1}}{(n+1)!}\Delta_{1,n+1,0}\nabla^{A}K_A^{(\mathbb L)n\pm}(x^A)\nn\\&+k^{(\mathbb S)n\pm}(x^A)-\vert\tau\vert\frac{\nabla^{A}K_A^{(\mathbb L,\tau)n\pm}(x^A)}{n+1+\delta_{n,-1}}\nn\\&-\vert\tau\vert \delta_{n,-1}\nabla^{A}K_A^{(\mathbb S)-1\pm}(x^A) \Big] + o( e^{-(n+3)\vert\tau\vert}),
\end{align}
and
\begin{align}
 &V_A^{(T)n}(\tau\to\pm\infty,x^A) = -\frac{1}{2}e^{(n+1)\vert\tau\vert}(1-\delta_{n,-1})\Big[n! K_A^{(\mathbb L)n\pm}(x^A)\nn\\&+(1-\delta_{n,0})\sum_{r=1}^{n}e^{-2r\vert\tau\vert}\sum_{k=1}^r \frac{k(-1)^{r}(n-k)! (r-1)!}{(k!)^2(r-k)!}\nn\\&\times\Big(\Delta_{k,1}\delta_A^B-\frac{2r}{(n+1)}\Delta_{1,k,1}\nabla_A\nabla^B\Big)K_{B}^{(\mathbb L)n\pm}(x^A)\Big]\nn\\&-
    \frac{1}{2}e^{-(n+1)\vert\tau\vert}\Big[\vert\tau\vert K_A^{(\mathbb L,\tau)n\pm}(x^A)
   - K_A^{(\mathbb S)n\pm}(x^A)  \nn
   \\&-(1-\delta_{n,-1})(-1)^{n+1}n!K_A^{(\mathbb L)n\pm}(x^A)\nn\\&-(1-\delta_{n,-1}-\delta_{n,0})\sum_{k=1}^{n} \frac{(-1)^{n+1}\;n!  }{(k!)^2}  \nn\\&\times \Big(\Delta_{k,1}\delta_A^B-2 \Delta_{1,k,1}\nabla_A\nabla^B\Big)K_B^{(\mathbb L)n\pm}(x^A)\nn\\&+(1-\delta_{n,-1})\frac{2(-1)^{n+1} }{(n+1)(n+1)!} \nn\\ 
    &\times\Delta_{1,n+1,1}\nabla_A\nabla^BK_B^{(\mathbb L)n\pm}(x^A)\Big]
    +o(e^{-(n+1)\vert\tau\vert}). 
\end{align}
We defined the various quantities such that the $\pm 1$ factor in the $\tau$ component is left explicit, such that the time reversal of $V_a$ absorbs this $\pm 1$ factor. From the antipodal matching conditions derived in Section \ref{sec:asymptInh}, we can deduce the antipodal relationships for the mode functions $\Gamma_{A}^{(\mathbb L,E)n\pm}(x^A)$, $\Gamma_{A}^{(\mathbb S,E)n\pm}(x^A)$, $\Gamma_{A}^{(\mathbb L,B)n\pm}(x^A)$ and $\Gamma_{A}^{(\mathbb S,B)n\pm}(x^A)$. Then, using the definitions \eqref{KAVect}-\eqref{hnVect}, we find the antipodal matching conditions for the asymptotic data
$K_A^{(\mathbb L)n\pm}(x^A)$, $K_A^{(\mathbb S)n\pm}(x^A) $, $K_A^{(\mathbb L,\tau)n\pm}(x^A) $ and $k^{(\mathbb S)n\pm}(x^A)$:
\begin{align}\label{matchV1}
   &K_A^{(\mathbb S)n+}(x^A)- O^{(q)n}_V(K_A^{(\mathbb L)n+}(x^A)) = (-1)^{n}\Upsilon^*\Big(K_A^{(\mathbb S)n-}(x^A) \nn\\
   &\qquad -  O^{(q)n}_V(K_A^{(\mathbb L)n-}(x^A))\Big), \quad n\ge0,\\&
   \epsilon^{AB}\nabla_A K_B^{(\mathbb S)-1,+}(x^A)- \epsilon^{AB}\nabla_AO^{(q)-1}_V(K_B^{(\mathbb L)-1,+}(x^A)) \nn\\&\quad=  
     -\Upsilon^*\Big(\epsilon^{AB}\nabla_AK_B^{(\mathbb S)-1,-}(x^A)\nn\\&\qquad-  \epsilon^{AB}\nabla_AO^{(q)-1}_V(K_B^{(\mathbb L)-1,-}(x^A))\Big),\\&
   \nabla^AK_A^{(\mathbb S)-1,+}(x^A) = \Upsilon^* \nabla^A K_A^{(\mathbb S)-1,-}(x^A),\\
   &K_A^{(\mathbb L)n+}(x^A)- O^{(p)n}_{VE}(\mathcal K_A^{(\mathbb S)n+}(x^A)) + O^{(p)n}_{VB}(K_A^{(\mathbb S)n+}(x^A)) \nn\\&\quad= (-1)^{n+1} \Upsilon^*\Big(K_A^{(\mathbb L)n-}(x^A)- O^{(p)n}_{VE}(\mathcal K_A^{(\mathbb S)n-}(x^A))\nn\\&\qquad + O^{(p)n}_{VB}(K_A^{(\mathbb S)n-}(x^A))\Big),\\
&k^{(\mathbb S)n+}(x^A)-\frac{1}{n+1}O^{(q)n}(\nabla^A K_A^{(\mathbb L)n+}(x^A))=(-1)^{n} \times  \nn \\ 
   &\qquad\Upsilon^*\Big(k^{(\mathbb S)n-}(x^A) -\frac{1}{n+1}O^{(q)n}(\nabla^A K_A^{(\mathbb L)n-}(x^A))\Big), \quad n\ge0, \\
   &k^{(\mathbb S)-1,+}(x^A)+\frac{1}{2}O^{(q)-1}(\nabla^A K_A^{(\mathbb S)-1,+}(x^A))= \nn \\ 
   &\qquad-\Upsilon^*\Big(k^{(\mathbb S)-1,-}(x^A) +\frac{1}{2}O^{(q)-1}(\nabla^A K_A^{(\mathbb S)-1,-}(x^A))\Big), \\&\nabla^A K_A^{(\mathbb L)n+}(x^A)-(n+1)O^{(p)n}(k^{(\mathbb S)n,+}(x^A))= (-1)^{n+1}\times \nn \\ 
   &\qquad \Upsilon^*\Big(\nabla^A K_A^{(\mathbb L)n-}(x^A)-(n+1)O^{(p)n}(k^{(\mathbb S)n-}(x^A))\Big),\\&
   K_A^{(\mathbb L,\tau)n+}(x^A)  = (-1)^{n+1}\Upsilon^*K_A^{(\mathbb L,\tau)n-}(x^A)\label{matchVder}
\end{align}
where $\mathcal K_A^{(\mathbb S)n\pm}(x^A) :=  K_A^{(\mathbb S)n\pm}(x^A)+2(1-\delta_{n,-1})(-1)^{n+1}n!K_A^{(\mathbb L)n\pm}(x^A)$ and the non-local operators $O^{(p)n}_{VE}$, $O^{(p)n}_{VB}$ and $O^{(q)n}_V$ acting on a vector $f_A(x^A)$ are defined as follows: 
\begin{align}
    O^{(p)n}_{VE}[f_A(x^A)]& := \sum_{\ell=1}^{+\infty}\sum_{m=-\ell}^{+\ell}(-1)^{\ell+n+1}\frac{(n+1)!}{2\Gamma(n+\ell+2)\Gamma(n-\ell+1)} \nn\\ 
    &\hspace{-0.3cm}\times \mathring V^{(E)\ell m}_A(x^A)\oint_{S^2}d\Omega' \overline{\mathring V^{(E)\ell m}_C(x^{A'})}f^C(x^{A'}),
    \end{align}
\begin{align}
    O^{(p)n}_{VB}[f_A(x^A)]& := \sum_{\ell=1}^{+\infty}\sum_{m=-\ell}^{+\ell}(-1)^{\ell+n+1}\frac{(n+1)!}{2\Gamma(n+\ell+2)\Gamma(n-\ell+1)} \nn\\ &\hspace{-0.3cm}\times \mathring V^{(B)\ell m}_A(x^A)\oint_{S^2}d\Omega' \overline{\mathring V^{(B)\ell m}_C(x^{A'})}f^C(x^{A'}),
    \end{align}
\begin{align}
    O^{(q)n}_V[f_A(x^A)]& :=\sum_{\ell=1}^{+\infty}\sum_{m=-\ell}^{+\ell} (-1)^{n+1}\Big[\frac{(-\ell)_{n+1}(\ell+1)_{n+1}}{(n+1)!}(H_{n+1} \nn\\ 
    &\hspace{-1.2cm}- 2 H_{\ell} )-(1-\delta_{n,-1})2n!\Big]  \mathring V^{(E)\ell m}_A(x^A)\nn\\&\times\oint_{S^2}d\Omega' \overline{\mathring V^{(E)\ell m}_C(x^{A'})}f^C(x^{A'})\nn\\-&\sum_{\ell=1}^{+\infty}\sum_{m=-\ell}^{+\ell} (-1)^{n+1}\frac{(-\ell)_{n+1}(\ell+1)_{n+1}}{(n+1)!}(H_{n+1} \nn\\ 
    &\hspace{-1.2cm}- 2 H_{\ell} )\mathring V^{(B)\ell m}_A(x^A)\oint_{S^2}d\Omega' \overline{\mathring V^{(B)\ell m}_C(x^{A'})}f^C(x^{A'}).
\end{align}
The operators $O^{(p)n}_{VE}$, $O^{(p)n}_{VB}$ annihilate all $\ell\ge n+1$ harmonics. There is an additional factor $-2 n!$ in the E part of the operator $O^{(q)n}_V$ due to the difference of the respective  $e^{-2(n+1)\vert\tau\vert}$ subleading asymptotic behavior of the E and B parity parts of a generic vector. Contrary to the scalar case, $O^{(q)n}_V$ does not annihilate all $\ell< n+1$ harmonics but only the magnetic part of the $\ell< n+1$ harmonics. Hence, $O^{(p)n}_{VE}O^{(q)n}_{V}\ne0$ (for $n\ge1$ since $O^{(p)-1}_{VE}=O^{(p)0}_{VE}=0$) and $O^{(p)n}_{VB}O^{(q)n}_{V}=0$. 
The antipodal matching conditions are generally non-local for the asymptotic data  $K_A^{(\mathbb S)n\pm}(x^A)$, $K_A^{(\mathbb L)n\pm}(x^A) $ and $k^{(\mathbb S)n\pm}(x^A)$ and are always local for $K_A^{(\mathbb L,\tau)n\pm}(x^A) $. If one assumes $\ell\ge n+1$, then the antipodal matching conditions become local for the leading asymptotic data $K_A^{(\mathbb L)n\pm}(x^A) $.

For inhomogeneous solutions, the procedure to extract antipodal matching conditions, which results in the existence of conserved quantities between $\tau =\infty$ and $\tau=-\infty$, goes as follows. First, one has to convert the source into scalars $\sigma^E_n$ and $\sigma^B_n$ as defined in Eqs \eqref{sigmaE} and \eqref{sigmaB}. This allows to compute $A_{n\ell m}^{(S,D)}$ and $B_{n\ell m}^{(S,D)}$, $D=E,B$, see Eqs. \eqref{ASE}, which can be substituted into the decomposition \eqref{VTNHdecomp}. We can then subtract these contributions to obtain a ``subtracted'' vector $\hat V^n_a$ that behaves asymptotically as a homogeneous solution. Then, we can identify modes  $\hat K_A^{(\mathbb S)n\pm}(x^A)$, $\hat K_A^{(\mathbb L)n\pm}(x^A) $, $\hat K_A^{(\mathbb L,\tau)n\pm}(x^A) $ and $\hat k^{(\mathbb S)n\pm}(x^A)$ that satisfy the antipodal matching conditions \eqref{matchV1}-\eqref{matchVder}. These antipodal relationships are equivalent to the conservation of the charges defined as follows. 
For $n\ge0$, the conserved charges are
\begin{subequations}\label{Q1A}
\begin{align}
    \mathcal Q_{Y}[\hat V^{(P)n}]&:=\oint_{S^2}d\Omega \hat V^{(P)n+}_A \overline{Y}^A\nonumber\\
    &=(-1)^{n}\oint_{S^2}d\Omega \hat V^{(P)n-}_A \Upsilon^*\overline{Y}^A,\\
    \mathcal Q_{Y}[\hat V^{(Q)n}]&:=\oint_{S^2}d\Omega \hat V^{(Q)n+}_A \overline{Y}^A\nonumber\\
    &=(-1)^{n+1}\oint_{S^2}d\Omega \hat V^{(Q)n-}_A \Upsilon^*\overline{Y}^A,
\end{align}    
\end{subequations}
where $Y^A(x^B)$ are arbitrary vectors on the sphere and
\begin{align}
    \hat V^{(P)n\pm}_A&:=\hat K_A^{(\mathbb S)n\pm}(x^A)- O^{(q)n}_V(\hat K_A^{(\mathbb L)n\pm}(x^A)), \\
    \hat V^{(Q)n\pm}_A&:=\hat K_A^{(\mathbb L)n\pm}(x^A)- O^{(p)n}_{VE}(\hat{\mathcal{K}}_A^{(\mathbb S)n\pm}(x^A))\nn \\&\quad+ O^{(p)n}_{VB}(\hat{K}_A^{(\mathbb S)n\pm}(x^A)). 
\end{align}
One could also use alternatively the data $\hat K_A^{(\mathbb L,\tau)n\pm}(x^A)$ or $\hat k^{(\mathbb S)n\pm}(x^A)$ to construct conserved quantities using the antipodal relationships but these would extract information already contained in $ \mathcal Q_{Y}[\hat V^{(P)n}]$, $ \mathcal Q_{Y}[\hat V^{(Q)n}]$. This is due to the relations \eqref{KAtauVectRel} and \eqref{hnVectRel}.  

For $n=-1$, we have the same amount of charges but they are more naturally expressed in terms of an arbitrary scalar on the sphere $T(x^A)$ as
\begin{subequations}\label{Q1B}
\begin{align}
    \mathcal Q_{T}[\hat v^{(P)D}]&:=\oint_{S^2}d\Omega \hat v^{(P)D+} \overline{T}=-\oint_{S^2}d\Omega \hat v^{(P)D-} \Upsilon^*\overline{T},\\
     \mathcal Q_{T}[\hat v^{(Q)D}]&:=\oint_{S^2}d\Omega \hat v^{(Q)D+} \overline{T}=\oint_{S^2}d\Omega \hat v^{(Q)D-} \Upsilon^*\overline{T},
\end{align}
\end{subequations}
with  $D=E,B$ and 
\begin{align}
    \hat v^{(P)E\pm} &:=\hat k^{(\mathbb S)-1,\pm}(x^A)+\frac{1}{2}O^{(q)-1}(\nabla^A \hat K_A^{(\mathbb S)-1,\pm}(x^A)), \\
    \hat v^{(P)B\pm} &:=  \epsilon^{AB}\nabla_A\hat K_B^{(\mathbb S)-1,\pm}(x^A)\nn\\&\qquad- \epsilon^{AB}\nabla_AO^{(q)-1}_V(\hat K_B^{(\mathbb L)-1,\pm}(x^A)), \\
    \hat v^{(Q)E\pm} &:=\nabla^A\hat K_A^{(\mathbb S)-1,\pm}(x^A), \\
    \hat v^{(Q)B\pm} &:= \epsilon^{AB}\nabla_A\hat K_B^{(\mathbb L)-1\pm}(x^A).
\end{align}

These conservation laws can also be recovered using the inner products built for transverse vectors as follows. In the generic case $n \geq 0$, we construct the vector KG inner products between $\hat V_a$ and transverse vectors on $dS_3$ that satisfy the homogeneous equation \eqref{generalEquationvectorsHarms}, $V^{(T)n,P}_a$, $V^{(T)n,Q}_a$, with, respectively, $p$ parity and $q$ parity. They converge for $\tau\to\pm\infty$ and the asymptotic values of $(\hat V^n,V^{(T)n,Q})_{KG}$ (resp. $(\hat V^n,V^{(T)n,P})_{KG}$) match with the charges $\mathcal Q_{Y}[\hat V^{(P)n}]$ (resp. $\mathcal Q_{Y}[\hat V^{(Q)n}]$) where $Y^A$ correspond to the asymptotic data of $V^{(T)n,Q}_a$ (resp.  $V^{(T)n,P}_a$). In the special case $n=-1$, the asymptotic values of the scalar KG inner product between, on the one hand, the scalars $\psi^D_{\hat V^n}$, $D=E,B$ built from $\hat V_a^n$ and, on the other hand, the scalars of either $p$ parity $\psi_n^{P}(\tau,x^A)$ or $q$ parity $\psi_n^{Q}(\tau,x^A)$ which satisfy Eq. \eqref{generalscalarequation}, match up to the charges \eqref{Q1B} once the function $T(x^A)$ is chosen as the asymptotic data of either $\psi_n^{Q}(\tau,x^A)$ or $\psi_n^{P}(\tau,x^A)$.

\section{Tensor harmonics} 

Finally, let us turn to tensor harmonics on dS$_3$. Antisymmetric tensors $T_{ab}=-T_{ba}$ reduce to vectors $V^c$ through $T_{ab}=\epsilon_{abc}V^c$ and we shall not consider them further. We consider a symmetric tensor field $T_{ab}=T_{ba}$  on $dS_3$  which satisfies an equation of the form  
\begin{equation}
    (\Box - \alpha)T_{ab} = S_{ab} \label{GeneralTensEqu}
\end{equation}
with $\alpha\in\mathbb R$ and $S_{ab}$ an arbitrary tensor on $dS_3$. Let $T_{1\;ab}$, $T_{2\;ab}$ be two complex symmetric tensors on $dS_3$, 
we define the tensor KG inner product 
\begin{equation}
     (T_1,T_2)_{KG}^T:=\oint_{S^2(\tau)} d\Omega n^a  \left(T_1^{bc} \mathcal{D}_a \overline{T_{2\;bc}} - \overline{T_{2\;bc}}\mathcal{D}_a T_1^{bc} \right). 
\end{equation}
This inner product is conserved when $T_{1\;ab}$ and $T_{2\;ab}$ satisfy the same equation \eqref{GeneralTensEqu} with $S_{ab}=0$.

\subsection{Tensor harmonics from scalars}

Let $\phi$, $\varphi$ be two scalars in $dS_3$. One can build a symmetric tensor as follows
\begin{equation}
    T^{(S)}_{ab} := \mathcal{D}_a \mathcal{D}_b \phi + \frac{1}{3}(\varphi - \Box\phi) q_{ab}.\label{Tscal}
\end{equation}
Such a tensor obeys $T^{(S)a}_a=\varphi$ and its symmetrized curl defined as 
\begin{equation}
    \text{Curl}(T_{ab}) :=  \epsilon_{(a}^{\;cd}\mathcal{D}_{\vert c}T_{d \vert b)}. 
\end{equation}
identically vanishes, 
\begin{align}\label{ZeroCurTab}
\text{Curl}(T_{ab}^{(S)})=0.     
\end{align}
After using commutation relations (see Eq. \eqref{CommunationRel}), we find
\begin{align}
    \Box T^{(S)}_{ab}&=\mathcal{D}_a\mathcal{D}_b (\Box + 6)\phi- \frac{1}{3}q_{ab}\Box(\Box + 6)\phi \nn\\ 
    &+\frac{1}{3}\Box \varphi q_{ab},\label{BoxTSab}\\
    \epsilon_a^{\;cd}\mathcal{D}_cT^{(S)}_{db} &= \frac{1}{3}\epsilon_{ab}^{\quad c}\mathcal{D}_c\Big( (\Box+3)\phi -\varphi\Big),\end{align}\begin{align}
    \mathcal{D}^aT^{(S)}_{ab} &= \frac{2}{3}  \mathcal{D}_b \Big(\Box+3\Big)\phi+\frac{1}{3}\mathcal D_b \varphi,\label{ScalTDiv}\\
    \mathcal{D}^a\mathcal{D}^bT^{(S)}_{ab} &= \frac{2}{3}  \Box \Big(\Box+3\Big)\phi+\frac{1}{3}\Box \varphi,\label{Scal2TDiv}  \\
    \text{Curl}( \text{Curl}(T_{ab}^{(S)}))&= (\Box-3)T_{ab}^{(S)}\nn\\ 
    & -\frac{3}{2} \mathcal{D}_{(a}\mathcal{D}^cT_{\vert c \vert b)}^{(S)}+\frac{1}{2}q_{ab}\mathcal{D}^c\mathcal{D}^dT^{(S)}_{cd}\nn\\&+\frac{1}{2}(\mathcal{D}_a\mathcal{D}_b \varphi-q_{ab}(\Box-2)\varphi)=0\label{Curlculpropscal}.
\end{align}
Given a trace and a double divergence, one can therefore construct a tensor \eqref{Tscal} where $\varphi$ is uniquely defined and $\phi$ is defined up to the ambiguity of adding the zero modes of the operator $\Box (\Box+3)$ acting on scalars.   Such zero modes are a linear combination of $\psi_{0}$ and  $\psi_1$, which are solutions to \eqref{generalscalarequation} for $n=0,1$ given explicitly in Eq. \eqref{solscal}. This also includes the inhomogeneous solutions $\phi_0^{(1)}$, $\phi_1^{(0)}$ to $\Box\phi_0^{(1)} = \psi_1$ and to $(\Box+3)\phi_1^{(0)} = \psi_0$ since particular solutions are $\phi_0^{(1)}=-\frac{1}{3}\psi_1$ and $\phi_1^{(0)}=\frac{1}{3}\psi_0$. 

Using scalar harmonics and tensors of the form \eqref{Tscal}, two classes of symmetric tensor harmonics can be built : pure trace and traceless harmonics. The pure trace harmonics are constructed as 
\begin{equation}
    T^{(T,c)\:n\ell m}_{ab} := \frac{1}{\sqrt{3}}q_{ab} \psi^{(c)}_{n\ell m}(\tau,x^A),\qquad c=p,q. 
\end{equation}
They satisfy the equation 
\begin{equation}
    \Box T^{(T,c)\:n\ell m}_{ab} = - \left[(n+1)^2-1\right]T^{(T,c)\:n\ell m}_{ab},\qquad c=p,q, 
\end{equation}
and their transformation under the antipodal map on the hyperboloid is 
\begin{align}
     & \Upsilon_{\mathcal{H}}^*[T^{(T,p)\:n\ell m}_{ab}] = (-1)^{n+1} T^{(T,p)\:n\ell m}_{ab},\\ 
     & \Upsilon_{\mathcal{H}}^*[T^{(T,q)\:n\ell m}_{ab}] = (-1)^{n} T^{(T,q)\:n\ell m}_{ab}.
\end{align}
The associated invariant inner product is equivalent to the scalar KG inner product : 
\begin{align}
(T^{(T,c)n\ell m },T^{(T,d)n\ell' m'})_{KG}^T= (\psi^{(c)}_{n\ell m},\psi^{(d)}_{n\ell' m'})_{KG}=\epsilon_{cd} \delta_{\ell \ell'}\delta_{m m'}.
\end{align}
Pure trace harmonics also obey the properties
\begin{align}
    \epsilon_a^{\;cd}\mathcal{D}_cT^{(T,c)\:n\ell m}_{db} &= -\frac{1}{\sqrt{3}} \epsilon_{ab}^{\quad c}\mathcal{D}_c \psi_{nlm}^{(c)},\\\mathcal{D}^b T^{(T,c)\:n\ell m}_{ab} &= \frac{1}{\sqrt{3}} \mathcal{D}_a \psi_{nlm}^{(c)},\\\mathcal{D}^a\mathcal{D}^b T^{(T,c)\:n\ell m}_{ab} &=- \frac{n(n+2)}{\sqrt{3}} \psi_{nlm}^{(c)}. 
\end{align}

Traceless tensor harmonics are constructed as follows:
\begin{equation}
    T^{(S,c)\:n\ell m}_{ab} :=\mathcal{D}_a \mathcal{D}_b \psi_{n\ell m}^{(c)}(\tau,x^A)  + \frac{1}{3}n (2 + n)q_{ab}\psi^{(c)}_{n\ell m}(\tau,x^A), \label{scalarTensorHarmonics}
\end{equation}
with $c=p,q$. They satisfy the equations
\begin{align}
    &\Box T^{(S,c)\:n\ell m}_{ab} = - \left[(n+1)^2-7\right]T^{(S,c)\:n\ell m}_{ab},\label{scalarTensorHarmonicsEquation}\\
    &\left(T^{(S,c)\:n\ell m}\right)^a_{\:a} = 0\label{scalarTensorHarmonicsTrace}.
\end{align}
Under the antipodal map on $dS_3$, they transform as 
\begin{align}
      &\Upsilon_{\mathcal{H}}^*[T^{(S,p)\:n\ell m}_{ab}] = (-1)^{n+1} T^{(S,p)\:n\ell m}_{ab},\\
      &\Upsilon_{\mathcal{H}}^*[T^{(S,q)\:n\ell m}_{ab}] = (-1)^{n} T^{(S,q)\:n\ell m}_{ab}.
\end{align}
Since 
\begin{equation}\label{LogTranslationEquation0}
    \mathcal{D}_a \mathcal{D}_b \psi_{000}^{(q)} = 0
\end{equation}
and 
\begin{equation}
    \mathcal{D}_a \mathcal{D}_b \psi_{1\ell m}^{(q)} + q_{ab} \psi_{1\ell m}^{(q)}  = 0, \qquad \vert m\vert\le \ell=0,1,\label{LogTranslationEquation}
\end{equation}
we are missing 5 solutions to the equations \eqref{scalarTensorHarmonicsEquation}-\eqref{scalarTensorHarmonicsTrace} which can be built as follows. First, using the inhomogeneous solution $M(\tau,x^A)$ defined in Eq. \eqref{psi0} which obeys $\Box M(\tau,x^A)=\psi_{000}^{(q)}(\tau,x^A)$, we can construct the following traceless tensor:
\begin{equation}\label{TSq0}
    T^{(S,q)000}_{ab} = \mathcal{D}_a \mathcal{D}_b  M(\tau,x^A) -\frac{1}{3}\psi_{000}^{(q)}(\tau,x^A) q_{ab}
\end{equation}
which obeys to Eqs. \eqref{scalarTensorHarmonicsEquation}-\eqref{scalarTensorHarmonicsTrace} with $n=0$. 

Second, we can use the inhomogeneous solutions to $(\Box+3)I_{\ell m}(\tau,x^A) = \psi_{1\ell m}^{(q)}(\tau,x^A)$ ($\ell=0,1$) and construct the traceless tensor 
\begin{align}\label{TSQ1lm}
    T^{(S,q)1\ell m}_{ab} &= \mathcal{D}_a \mathcal{D}_b  I_{\ell m}(\tau,x^A) \nn\\ 
    &+ \Big(I_{\ell m}(\tau,x^A)-\frac{1}{3}\psi_{1\ell m}^{(q)}(\tau,x^A)\Big) q_{ab},\quad \vert m\vert\le \ell\le1 
\end{align}
with 
\begin{align}
    I_{00}(\tau,x^A) &= \frac{Y_{00}(x^A)}{8}\Big(\sinh\tau-2 \tau  \cosh\tau+\tau \sech\tau\Big),\label{I00}\\
    I_{1m}(\tau,x^A)&=-\frac{Y_{1m}(x^A)}{8} \sinh\tau \Big(\tau (2+\sech^2\tau)+\tanh\tau\Big).\label{I1m}
\end{align}
This tensor satisfies Eqs. \eqref{scalarTensorHarmonicsEquation}-\eqref{scalarTensorHarmonicsTrace} with $n=1$. 

Traceless tensor harmonics admit the following divergence, curl and double divergence:
\begin{align}
    \epsilon_a^{\;cd}\mathcal{D}_cT^{(S,c)\:n\ell m}_{db} &= -\frac{1}{3}\Big(n-1-\frac{1}{4}\delta_{c,q}\delta_{n,1}(\delta_{\ell,0}+\delta_{\ell,1})\Big) \nn\\ 
    &\times (n+3)\epsilon_{ab}^{\quad c} V_c^{(L,c)n\ell m},\\\mathcal{D}^bT^{(S,c)\:n\ell m}_{ab} &= -\frac{2}{3} \Big(n-1-\frac{1}{4}\delta_{n,1}\delta_{c,q}(\delta_{\ell,0}+\delta_{\ell,1})\Big) \nn\\ 
    &\times (n+3) V_a^{(L,c)n\ell m},\\
    \mathcal{D}^a\mathcal{D}^bT^{(S,c)\:n\ell m}_{ab} &= -\frac{2}{3}\Big(n-1-\frac{1}{4}\delta_{n,1}\delta_{c,q}(\delta_{\ell,0}+\delta_{\ell,1})\Big)\nn\\ 
    &\times (n+3)\mathcal{D}^aV_a^{(L,c)n\ell m}.   
\end{align}

Their normalization under the KG inner product is
\begin{align}\label{TensKGScal}
(T^{(S,c)n\ell m },T^{(S,d)n\ell' m'})_{KG}^T=&\epsilon_{cd} \delta_{\ell \ell'}\delta_{m m'}\frac{2}{3}\Big(n-1 \nn\\ 
&\hspace{-3cm}-\frac{1}{4}\delta_{n,1}(\delta_{\ell,0}+\delta_{\ell,1})\Big)  \Big(n-\frac{1}{2}\delta_{n,0}\delta_{\ell0}\Big)(n+2)(n+3). 
\end{align}
It vanishes for $n=0$, $\ell\ge1$ and $n=1$, $\ell\ge2$. 

Traceless tensor harmonics are orthogonal to pure trace harmonics under the tensor KG inner product:
\begin{equation}
    (T^{(T,c)n\ell m },T^{(S,d)n'\ell' m'})_{KG}^T = 0. 
\end{equation}




\subsection{Tensor harmonics from transverse vectors}
Let $V^{(T)}_a$ be a transverse vector in $dS_3$. One can build a symmetric tensor as follows 
\begin{equation}
    T^{(V)}_{ab} :=  \mathcal{D}_{(a}V^{(T)}_{b)} .\label{VectorTensor}
\end{equation}
Using the commutation relations \eqref{CommunationRel}, we find 
\begin{align}
    \epsilon_a^{\;cd} \mathcal{D}_c T_{db}^{(V)}&= \frac{1}{2}\mathcal{D}_{(a} \text{Curl}(V^{(T)}_{b)}) + \frac{1}{4}  \epsilon_{a b}^{\;c}(\Box +2)V^{(T)}_c , \\
    \mathcal{D}^b T^{(V)}_{ab} &= \frac{1}{2}(\Box+2)V^{(T)}_a\label{divV},\\
    \mathcal{D}^a\mathcal{D}^b T^{(V)}_{ab}&=0\label{VectorTensorDoubleDiv},\\
    T^{(V)\;a}_{a}&=0\label{VectorTensorTrace},\\
    \mathcal{D}_{(a}\mathcal{D}^cT_{\vert c \vert b)}^{(V)} &= \frac{1}{2}(\Box-2)T_{ab}^{(V)}\label{VectorTensorProp} ,\\
    \Box T^{(V)}_{ab} &= \mathcal{D}_{(a} (\Box+4)V^{(T)}_{b)}\label{BoxTVab}\\
     \text{Curl}(\text{Curl}(T_{ab}^{(V)}))&=\frac{1}{4}\mathcal{D}_{(a} \text{Curl}(\text{Curl}(V^{(T)}_{b)}))\label{CurlcurlTVab}.
\end{align}
Given a traceless tensor with vanishing double divergence, one can construct the transverse vector $V^{(T)}_a$ from Eq. \eqref{divV} with the ambiguity of adding the zero modes of the operator $\square+2$ acting on transverse vectors. Such zero modes are entirely spanned by the solutions $V_a^{(T)1}$ which are explicitly given by Eq. \eqref{expVT} with $n=1$.

Using transverse vector harmonics, we can construct tensor harmonics of the form \eqref{VectorTensor}. We define
\begin{equation}\label{tvh}
    T^{(V,D,c)\:n\ell m}_{ab} := \mathcal{D}_{(a}V_{b)}^{(D,c)\: n\ell m},\qquad \text{$D=E,B$ \text{and} $c=p,q$}.
\end{equation}
Because there is only one $\ell=0$ transverse vector, the only $\ell=0$ tensor vector harmonics is $T^{(V,E,p)\:-100}_{ab}$. 

The harmonics \eqref{tvh} obey the equation
\begin{equation}
    \Box T^{(V)\:n\ell m}_{ab} = - \left[(n+1)^2-6\right]T^{(V)\:n\ell m}_{ab}\label{TensorVectHarmEqu}
\end{equation}
and their transformation law under the antipodal map is
\begin{align}
      &\Upsilon_{\mathcal{H}}^*[T_{ab}^{(V,E,p)\;n\ell m}] = (-1)^{n} T_{ab}^{(V,E,p)\;n\ell m},\\
      &\Upsilon_{\mathcal{H}}^*[T_{ab}^{(V,E,q)\;n\ell m}] = (-1)^{n+1} T_{ab}^{(V,E,q)\;n\ell m}, \\ 
&      \Upsilon_{\mathcal{H}}^*[T_{ab}^{(V,B,p)\;n\ell m}] = (-1)^{n} T_{ab}^{(V,B,p)\;n\ell m},\\  &\Upsilon_{\mathcal{H}}^*[T_{ab}^{(V,B,q)\;n\ell m}] = (-1)^{n+1} T_{ab}^{(V,B,q)\;n\ell m}.
\end{align}

Because the $n=1,\ell=1$ vectors satisfy the Killing equation \eqref{KillingEquation}, we are missing 6 solutions, \emph{i.e.} $T_{ab}^{(V,D,q)\;11 m}=0$ for $D=E,B$, $m=\pm 1,0$. These missing solutions are built as follows. The inhomogeneous transverse solutions to the equations $(\Box+2)\Xi^{(D)m}_{a}=V_{a}^{(D,q)\: 11 m}$  for $D=E,B$, $m=\pm 1,0$ allow to construct 6 tensors satisfying Eq. \eqref{TensorVectHarmEqu} with $n=1$. We define
\begin{equation}\label{tvhl1}
    T_{ab}^{(V,D,q)\;11 m}:= \mathcal{D}_{(a}\Xi_{b)}^{(D)m},\qquad D=E,B
\end{equation}
with 
\begin{align}
    \Xi_{\tau}^{(E)m} &:= \frac{\tanh\tau \left(\tau +\tau  \tanh ^2\tau+\tanh\tau+2 \tau  \sech^2\tau\right)}{8 \sqrt{2}}Y_{1m}(x^A),\\
     \Xi_{A}^{(E)m} &:= -\frac{1}{32} \sech^2\tau  \Big(3 \tau +3 \sinh (2 \tau )+\sinh (4 \tau ) \nn\\ 
     &+2 \tau  \cosh (2 \tau )+\tau  \cosh (4 \tau )\Big)\mathring V^{(E)1m}_A(x^A),
\end{align}
and 
\begin{align}
    \Xi_{\tau}^{(B)m} &:=0,\\
     \Xi_{A}^{(B)m} &:= -\frac{1}{16} \tanh \tau \Big(4 \tau +\sinh (2 \tau )+2 \tau  \cosh (2 \tau )\Big)\mathring V^{(B)1m}_A(x^A).
\end{align}

The vector harmonic tensors are traceless and have a zero double divergence, see Eqs. \eqref{VectorTensorDoubleDiv} and \eqref{VectorTensorTrace}. Their divergence is 
\begin{equation}\label{divTVD}
    \mathcal{D}^aT^{(V,D,c)\:n\ell m}_{ab} = -\frac{1}{2}(n-1-\frac{1}{4}\delta_{n, 1}\delta_{\ell 1}\delta_{cq})(n+3)V_{b}^{(D,c)\: n\ell m},
\end{equation}
for $D=E,B$ and $c=p,q$ while their curl is 
\begin{align}
    \epsilon_a^{\;cd} \mathcal{D}_c T_{db}^{(V,E,p)\:n\ell m} =& \frac{1}{2}(n+1) T^{(V,B,p)\:n\ell m}_{ab} \nn\\ 
    &\hspace{-1cm}-\frac{1}{4}(n-1)(n+3)\epsilon^{\quad c}_{ab}V^{(E,p)n\ell m}_c,\\\epsilon_a^{\;cd} \mathcal{D}_c T_{db}^{(V,E,q)\:n\ell m} =& -\frac{1}{2}(n+1) T^{(V,B,q)\:n\ell m}_{ab}\nn\\ 
    &\hspace{-1cm}-\frac{1}{4}(n-1-\frac{1}{4}\delta_{n, 1}\delta_{\ell 1})(n+3)\epsilon^{\quad c}_{ab}V^{(E,q)n\ell m}_c,
    \end{align}
\begin{align}
    \epsilon_a^{\;cd} \mathcal{D}_c T_{db}^{(V,B,p)\:n\ell m} =& -\frac{1}{2}(n+1+\delta_{n,-1}\sqrt{\ell(\ell+1)}) T^{(V,E,p)\:n\ell m}_{ab}\nn\\ 
    &\hspace{-1cm}-\frac{1}{4}(n-1)(n+3)\epsilon^{\quad c}_{ab}V^{(B,p)n\ell m}_c,\\\epsilon_a^{\;cd} \mathcal{D}_c T_{db}^{(V,B,q)\:n\ell m} =& \frac{1}{2}(n+1-\delta_{n,-1}\sqrt{\ell(\ell+1)}) T^{(V,E,q)\:n\ell m}_{ab} \nn\\ 
    &\hspace{-1cm}-\frac{1}{4}(n-1-\frac{1}{4}\delta_{n, 1}\delta_{\ell 1})(n+3)\epsilon^{\quad c}_{ab}V^{(B,q)n\ell m}_c . 
\end{align}

Under the KG inner product, we get
\begin{align}\label{TensKGVect}
(T^{(V,D,c)n\ell m },T^{(V,D',d)n\ell' m'})_{KG}^T &=\epsilon_{cd} \delta_{\ell \ell'}\delta_{D D'}\delta_{m m'}\frac{1}{2} (n-1 \nn \\ 
&\hspace{-1.5cm}-\frac{1}{4}\delta_{n, 1}\delta_{\ell 1})(n+3)(1-\delta_{n,-1}\delta_{DE}). 
\end{align}
It vanishes for $n=-1$, $D=E$ and for $n=1$, $\ell\ge2$.

Since $V_{a}^{(E,c)\: -1\ell m}\propto \mathcal{D}_a \psi_{0\ell m}^{(c)}$, see Eqs.  \eqref{TransverseVectLongVect1}-\eqref{TransverseVectLongVect2} and \eqref{eq:75}, the tensor harmonics $ T_{ab}^{(V,E,c)\:-1\ell m}$ are related to  $ T_{ab}^{(S,c)\:0\ell m}$ as follows 
\begin{align}
    T_{ab}^{(V,E,p)\:-1\ell m} &=\frac{1}{\delta_{\ell,0}+1}T_{ab}^{(S,p)\:0\ell m} ,\label{TVEToTSp}\\ T_{ab}^{(V,E,q)\:-1\ell m} &= \frac{1}{\ell(\ell+1)}T_{ab}^{(S,q)\:0\ell m},\quad  \ell\ge1. \label{TVEToTSq}
\end{align}

\subsection{SDT tensor harmonics}
\label{SDTsec}

Finally, let us study the symmetric divergence-free and traceless (SDT) tensor harmonics on $dS_3$. In the literature \cite{Higuchi:1986wu}, they are defined as
\begin{align}
\mathcal{D}^a T^{(SDT)n}_{ab} = 0, \quad \left(T^{(SDT)n}\right)^a_{\quad a} = 0, \nn \\ 
\Box T^{(SDT)n}_{ab} = - \left[(n+1)^2-3\right] T^{(SDT)n}_{ab}.
\label{generalSDTTensorEquations}
\end{align}

In the Beig-Schmidt expansion such wave operators appear at order $n+2$ in the expansion \cite{beig1982einstein}. Such tensors have been studied in detail in \cite{Compere:2011db} but the proof implicitly assumed $\ell \ge n+1$ in their Eq. (A.110), see our Eqs. \eqref{psi(p)nlm}-\eqref{psi(q)nlm} as a comparison. We generalize this analysis to arbitrary $n\ge -1$, $\ell \geq 0$. We further derive the asymptotic properties of tensor harmonics and their transformation law under the antipodal map.

In order to find the general solution to Eq. \eqref{generalSDTTensorEquations}, we start with the decomposition in spherical harmonics 
\begin{align}
    T_{\tau\tau}^{\ell m} &= f_1(\tau) Y_{\ell m}(\theta,\phi), \label{generaltensortautau}\\
    T_{\tau A}^{\ell m} &= f_2(\tau) \mathring V_{A}^{(E)\ell m}(\theta,\phi) + f_3(\tau) \mathring{V}_{A}^{(B)\ell m}(\theta,\phi) \label{generaltensorAtau} \\
    T_{AB}^{\ell m} &= f_4(\tau) \mathring T_{AB}^{(E)\ell m}(\theta,\phi) + f_{5}(\tau) \mathring  T^{(B)\ell m}_{AB}(\theta,\phi) \nonumber \\
    &+ f_6(\tau)\mathring T_{AB}^{(S)\ell m}(\theta,\phi)  \label{generaltensorAB},
\end{align}
where $\mathring T_{AB}^{(E)\ell m}(\theta,\phi)$, $\mathring T^{(B)\ell m}_{AB}(\theta,\phi)$ and $\mathring T^{(S)\ell m}(\theta,\phi)$ are the normalized tensor harmonics on the sphere  
\begin{align}
    \mathring T_{AB}^{(E)\ell m}(\theta,\phi):=&\sqrt{\frac{2 (\ell-2)!}{(\ell+2)!}}\left(\nabla_A\nabla_B+\frac{\ell(\ell+1)}{2} \gamma_{AB}\right) Y_{\ell m}(\theta,\phi),
    \end{align}
\begin{align}
\mathring T_{AB}^{(B)\ell m}(\theta,\phi):=&\sqrt{\frac{2 (\ell-2)!}{(\ell+2)!}}\nabla_{(A}\epsilon_{B)C}\nabla^C Y_{\ell m}(\theta,\phi),\\
\mathring T_{AB}^{(S)\ell m}(\theta,\phi):=&\frac{1}{\sqrt{2}}\gamma_{AB} Y_{\ell m}(\theta,\phi).
\end{align}
$\mathring T_{AB}^{(E)\ell m}(\theta,\phi)$, $\mathring T^{(B)\ell m}_{AB}(\theta,\phi)$ are defined only for $\ell\ge2$ and are taken to vanish for $\ell=0,1$ in Eq. \eqref{generaltensorAB}.
When imposing the trace condition, we find 
\begin{equation}
    f_1(\tau) = \sqrt{2} \sech^2\tau f_6(\tau).
\label{f1functionf6}
\end{equation}

There is only one solution to the divergence condition for $\ell = 0$. It requires $n=-1$. The solution is a traceless tensor harmonic built from the scalar $\psi_{100}^{(p)}$ : 
\begin{equation}
    T_{ab}^{(E,p)-100} := \mathcal{D}_a\mathcal{D}_b \psi_{100}^{(p)}+ q_{ab} \psi_{100}^{(p)}.\label{TEminus100}
\end{equation}

As in \cite{Compere:2011db}, we have proven that 
\begin{lemma}\label{lemma:T1}
Let $T_{ab}$ be a SDT tensor field on $dS_3$ satisfying
\begin{align}
\left(\Box-\alpha\right)T_{ab}=0.
\end{align}
If $\alpha\neq 3$, then $T_{ab}$ has no $\ell=0$ harmonic.
\end{lemma}
After analysis (see also \cite{Compere:2011db}), there exist 9 independent SDT tensors for $\ell=1$.  Three of them arise for $n=-1$. These three tensors are traceless tensor harmonics constructed from the scalars $\psi^{(p)}_{11m}$, $m=-1,0,1$ as
\begin{equation}
    T_{ab}^{(E,p)-11m} := \mathcal{D}_a\mathcal{D}_b \psi^{(p)}_{11m} + q_{ab} \psi^{(p)}_{11m},\qquad m=-1,0,1. \label{TEminus11m}
\end{equation}
They obey the equation 
\begin{equation}
    \left(\Box-3\right)T_{ab}^{(E,p)-11m} = 0,\qquad m=-1,0,1 ,
\end{equation}

The 6 other $\ell=1$ SDT harmonic tensors have $n=0$. One can construct them from $n=1$ $\ell =1$ vector harmonics with $p$ parity as
\begin{align}
    T_{ab}^{(E,p)01m}:=\mathcal{D}_{(a}V^{(E,p)11m}_{b)}\label{TE01m},\\
    T_{ab}^{(B,p)01m}:=\mathcal{D}_{(a}V^{(B,p)11m}_{b)}. \label{TB01m}
\end{align} 
They satisfy the following equation of motion 
\begin{equation}
    (\Box-2)T_{ab}^{(D,p)01m} = 0, \qquad D=E,B,
\end{equation}
i.e., Eq. \eqref{generalSDTTensorEquations} with $n=0$. 
This derivation proves the following lemma : 
\begin{lemma}\label{lemma:T2}
Let $T_{ab}$ be a SDT tensor field on $dS_3$ satisfying
\begin{align}
\left(\Box-\alpha\right)T_{ab}=0.
\end{align}
If $\alpha\neq2, 3$, then $T_{ab}$ has no $\ell=0,1$ harmonic.
\end{lemma}

For $\ell \ge 2$, the divergence-free conditions lead to the relations 
\begin{align}
f_2(\tau)= & -\frac{\sqrt{2}}{\sqrt{\ell(\ell+1)}}\left(\tanh \tau +\partial_\tau \right)f_6(\tau), \\
f_4(\tau)= & \sqrt{\frac{(\ell-2)!}{(\ell+2)!}}\left[\left(\ell+\ell^2+2 \cosh 2 \tau\right) +2 \cosh^2 \tau \right.\nn\\ 
&\times \left. \left(3 \tanh \tau \partial_\tau +\partial^2_\tau\right)\right]f_6(\tau), \\
f_5(\tau)= & -\frac{\sqrt{2}}{\sqrt{(\ell-1)(\ell+2)}} \cosh^2 \tau\left(2 \tanh \tau + \partial_\tau \right)f_3(\tau) .
\end{align}
The SDT tensor harmonics $\ell \ge 2$ are determined by 2 functions $f_3(\tau)$ and $f_6(\tau)$. These lead to two families of SDT tensors characterized by their parity over the sphere. We define the E parity tensors
\begin{equation}
  T^{(E)\ell m}_{ab}(f) := T^{\ell m}_{ab}\left(f_3(\tau)=0,f_6(\tau) = \sqrt{\frac{(\ell+2)!}{2(\ell-2)!}} f(\tau)\right)
\end{equation}
and the B parity tensors
\begin{align}
    T^{(B)\ell m}_{ab}(\tilde f) :=  T^{\ell m}_{ab} &\Big(f_3(\tau)=\sqrt{(\ell-1)(\ell+2)}\tilde f(\tau) \nn \\ 
    & ,f_6( \tau) = 0 \Big).
\end{align}

Explicitly, we get for $\ell\ge2$:
\begin{align}
    T^{(E)\ell m}_{\tau\tau} (f)&=\sqrt{(\ell-1)\ell(\ell+1)(\ell+2)} \sech^2\tau f(\tau) Y_{\ell m}(\theta,\phi),\label{TEttSDT}\\
    T^{(E)\ell m}_{\tau A}(f) &= -\sqrt{(\ell-1)(\ell+2)}\left(\tanh \tau +\partial_\tau \right)f(\tau) \mathring V_{A}^{(E)\ell m}(\theta,\phi)  ,\\
    T^{(E)\ell m}_{AB}(f) &= \frac{1}{\sqrt{2}}\left[\left(\ell+\ell^2+2 \cosh 2 \tau\right) +2 \cosh^2 \tau \right. \nn\\ 
    &\left. \times \left(3 \tanh \tau \partial_\tau +\partial^2_\tau\right)\right]f(\tau)\mathring T_{AB}^{(E)\ell m}(\theta,\phi) \nonumber \\&+ \frac{1}{\sqrt{2}}\sqrt{(\ell-1)\ell(\ell+1)(\ell+2)}f(\tau)\mathring T_{AB}^{(S)\ell m}(\theta,\phi),
\end{align}
and
\begin{align}
    T^{(B)\ell m}_{\tau\tau}(\tilde f) &= 0, \\
    T^{(B)\ell m}_{\tau A}(\tilde f) &= \sqrt{(\ell-1)(\ell+2)}\tilde f(\tau) \mathring{V}_{A}^{(B)\ell m}(\theta,\phi), \\
    T^{(B)\ell m}_{AB}(\tilde f) &= -\sqrt{2} \cosh^2 \tau\left(2 \tanh\tau +\partial_\tau \right) \tilde f(\tau) \mathring T_{AB}^{(B)\ell m}(\theta,\phi)\label{TBABSDT} .
\end{align}
Upon imposing Eqs. \eqref{generalSDTTensorEquations}, we find that $f$ and $\tilde f$ must obey Eq. \eqref{generalscalarequation}. Hence, we distinguish 4 families of $\ell\ge2$ transverse tensors. For $n\ge1$, we define  :
\begin{equation}
    T^{(E,c)n\ell m}_{ab}:=\frac{1}{(n+1)\sqrt{2n(n+2)}} T^{(E)\ell m}_{ab}(f=\psi^{(c)}_{n\ell}(\tau)),
\end{equation}
for $c=p,q$ and 
\begin{align}
     T^{(B,p)n\ell m}_{ab} &:=\frac{1}{\sqrt{2n(n+2)}} T^{(B)\ell m}_{ab}(f=\psi^{(p)}_{n\ell}(\tau)),\nn \\  T^{(B,q)n\ell m}_{ab} &:=- \frac{1}{\sqrt{2n(n+2)}}T^{(B)\ell m}_{ab}(f=\psi^{(q)}_{n\ell}(\tau)),
\end{align}
with $\psi^{(c)}_{nl}(\tau)$ being the harmonics functions defined in Eqs. \eqref{psi(p)nlm} and \eqref{psi(q)nlm}. 

A particular case of SDT tensor harmonics occurs for $n=-1$. In that case, the tensors $T_{ab}^{(E)\:\ell m}(f=\psi^{(c)}_{-1\ell})$ are proportional to a scalar $n=1$ tensor harmonic \eqref{scalarTensorHarmonics}: 
\begin{align}
    T_{ab}^{(E)\:\ell m}(f=\psi^{(p)}_{-1\ell }) &= \sqrt{\frac{(\ell+2)!}{(\ell-2)!}}(\mathcal{D}_a \mathcal{D}_b\psi^{(p)}_{1\ell m} + q_{ab}\psi^{(p)}_{1\ell m}),\\
 T_{ab}^{(E)\:\ell m}(f=\psi^{(q)}_{-1\ell }) &=\sqrt{\frac{(\ell-2)!}{(\ell+2)!}}(\mathcal{D}_a \mathcal{D}_b\psi^{(q)}_{1\ell m} + q_{ab}\psi^{(q)}_{1\ell m}).
\end{align}
Furthermore, starting from Eqs. \eqref{TEttSDT}-\eqref{TBABSDT} and imposing the condition $\text{Curl}(T)_{ab}=0$, we find that the only solutions are a linear combination of  
\begin{equation}
    f(\tau)=\psi^{(c)}_{-1 \ell}(\tau),\qquad \tilde f(\tau)=0,
\end{equation}
for $c=p,q$, $\ell \geq 2$. As a consequence and using Eqs. \eqref{TEminus100}-\eqref{TEminus11m} for low harmonics, we have the following lemma already stated in \cite{Ashtekar:1978zz} and derived as Lemma 1 in \cite{Compere:2011db}: 
\begin{lemma}\label{lemma:T3}
Any symmetric, traceless, divergence-free and curl-free tensor $t_{ab}$ on dS$_3$ obeys 
\begin{equation}
    (\Box-3)t_{ab} = 0
\end{equation}
and is a linear combination of the tensors  $T_{ab}^{(E,c)\:-1\ell m}$. Such a tensor can be expressed in terms of a scalar $\Phi$ obeying $(\Box + 3)\Phi=0$ as 
\begin{equation}
    t_{ab}=\mathcal{D}_a\mathcal{D}_b \Phi + q_{ab} \Phi .
\end{equation}    
The scalar $\Phi$ is ambiguous under a shift by a linear combination of harmonics $\psi^{(q)}_{100}$, $\psi^{(q)}_{11m}$, $m=-1,0,1$.
\end{lemma}
The latter ambiguity is due to Eqs. \eqref{LogTranslationEquation0}-\eqref{LogTranslationEquation}. 
For $n=-1$, we therefore define the following harmonics for $\ell\ge2$
\begin{align}
   T_{ab}^{(E,p)\:-1\ell m}&:= \frac{1}{\sqrt{(\ell-1)\ell(\ell+1)(\ell+2)}}T_{ab}^{(E)\:\ell m}(f=\psi^{(p)}_{-1\ell });\\
   T_{ab}^{(E,q)\:-1\ell m}&:= \frac{1}{\sqrt{(\ell-1)\ell(\ell+1)(\ell+2)}}T_{ab}^{(E)\:\ell m}(f=\psi^{(q)}_{-1\ell });\\
   T_{ab}^{(B,p)\:-1\ell m}&:= \frac{1}{\sqrt{2}}T_{ab}^{(B)\:\ell m}(f=\psi^{(p)}_{-1\ell });\\
   T_{ab}^{(B,q)\:-1\ell m}&:= \frac{1}{\sqrt{2}}T_{ab}^{(B)\:\ell m}(f=\psi^{(q)}_{-1\ell }). 
\end{align}
In the other particular case $n=0$, the SDT harmonics (which have $\ell \geq 2$) are derived from the $n=1$ vector harmonics as
\begin{align}\label{propTD}
    T_{ab}^{(E)\:\ell m}(f=\psi^{(p)}_{0\ell }) &:=2 \sqrt{(\ell+2)(\ell-1)}\mathcal{D}_{(a}V_{b)}^{(E,p)\: 1\ell m};\\
    T_{ab}^{(E)\:\ell m}(f=\psi^{(q)}_{0\ell }) &:=- \frac{2}{\sqrt{(\ell+2)(\ell-1)}}\mathcal{D}_{(a}V_{b)}^{(E,q)\: 1\ell m};\\
    T_{ab}^{(B)\:\ell m}(f=\psi^{(p)}_{0\ell }) &:=2 \sqrt{(\ell+2)(\ell-1)}\mathcal{D}_{(a}V_{b)}^{(B,p)\: 1\ell m};\\
    T_{ab}^{(B)\:\ell m}(f=\psi^{(q)}_{0\ell }) &:=\frac{2}{\sqrt{(\ell+2)(\ell-1)}} \mathcal{D}_{(a}V_{b)}^{(B,q)\: 1\ell m}.
\end{align}
By construction, the vector harmonics $V_{a}^{(D,c)}$, $D=E,B$, $c=p,q$, are transverse. This leads to the following lemma: 
\begin{lemma}\label{lemma:T4}
Any symmetric tensor $T_{ab}$ on $\text{dS}_3$ obeying
\begin{align}
  (\square-2)T_{ab}=0,\qquad T^a_a=0, \qquad \mathcal D^b T_{ab}=0 
\end{align}
can be written as 
\begin{equation}
    T_{ab} = \mathcal D_{(a} V_{b)}
\end{equation}
where $V_a$ obeys  $(\square+2)V_a =0$ and $\mathcal D_a V^a=0$. The vector $V_a$ is ambiguous under a shift by a Killing vector $\chi^a$.      
\end{lemma}
The latter ambiguity is due to Killing equations \eqref{KillingEquation}. 

For $n=0$, we thus define for $\ell\ge2$
\begin{align}
   T_{ab}^{(E,p)\:0\ell m}&:= \frac{1}{\sqrt{(\ell-1)(\ell+2)}}T_{ab}^{(E)\:\ell m}(f=\psi^{(p)}_{0\ell });\\
   T_{ab}^{(E,q)\:0\ell m}&:= -\frac{1}{\sqrt{(\ell-1)(\ell+2)}}T_{ab}^{(E)\:\ell m}(f=\psi^{(q)}_{0\ell });\\
   T_{ab}^{(B,p)\:0\ell m}&:= \frac{1}{\sqrt{(\ell-1)(\ell+2)}}T_{ab}^{(B)\:\ell m}(f=\psi^{(p)}_{0\ell });\\
   T_{ab}^{(B,q)\:0\ell m}&:= \frac{1}{\sqrt{(\ell-1)(\ell+2)}}T_{ab}^{(B)\:\ell m}(f=\psi^{(q)}_{0\ell })  .
\end{align}

The transformation laws of the SDT tensor harmonics under the antipodal map  are 
\begin{align}
      \Upsilon_{\mathcal{H}}^*[T_{ab}^{(E,q)\:n\ell m}] &= (-1)^{n} T_{ab}^{(E,q)\:n\ell m},\\  \Upsilon_{\mathcal{H}}^*[T_{ab}^{(E,p)\:n\ell m}] &= (-1)^{n+1} T_{ab}^{(E,p)\:n\ell ,m},\label{SDTHarmonicsParityE}\\
      \Upsilon_{\mathcal{H}}^*[T_{ab}^{(B,q)\:n\ell m}] &= (-1)^{n} T_{ab}^{(B,q)\:n\ell m},\\  \Upsilon_{\mathcal{H}}^*[T_{ab}^{(B,p)\:n\ell m}] &= (-1)^{n+1} T_{ab}^{(B,p)\:n\ell m}.\label{SDTHarmonicsParityB}
\end{align}
The curl maps the E harmonics to the B harmonics and vice-versa: 
\begin{align}
    \epsilon_a^{\;cd} \mathcal{D}_c T_{db}^{(E,p)\:n\ell m} &=  (n+1) T^{(B,p)\:n\ell m}_{ab},\\ \epsilon_a^{\;cd} \mathcal{D}_c T_{db}^{(B,p)\:n\ell m} &= -\Big(n+1+\delta_{n,-1}\sqrt{\frac{(\ell+2)!}{2(\ell-2)!}}\Big) T_{ab}^{(E,p)\:n\ell m},\\
    \epsilon_a^{\;cd} \mathcal{D}_c T_{db}^{(E,q)\:n\ell m} &=  -(n+1) T^{(B,q)\:n\ell m}_{ab},\\ \epsilon_a^{\;cd} \mathcal{D}_c T_{db}^{(B,q)\:n\ell m} &=  \Big(n+1-\delta_{n,-1}\sqrt{\frac{(\ell+2)!}{2(\ell-2)!}}\Big)T_{ab}^{(E,q)\:n\ell m}.
\end{align}

In summary, the general SDT solution to Eq. \eqref{generalSDTTensorEquations} can be decomposed as 
\begin{align}
    T_{ab}^{(SDT) n} &= \delta_{n, -1} \sum_{\ell=0}^{1}\sum_{m=-\ell}^\ell a_{\ell m} T^{(E,p)-1\ell m}_{ab} \nn\\ 
    &+ \delta_{n,0} \sum_{m=-1}^1 \left(a^{(E)}_m T^{(E,p)01m}_{ab}+a^{(B)}_m T^{(B,p)01m}_{ab} \right)\nn \\
    &+ \sum_{\ell \ge2, m} \left( a^{(E,p)}_{\ell m}T_{ab}^{{(E,p)n\ell m}} +a^{(E,q)}_{\ell m}T_{ab}^{{(E,q)n\ell m}} \right.\nn  \\ 
     & \left. +a^{(B,p)}_{\ell m}T_{ab}^{{(B,p)n\ell m}} +a^{(B,q)}_{\ell m}T_{ab}^{{(B,q)n\ell m}} \right).
\end{align}

The inner products between the tensor basis $T^{(E,p)n\ell m}$, $T^{(E,q)n\ell m}$, $T^{(B,p)n\ell m}$ and $T^{(B,q)n\ell m}$ $\forall \ell \geq 2$, $\forall \vert m \vert \le \ell$ can be obtained as 
\begin{align}
(T^{(D,c)n\ell m},T^{(D',d)n\ell' m'})_{KG}^T &=\epsilon_{cd}\delta_{\ell\ell'}\delta_{mm'}\delta_{D,D'}\nn\\ 
&\times (1-\delta_{n,-1}\delta_{DE}-\delta_{n,0}).
\end{align}
These inner products are zero for $n=0$. In that case the tensors can be expressed in terms of vector harmonics. For $n=-1$, the E parity tensors are determined in terms of $n=1$ scalar tensor harmonics, which also have a vanishing inner product.

Furthermore, any SDT tensor defines a divergence-free vector as follows 
\begin{equation}
    \xi_a^T := \cosh\tau \, n^b T_{ab}. \label{XiaT}
\end{equation}
If $T_{ab}$ satisfies Eq. \eqref{GeneralTensEqu}, then $\xi_a^T$ satisfies
\begin{equation}
    (\Box - \alpha+1)\xi_a^T = \xi_a^S \label{EquationXiaT}
\end{equation}
with $\xi_a^S$ the vector associated to $S_{ab}$ in \eqref{GeneralTensEqu}, i.e. $\xi_a^S=\cosh\tau \, n^b S_{ab}$. 
The associated vector KG inner product with vectors defined from the $\ell \geq 2$ SDT tensor harmonics is \begin{align}
    &( \xi^{(D,c)n\ell m},\xi^{(D',d)n\ell' m'} )_{KG}^V = -\frac{(\ell-1)(\ell+2)}{2n(n+2)}\nn\\
    & \hspace{1.5cm}\times (T^{(D,c)n\ell m},T^{(D',d)n\ell' m'} )_{KG}^T ,\qquad n\ne0,\label{xixiKGprod}\\&
   ( \xi^{(D,c)0\ell m},\xi^{(D',d)0\ell' m'} )_{KG}^V =\delta_{\ell\ell'}\delta_{mm'}\delta_{DD'}\epsilon_{cd},
\end{align}
for $D',D=E,B$, $c,d=p,q$ and we defined $\xi^{(D,c)n\ell m}:= \cosh\tau \, n^b T_{ab}^{(D,c)n\ell m}$. 

For $n=0$, this vector formulation allows to have a well-defined inner product for $\ell\ge2$. For $n=-1$, the vector inner product \eqref{xixiKGprod} is well-defined for B parity vectors but not for E parity vectors. This is because $\xi^{(E,d)-1\ell' m'} $ corresponds to the $n=-1$ E parity transverse vector harmonics (see Eq.\eqref{TransverseVectLongVect1}). Hence, one must use the scalar $\psi^{E(c)-1\ell m}_\xi = \cosh\tau\,  n^a\xi^{(E,c)-1\ell m}_a$ which has the following normalization under the scalar KG inner product  
\begin{equation}
    ( \psi^{E(c)-1\ell m}_\xi,\psi^{E(d)-1\ell' m'}_\xi )_{KG}=\delta_{\ell\ell'}\delta_{mm'}\epsilon_{cd}, \quad \ell\ge2. 
\end{equation}
For $\ell=0,1$ harmonics, the SDT tensors $T^{(D,p)01 m}=T^{(V,D,p)11m}$ with $\vert m\vert\le1$ and $T^{(E,p)-1\ell m}=T^{(S,p)1\ell m}$ with $\ell=0,1$ and $\vert m\vert\le\ell$ admit a well-defined tensor inner products with, respectively, $T^{(V,D,q)11m}$ and $T^{(S,q)1\ell m}$; see Eqs. \eqref{TensKGScal}-\eqref{TensKGVect}. Equivalently, the vectors $\xi^{(D,p)01m}$ admit a well-defined vector KG inner product with the vector harmonics $V^{(D,q)01m}$ for $D=E,B$, and $\vert m\vert\le1$ and the scalars $\psi^{E(p)-1\ell m}_\xi$ admits a well-defined scalar KG inner product with $\psi^{(q)}_{-1\ell m}$ for $\ell=0,1$ and $\vert m\vert\le\ell$.


The tensor KG inner product, the vector KG inner product with vectors built from tensors and the scalar inner product with scalars built from these vectors are three ways of defining conserved quantities with SDT tensors satisfying Eq. \eqref{generalSDTTensorEquations}. In the general case $n \geq 1$ they are equivalent, while in the special case $n=0$ the tensor KG product vanishes and in the special case $n=-1$ both the tensor and the $E$ parity vector KG products vanish. 

In addition to these charges, a special role is played by the $\ell=0,1$ harmonics in relationship to the Killing and conformal Killing vectors of $dS_3$  \cite{Compere:2011db}. We recall the result as follows. Let us define the following charge for any tensor $T_{ab}$ and vector $V^a$, 
\begin{equation}\label{defQ}
    Q_{T,V}(\tau) := \oint_{S^2(\tau)} d\Omega\;n^a T_{ab}\overline{V^b}.
\end{equation}
For any symmetric divergence-free tensor $ Q_{T,\chi}$ is conserved if $\chi^b$ is a Killing vector of $dS_3$. Here overbars denote complex conjugation which is necessary for complex $m=\pm 1$ modes. The definition of the Lorentz charges at spatial infinity for asymptotically flat spacetime correspond to a charge of the form \eqref{defQ} (see e.g. \cite{Compere:2011ve}). In addition, for any SDT tensor, $Q_{T,V}$ is conserved if $V^a$ is a conformal Killing vector of $dS_3$. This structure matches with the logarithmic translations charges at spatial infinity for asymptotically flat spacetime \cite{Compere:2011ve}. 

Lemmae \ref{lemma:T1} and \ref{lemma:T2} lead to the following lemma that generalizes Lemma 5 of \cite{Compere:2011db}:
\begin{lemma}\label{lemma:T5}
Let $T_{ab}$ be a SDT tensor satisfying $(\Box-\alpha)T_{ab} = 0$, $\alpha\in\mathbb R$. If 
$(\Box-3)T_{ab} \ne 0$, 
then
$    Q_{T,\hat\chi} =0 
$ for any proper conformal Killing vector $\hat\chi^a$. If $(\Box-2)T_{ab} \ne 0$  then $Q_{T,\chi} =0$ for any Killing vector $\chi^a$.
\end{lemma}
Indeed, for any $\ell\ge2$ harmonics, $Q_{T,\chi}=Q_{T,\hat\chi}=0$ from the orthogonality of spherical harmonics. For $\ell=0,1$ harmonics, only SDT tensors satisfying Eq. \eqref{generalSDTTensorEquations} with $n=-1$, $n=0$ possess $\ell=0,1$ harmonics (see Lemmae \ref{lemma:T1} and \ref{lemma:T2}). Let us compute explicitly $Q_{T,\chi}$ and $ Q_{T,\hat\chi}$ in that case: 
\begin{align}
    Q_{T^{(D',p),0\ell m},V^{(D,q)11m'}} &=-\frac{1}{2}\delta_{\ell1}\delta_{m,m'}\delta_{D'D}, \\
    Q_{T^{(E,p),-1\ell m},V^{(D,q)11m'}} &=0,\\
    Q_{T^{(E,p),-1\ell m},V^{(L,q)1\ell' m'}} &= -2 \delta_{\ell\ell'}\delta_{m,m'},\\ 
    Q_{T^{(D,p),0\ell m},V^{(L,q)1\ell' m'}} &= 0, 
\end{align}
with $\ell=0,1$, $\vert m\vert\le\ell$ and $D=E,B$.

This proves Lemma \ref{lemma:T5}.  Lemmae \ref{lemma:T4} and \ref{lemma:T5} imply Lemma 3 of \cite{Compere:2011db} after using the properties \eqref{curlpV}.

\subsection{Additional properties of tensors}\label{sec:PropertiesTensors}
Let us derive more general features of tensors on $dS_3$.

First, there are two equivalence relationships between either the vector KG inner product or the scalar KG inner product and the charges \eqref{defQ} associated with a vector. For any symmetric tensor $T^{(V)}_{ab}$ built from a transverse vector $V^{(T)}_a$ (see Eq.\eqref{VectorTensor}) and for any Killing vector $\chi^a$, we have 
\begin{align}
-(V^{(T)},\chi)_{KG}^V   
 &=\oint_{S^2(\tau)} d\Omega n_a\left(\overline\chi^b \mathcal{D}^a V^{(T)}_b-V^{(T)}_b \mathcal{D}^a \overline\chi^b\right)\nn\\
 &= \oint_{S^2(\tau)} d\Omega n_a\left(\overline\chi^b \mathcal{D}^a V^{(T)}_b+V^{(T)}_b \mathcal{D}^b \overline\chi^a\right) \nn\\
 &=  \oint_{S^2(\tau)} d\Omega n_a\left(\overline\chi^b \mathcal{D}^a V^{(T)}_b+\mathcal{D}^b (V^{(T)}_b  \overline\chi^a) \right)\nn\\
  &=  \oint_{S^2(\tau)} d\Omega n_a\left(\overline\chi^b \mathcal{D}^a V^{(T)}_b+\mathcal{D}^b (V^{(T)a}  \overline\chi_{b}) \right)\nn \\
    &= \oint_{S^2(\tau)} d\Omega n_a\left(\overline\chi^b \mathcal{D}^a V^{(T)}_b+\mathcal{D}^b V^{(T)a}  \overline\chi_{b} \right)\nn \\
     &=2 \oint_{S^2(\tau)} d\Omega n^a\overline\chi^b T^{(V)}_{ab}= 2 Q_{T^{(V)},\chi}. 
 \end{align}
We used Killing's equations in the second step and, in the fourth step, we added a total derivative. In the final step we used the definition \eqref{defQ}. 

Similarly, let us consider the class of tensors that can be written as 
\begin{equation}
    k_{ab}=\mathcal{D}_a \mathcal{D}_b\phi -(\Box+2)q_{ab}\phi\label{kabtensor}
\end{equation}
for an arbitrary $\phi$. Such tensors satisfy 
\begin{equation}
    k^a_a=-2 (\Box+3)\phi,\quad\mathcal{D}^bk_{ab} = 0,\quad \text{Curl}(k_{ab})=0. 
\end{equation}
Given a proper conformal Killing vector $\hat\chi^a=\mathcal{D}^a H$, we have 
\begin{align}
    Q_{k,\hat\chi} &= \oint_{S^2(\tau)} d\Omega\;n^a k_{ab}\overline{\hat\chi^b}\\
    &= \oint_{S^2(\tau)} d\Omega\;n^a(\mathcal{D}_a \mathcal{D}_b\phi -(\Box+2) q_{ab}\phi)\mathcal{D}^b\overline H\\&=\oint_{S^2(\tau)} d\Omega\;n^a\Big(-(\Box+2)\phi\mathcal{D}_a \overline H + \mathcal{D}_b \overline H \mathcal{D}_a\mathcal{D}^b \phi \Big)\\&=2\oint_{S^2(\tau)} d\Omega\;n^a\Big(-\phi \mathcal{D}_a \overline H+\mathcal{D}_a\phi\overline  H-\mathcal{D}^b(\mathcal{D}_{[a}\overline H\mathcal{D}_{b]}\phi)\Big)\\&=-2(\phi,H)_{KG}
\end{align}
where we used that $H$ satisfies Eq.  \eqref{HlogEquation}. This generalizes Eq. (233) of  \cite{Compere:2023qoa}.

Let us finally derive two additional lemmae. We have 
\begin{lemma}\label{lemma:T6}A symmetric tensor $T_{ab}$ of $dS_3$ obeys 
\begin{equation}
    \text{Curl}( T_{ab}) =0
\end{equation}
if and only if it can be written in the form \eqref{Tscal}. If one further imposes 
\begin{equation}
    (\Box + (n+1)^2-7) T_{ab}=0, \quad T^{a}_a = 0,\label{eqS1}
\end{equation}
then $T_{ab}$ admits the decomposition \begin{align}
     T_{ab} &= \sum_{\ell, m} \bigg( a_{\ell m} T^{(S,p)n\ell m}_{ab} + b_{\ell m}T^{(S,q)n\ell m}_{ab} \bigg) 
\end{align}
where $a_{\ell m} $ , $b_{\ell m}$ are constants and $T^{(S,c)n\ell m}_{ab}$ ($c=p,q$) are the scalar tensor harmonics defined in Eqs.\eqref{scalarTensorHarmonics}, \eqref{TSq0} and \eqref{TSQ1lm}. 
\end{lemma}
Let us prove this lemma. Let $T_{ab}$ be a tensor in $dS_3$. It can be decomposed as follows \cite{Deser:1967zzb} 
\begin{equation}
    T_{ab} = T_{ab}^{(S)}+\mathcal{D}_{(a}V_{b)} + \bar T_{ab}\label{tensordecomp}
\end{equation}
where $V_a$ is a transverse vector, $T_{ab}^{(S)}$ is a tensor of the form \eqref{Tscal} and $\bar T_{ab}$ is an SDT tensor. 

Let us focus on the second term. If $T_{ab} = \mathcal{D}_{(a}V_{b)}$, then 
\begin{equation}
    \text{Curl}( T_{ab}) =\frac{1}{2}\mathcal{D}_{(a}\text{Curl}( V_{b)})\label{eq343}
\end{equation}
In the previous section (see Lemma \ref{lemma:V2}), we have seen that the only transverse vectors that satisfy $\text{Curl}(V_{a})=0$ are also longitudinal and can be written as  $V_{a}=\mathcal{D}_a\phi$ with $\phi$ a scalar that satisfies Eq. \eqref{generalscalarequation} with $n=0$ ($\Box\phi=0$). If $\text{Curl}(V_{a})$ is a Killing vector, it implies that $V_{a}$ is also a Killing vector. In that case, $\mathcal{D}_{(a}V_{b)}=0$. 
Hence, the tensor $T_{ab}$ is 
\begin{equation}
    T_{ab} = \mathcal{D}_a\mathcal{D}_b \phi
\end{equation}
which corresponds to a combination of tensor harmonics built from scalars that satisfy Eq. \eqref{eqS1} with $n=0$.

Regarding the third term, we can use Lemma \ref{lemma:T3} that implies that any SDT tensor such that $\text{Curl}( T_{ab})=0$ can be written as  
\begin{equation}
    \mathcal{D}_a\mathcal{D}_b \phi + q_{ab}\phi
\end{equation}
with $\phi$, a scalar satisfying \eqref{generalscalarequation} with $n=1$. 
We have so far shown that the condition $\text{Curl}( T_{ab})=0$
imposes that $T_{ab}$ is a tensor of the form \eqref{Tscal}. The converse is true from Eq. \eqref{ZeroCurTab}. Further imposing the traceless condition and the equation of motion, we find that  
\begin{equation}
    T_{ab}=\mathcal{D}_a\mathcal{D}_b \phi - \frac{1}{3}q_{ab} \Box \phi
\end{equation}
with $\phi$ satisfying Eq. \eqref{generalscalarequation} or proportional to the solutions \eqref{psi0}, \eqref{I00}-\eqref{I1m}, which proves the lemma.

The next lemma is 
\begin{lemma}\label{lemma:T8}
A symmetric tensor $T_{ab}$ on $dS_3$ satisfies 
\begin{align} 
    &\mathcal{D}_{(a}\mathcal{D}^cT_{cb)} = \frac{1}{2}(\Box-2)T_{ab},\quad T^a_a=0,
\end{align}
if and only if there exists a transverse vector $V_{a}$ such that $T_{ab}=\mathcal{D}_{(a}V_{b)}$.
If we further impose 
\begin{equation}
    (\Box+\left[(n+1)^2-6\right] )T_{ab} = 0
\end{equation}
then $T_{ab}$ admits as general solution 
\begin{align}
     T_{ab} &= \delta_{n,-1} a_0T^{(V,E,p)-100}_{ab}+\sum_{\ell\ge1, m} \bigg( a_{\ell m}^E T^{(V,E,p)n\ell m}_{ab}  \nn\\
     &+ b^E_{\ell m}T^{(V,E,q)n\ell m}_{ab}+ a^B_{\ell m}T^{(V,B,p)n\ell m}_{ab} + b^B_{\ell m}T^{(V,B,q)n\ell m}_{ab}\bigg) 
\end{align}
where $a_0$, $a_{\ell m}^E $ , $b^E_{\ell m}$, $a_{\ell m}^B $ , $b^B_{\ell m}$ are constants and $T^{(V,D,c)n\ell m}_{ab}$, $D=E,B$ and $c=p,q$ are the vector tensor harmonics defined in Eqs. \eqref{tvh}-\eqref{tvhl1}.  
\end{lemma}
The proof starts from the decomposition \eqref{tensordecomp}. Let us first assume $T_{ab}=\bar T_{ab}$. Imposing the conditions $\mathcal{D}_{(a}\mathcal{D}^cT_{cb)} = \frac{1}{2}(\Box-2)T_{ab}=0$ and using Lemma \ref{lemma:T4}, we get that 
\begin{equation}
    T_{ab} = \mathcal{D}_{(a}V_{b)}
\end{equation}
where $V_a$ is a transverse vector. 

If we next assume $T_{ab}=T^{(S)}_{ab}$ (see Eq.  \eqref{Tscal}), the traceless condition imposes $\varphi=0$ and the condition $\mathcal{D}_{(a}\mathcal{D}^cT_{cb)} = \frac{1}{2}(\Box-2)T_{ab}$ becomes 
\begin{equation}
   \mathcal{D}_a\mathcal{D}_b \Box \phi  + q_{ab} (\Box+4)\Box \phi= 0,
\end{equation}
which imposes $\Box\phi=H$ with $\mathcal{D}_a \mathcal{D}_b H + q_{ab}H=0$. The generic solution is $\phi=\tilde \phi -\frac{H}{3}$ with $\Box\tilde \phi =0 $. But then $T_{ab}$ is written as 
\begin{equation}
     T_{ab} = \mathcal{D}_a\mathcal{D}_b \tilde\phi.
\end{equation}
 As shown in the previous section (see Lemma \ref{lemma:V2}), the vector $V_a=\mathcal{D}_a \tilde \phi$ is transverse for $\Box\tilde \phi=0$. This proves that, for any traceless tensor $T_{ab}$ such that $\mathcal{D}_{(a}\mathcal{D}^cT_{cb)} = \frac{1}{2}(\Box-2)T_{ab}$, there exists a transverse vector $V_a$ such that $ T_{ab} = \mathcal{D}_{(a}V_{b)}$. The converse follows from Eqs. \eqref{divV},\eqref{VectorTensorTrace} and \eqref{VectorTensorProp}. Then, if we impose the equation $(\Box+\left[(n+1)^2-6\right] )T_{ab} = 0$, the complete set of solutions is given by a combination of tensor harmonics built from transverse vectors, see Eqs. \eqref{tvh}-\eqref{tvhl1}.

\subsection{Inhomogeneous solutions}

Let $T_{ab}^n$ be a tensor satisfying the equations 
\begin{align}
    (\Box+(n+1)^2-3)T^{n}_{ab}&= S^{n}_{ab}(\tau,x^A),\label{generalInhomTensEqu1} \\\quad \mathcal{D}^aT^{n}_{ab}&= S^{n}_{a}(\tau,x^A),\\\quad T_{a}^{n\;a}&= S^n(\tau,x^A), \label{generalInhomTensEqu3}
\end{align}
with $S^{n}_{ab}(\tau,x^A)$, $S^{n}_a(\tau,x^A)$ and $S^n(\tau,x^A)$ being arbitrary sources. Such sets of equations appear in the resolution of Einstein's equations at spatial infinity 
\cite{beig1982einstein}. We would like to find the generic solution $T_{ab}^n$ to Eqs. \eqref{generalInhomTensEqu1}-\eqref{generalInhomTensEqu3}. This will lead to the construction of SDT tensors from $T_{ab}^n$. 

By consistency, the sources obey the following relations 
\begin{align}
    (\Box+(n+1)^2+1)S^{n}_{a}&=\mathcal{D}^b S^{n}_{ab}+2\mathcal{D}_a S^n, \\(\Box+(n+1)^2+3)\mathcal{D}^aS^{n}_{a}&=\mathcal{D}^a\mathcal{D}^bS^{n}_{ab}+2 \Box S^n, \\(\Box+(n+1)^2-3)S^{n} &= S^{n\;a}_a. 
\end{align}
For $n \neq 0$, we define the following tensor
\begin{align}
    X_{ab}^{n}:=&\frac{1}{2}(\Box-2)T^{n}_{ab}- \mathcal{D}_{(a}\mathcal{D}^cT^{n}_{b)c} + \frac{1}{3} q_{ab} \mathcal{D}^c \mathcal{D}^d T^{n}_{cd}\nn \\ 
    &- \frac{1}{6}q_{ab}(\Box-2) T^{n\;a}_a\nn\\
   =&\frac{1}{2}(1-(n+1)^2)T_{ab}^{n}+\frac{1}{2} S_{ab}^{n}- \mathcal{D}_{(a}S^{n}_{b)}+ \frac{1}{3} q_{ab} \mathcal{D}^cS^{n}_c \nn \\ 
   &-  \frac{1}{6} q_{ab}  (\Box-2)S^{n}\label{eq321}
\end{align}
which satisfies the equation 
\begin{align}
    (\Box+(n+1)^2-3)X^{n}_{ab}&= S^{(X)n}_{ab},\\ \mathcal{D}^bX^{n}_{ab}&=\mathcal D_a S^{(X)n}, \\ X_{a}^{n\;a}&= 0,
\end{align}
with 
\begin{align}
     S^{(X)n}&:=- \frac{1}{6} (\mathcal{D}^cS^{n}_c+(\Box+4)S^{n}), \\
     S^{(X)n}_{ab} &:=\frac{1}{2}(\Box-2)S^{n}_{ab}- \mathcal{D}_{(a}\mathcal{D}^cS^{n}_{b)c}+ \frac{1}{3} q_{ab} \mathcal{D}^c\mathcal{D}^d S_{cd}^{n}\nn\\&-\frac{1}{6} q_{ab} (\Box-2)S^{n\;a}_a- 2 \mathcal{D}_a \mathcal{D}_b S^{n} +\frac{2}{3} q_{ab} \Box S^{n} . 
\end{align}
Since $n \neq 0$, we can use Eq. \eqref{eq321} to solve for $T^{n}_{ab}$ in terms of $X_{ab}^{n}$. 

We can further decompose $X_{ab}^{n}$ with the following method. For $n>0$, we define the tensor 
\begin{align}
    Y^{(SDT)n}_{ab} &:= \text{Curl}(\text{Curl}(X_{ab}^n))\nn\\=& (\Box-3) X^{n}_{ab} -\frac{3}{2} \mathcal{D}_{(a}S^{(X)n}_{b)}+\frac{1}{2} q_{ab}\mathcal{D}^{c}S^{(X)n}_{c}\nn\\=&-(n+1)^2 X^{n}_{ab}+S_{ab}^{(X)n} -\frac{3}{2} \mathcal{D}_{(a}S^{(X)n}_{b)}+\frac{1}{2} q_{ab}\mathcal{D}^{c}S^{(X)n}_{c}\label{YSDT}
\end{align}
Since $n \neq -1$, we can solve for $X_{ab}^{n}$ in terms of $Y_{ab}^{(SDT)n}$. By construction, $Y_{ab}^{(SDT)n}$ is a SDT tensor which obeys
\begin{align}
(\Box+(n+1)^2-3)Y_{ab}^{(SDT)n}&=S^{(SDT)n}_{ab},\\ \mathcal{D}^bY_{ab}^{(SDT)n}&=0, \\ Y_{a}^{(SDT)n\, a}&= 0,   
\end{align}
with $S^{(SDT)n}_{ab} :=  \text{Curl}(\text{Curl}(S^{(X)n}_{ab}))$. Indeed, we can decompose $X^{n}_{ab}$ in terms of an auxiliary SDT tensor $\bar X_{ab}^{n}$ and a traceless scalar part as follows
\begin{equation}
    X_{ab}^{n} = \bar X_{ab}^{n} +\mathcal{D}_a\mathcal{D}_b\phi^{(S)n}_1 -\frac{1}{3}\Box\phi^{(S)n}_1  q_{ab}\label{XabDecomposition}
\end{equation}
where the scalar $\phi^{(S)n}_1 $ is defined non-locally from 
\begin{equation}
   (\Box +3)\phi^{(S)n}_1   =-\frac{1}{4} \Big( \mathcal{D}^cS^{n}_c+(\Box+4)S^{n}\Big):= S^{(\phi)n} . \label{EquPhiXab}
\end{equation}
This equation is obtained from   $\mathcal{D}^bX_{ab}=\frac{2}{3} \mathcal{D}_a (\Box+3)\phi^{(S)n}_1 $ after using Eqs. \eqref{XabDecomposition} and \eqref{ScalTDiv}   \footnote{Technically, the equation is $(\Box +3)\phi^{(n)}=S^{(n)\phi}+ C$, where $C$ is a constant. However, this constant is not relevant since $\mathcal{D}_a\mathcal{D}_b C -\frac{1}{3}\Box\;C  q_{ab}=0$.}. Because of the decomposition \eqref{XabDecomposition} and the fact that the symmetrized curl of a tensor constructed from a scalar is zero, the tensor $Y_{ab}^{(SDT)n}$ is SDT. In summary, the tensor $T_{ab}^{n}$ for $n \neq -1,0$ can be reduced to the SDT tensor $Y_{ab}^{(SDT)n}$, using local expressions of the sources in Eqs. \eqref{generalInhomTensEqu1}-\eqref{generalInhomTensEqu3}. The non-locally defined auxiliary SDT tensor $\bar X_{ab}^{n}$ is only used for the proof and is not used in the construction of $Y_{ab}^{(SDT)n}$. 

Let us now discuss the special cases $n=0,-1$. For $n=-1$, Eqs. \eqref{generalInhomTensEqu1}-\eqref{generalInhomTensEqu3} are invariant under the shift $T_{ab}\to T_{ab}+\mathcal{D}_{a}\mathcal{D}_b \phi + q_{ab}\phi$ with $\phi$ a scalar that satisfies $(\Box+3)\phi = 0$. Also, the construction \eqref{YSDT} does not allow to reconstruct $X_{ab}^{n=-1}$ (and therefore $T_{ab}^{n=-1}$). Instead, we use the decomposition \eqref{XabDecomposition} to build an SDT tensor. It is non-algebraic since it requires to find the solution to the normally hyperbolic partial differential equation \eqref{EquPhiXab}. We define the following SDT tensor 
\begin{equation}
Y_{ab}^{(SDT)n=-1}:= X_{ab}^{n=-1} - \mathcal{D}_a\mathcal{D}_b\phi^{(S)n=-1}_1 +\frac{1}{3}\Box\;\phi^{(S)n=-1}_1  q_{ab}\label{YSDTm1}
\end{equation}
with $\phi^{(S)n=-1}_1$ satisfying Eq. \eqref{EquPhiXab} with $n=-1$ and $X_{ab}^{n=-1}$ defined from Eq. \eqref{eq321} with $n=-1$.

For $n=0$, Eqs. \eqref{generalInhomTensEqu1}-\eqref{generalInhomTensEqu3} are invariant under the shift $T_{ab}\to T_{ab}+\mathcal{D}_{(a}V_{b)}$ with $V_a$ a transverse vector that obeys $(\Box+2)V_a = 0$. The construction \eqref{eq321} is not adapted for $n=0$ since it does not allow to reconstruct $T_{ab}^{n=0}$. Instead, we construct the tracefree tensor
\begin{align}
    X^{0}_{ab} &:= \text{Curl}(\text{Curl}(T^{0}_{ab})) \nn\\=&(\Box-3) T^{0}_{ab} + \frac{1}{2} (\mathcal{D}_a\mathcal{D}_b S^{0} -  q_{ab}(\Box-2) S^{0}) \nn\\ 
    &-\frac{3}{2} \mathcal{D}_{(a}S^{0}_{b)}+\frac{1}{2} q_{ab}\mathcal{D}^{c}S^{0}_{c} \nn \\=&-T^{0}_{ab}+S_{ab}^{0} + \frac{1}{2} (\mathcal{D}_a\mathcal{D}_b S^{0} -  q_{ab}(\Box-2) S^{0}) \nn\\ 
    &-\frac{3}{2} \mathcal{D}_{(a}S^{0}_{b)}+\frac{1}{2} q_{ab}\mathcal{D}^{c}S^{0}_{c}\label{CurlcurlT0}
\end{align}
which satisfies
\begin{align}
   \mathcal{D}^b X^{0}_{ab} &= \frac{1}{4}\Big((\Box-2)S_a^{0}- \mathcal{D}_a \mathcal{D}^c S_c^{0}\Big),\label{DivX0} \\
    \mathcal{D}^a\mathcal{D}^b X^{0}_{ab} &= 0,\quad X^{0\;a}_a=0.\label{DoubleDivX0}
\end{align}
Because $X^{0}_{ab}=\text{Curl}(\text{Curl}(T^{0}_{ab}))$, it can be decomposed as  
\begin{equation}
     X_{ab}^{0} = Y^{(SDT)0}_{ab} + \mathcal{D}_{(a}W^{(T)1}_{b)}\label{YSDT0}
\end{equation}
with $Y^{(SDT)0}_{ab}$ a SDT tensor and $W^{(T)1}_a$ a transverse vector satisfying an equation of the form Eq. \eqref{GeneralNHDFVectEqu0} with $n=1$. This can be proven starting from the decomposition \eqref{tensordecomp} :
\begin{equation}
    T_{ab}^{0} =\bar T^{0}_{ab} + \mathcal{D}_{(a}\tilde W^{(T)1}_{b)}+ \mathcal{D}_a\mathcal{D}_b \phi+\frac{1}{3}(\varphi-\Box\phi)q_{ab}
\end{equation}
where $\bar T^{0}_{ab} $ is an SDT tensor, $\tilde W^{(T)1}_a$ is a transverse vector and $ \phi$, $ \varphi$ are scalars. Then, we use Eqs. \eqref{Curlculpropscal} and \eqref{CurlcurlTVab}, which leads to  
\begin{equation}
    X_{ab}^{0} =\text{Curl}(\text{Curl}(\bar T^{0}_{ab})) + \frac{1}{4}\mathcal{D}_{(a}\text{Curl}(\text{Curl}(\tilde W^{(T)1}_{b)})).
\end{equation}

Since $[\text{Curl}\,\text{Curl},\mathcal{D}^a]V_a=0$, $[\text{Curl}\,\text{Curl},\mathcal{D}^a]T_{ab}=0$ and $\text{Curl}\,\text{Curl}(T^a_a)=0$ for any transverse vector $V_a$ and SDT tensor $T_{ab}$, we have that $Y^{(SDT)0}_{ab}:= \text{Curl}(\text{Curl}(\bar T^{0}_{ab}))$ is SDT and $ W^{(T)1}_{b}:=\frac{1}{4} \text{Curl}(\text{Curl}(\tilde W^{(T)1}_{b}))$ is transverse. 

Using Eq. \eqref{DivX0}, $W^{(T)1}_a$ must satisfy the equation :
\begin{align}
    (\Box + 2)W^{(T)1}_a &=\frac{1}{2}\Big((\Box-2)S_a^{0}- \mathcal{D}_a \mathcal{D}^c S_c^{0}\Big)
    := S^{(W)}_a . \label{SourcedW}
\end{align}
The vector $W^{(T)1}_a$ is therefore defined non-locally. 

In summary, for all cases $n+1 \in \mathbb N$, we built a SDT tensor $Y_{ab}^{(SDT)n}$ from a tensor $T^n_{ab}$ satisfying Eqs. \eqref{generalInhomTensEqu1}-\eqref{generalInhomTensEqu3}. Using the relations \eqref{BoxTSab}, \eqref{BoxTVab} along with the fact that $[\text{Curl},\Box]T_{ab}=0$ for any tensor $T_{ab}$ in $dS_3$, we find that the SDT tensor $Y_{ab}^{(SDT)n}$ obeys the equations 
\begin{align}
    (\Box+(n+1)^2-3)Y_{ab}^{(SDT)n}&= S^{(SDT)n}_{ab},\label{generalInhomTensEquBis1}\\\mathcal{D}^{a}Y_{ab}^{(SDT)n}&= 0,\\ Y_{a}^{(SDT)n\;a}&= 0,\label{generalInhomTensEquBis3}
\end{align}
with 
\begin{equation}\label{Kab}
    S^{(SDT)n}_{ab}:=\left\{
    \begin{array}{ll}
        S^{(X)n=-1}_{ab}\\[6pt]-\mathcal{D}_a\mathcal{D}_b S^{(\phi)n=-1}+\frac{1}{3} q_{ab}\Box S^{(\phi)n=-1}& \mbox{for } n=-1;\\&\\ 
       \text{Curl}\, \text{Curl}(S^{0}_{ab})-\mathcal{D}_{(a} S^{(W)}_{b)} & \mbox{for } n=0; \\&\\
       \text{Curl}\, \text{Curl}(S^{(X)n}_{ab})  & \mbox{for } n\ge 1.
    \end{array}
    \right.
\end{equation}

This set of equations is then equivalent to the set of equations for the transverse vector $T_a^{(T)n}$  (see Eqs. \eqref{XiaT}-\eqref{EquationXiaT})
\begin{equation}
    (\Box+(n+1)^2-2)T_a^{(T)n}= R^{(T)n}_{a},\quad \mathcal{D}^aT^{(T)n}_{a}= 0\label{generalInhomTenstoVectEqu}
\end{equation}
with 
\begin{align}
    T^{(T)n}_{a}:= \cosh\tau \, n^b Y^{(SDT)n}_{ab},\\
    R^{(T)n}_{a}:= \cosh\tau \, n^b S^{(SDT)n}_{ab}. 
\end{align}
As in the previous section, Eq. \eqref{generalInhomTenstoVectEqu} is solved by defining the scalars
\begin{align}
    \theta^{E}_{n} &:= \cosh\tau\, n^a T^{(T)n}_{a}, \quad\theta^{B}_{n}:= \cosh\tau \, n^a \text{Curl}(T^{(T)n}_{a}), \\
    \rho^{E}_n &:= \cosh\tau \, n^a R^{(T)n}_{a}, \quad\rho^{B}_n := \cosh\tau \, n^a \text{Curl}(R^{(T)n}_{a}),\label{kappaEB}
\end{align}
which satisfy Eq. \eqref{InhomognenousScaleq}. 
The solutions are 
\begin{align}
    \theta^E_{n} &= \sum_{\ell,m} \left(A^{(E)}_{n\ell m }+A^{(S,E)}_{n\ell m}(\tau;\bar\tau)\right) \psi^{(p)}_{n\ell m}(\tau,x^A) \nn\\ 
    &+\left(B^{(E)}_{n\ell m }+B^{(S,E)}_{n\ell m}(\tau;\bar\tau)\right) \psi^{(q)}_{n\ell m}(\tau,x^A),\label{psiENHSolT}
    \end{align}
    \begin{align}
    \theta^B_{n} &= \sum_{\ell,m} \left(A^{(B)}_{n\ell m }+A^{(S,B)}_{n\ell m}(\tau;\bar\tau)\right) \psi^{(p)}_{n\ell m}(\tau,x^A)\nn\\
    &+\left(B^{(B)}_{n\ell m }+B^{(S,B)}_{n\ell m}(\tau;\bar\tau)\right) \psi^{(q)}_{n\ell m}(\tau,x^A),\label{psiBNHSolT}
\end{align}
with $A^{(D)}_{n\ell m }$ and $B^{(D)}_{n\ell m }$ constants and
\begin{align}
    A^{(S,D)}_{n\ell m}(\tau;\bar\tau)&= \int_{\bar\tau}^\tau d\tau' \oint_{S^2(\tau')} d\Omega \, \rho^D_n(\tau',x^A)\overline{\psi^{(q)}_{n\ell m}(\tau',x^A)},\label{ASETensorbeg}\\
     B^{(S,D)}_{n\ell m}(\tau;\bar\tau)&= -\int_{\bar\tau}^\tau d\tau' \oint_{S^2(\tau')} d\Omega \,\rho^D_n(\tau',x^A)\overline{\psi^{(p)}_{n\ell m}(\tau',x^A)},\label{BSETesnorend}
\end{align}
with $D=E,B$. Using the KG inner product on the scalars, we can single out the coefficients 
\begin{align}
(\theta^E_n,\psi^{(q)}_{n\ell m})_{KG} &=A^{(E)}_{n\ell m }+A^{(S,E)}_{n\ell m}(\tau;\bar\tau),\\ 
(\theta^E_n,\psi^{(p)}_{n\ell m})_{KG} &=-B^{(E)}_{n\ell m }-B^{(S,E)}_{n\ell m}(\tau;\bar\tau), \\ 
(\theta^B_n,\psi^{(q)}_{n\ell m})_{KG} &=A^{(B)}_{n\ell m }+A^{(S,B)}_{n\ell m}(\tau;\bar\tau),\\
(\theta^B_n,\psi^{(p)}_{n\ell m})_{KG} &=-B^{(B)}_{n\ell m }-B^{(S,B)}_{n\ell m}(\tau;\bar\tau).
\end{align}

Using the definitions \eqref{TEttSDT}-\eqref{TBABSDT}, we can then construct the SDT tensor solution to Eqs. \eqref{generalInhomTensEquBis1}-\eqref{generalInhomTensEquBis3} as follows 
\begin{widetext}
\begin{align}
    Y_{ab}^{(SDT)n} &= \delta_{n, -1} \sum_{\ell=0}^{1}\sum_{m=-\ell}^\ell a_{\ell m} T^{(E,p),-1\ell m}_{ab}
    + \delta_{n,0} \sum_{m=-1}^1 \left(a^{(E)}_m T^{(E,p),01m}_{ab}+a^{(B)}_m T^{(B,p),01m}_{ab} \right) + \sum_{l\ge2, m} \Big(T^{(E)\ell m}_{ab}\Big[f=\frac{(\theta_n^E,\psi_{n\ell m}^{(q)})\psi_{n\ell}^{(p)}(\tau)}{\sqrt{\ell(\ell+1)(\ell+2)(\ell-1)}}\Big] \nn \\&-T^{(E)\ell m}_{ab}\Big[f=\frac{(\theta^E_n,\psi_{n\ell m}^{(p)})\psi_{n\ell}^{(q)}(\tau)}{\sqrt{\ell(\ell+1)(\ell+2)(\ell-1)}}\Big]
    -T^{(B)\ell m}_{ab}\Big[\tilde f =\frac{(\theta_n^B,\psi_{n\ell m}^{(q)})\psi_{n\ell}^{(p)}(\tau)}{\sqrt{\ell(\ell+1)(\ell+2)(\ell-1)}}\Big]+T^{(B)\ell m}_{ab}\Big[\tilde f =\frac{(\theta^B_n,\psi_{n\ell m}^{(p)})\psi_{n\ell}^{(q)}(\tau)}{\sqrt{\ell(\ell+1)(\ell+2)(\ell-1)}}\Big]\Big),\label{SDTNHdecomp}
\end{align}
\end{widetext}
where $a_{\ell m}$, $a^{(E)}_m$, $a^{(B)}_m$ are constants. A generic solution to the system \eqref{generalInhomTensEqu1}-\eqref{generalInhomTensEqu3} can be built from \eqref{SDTNHdecomp}, using Eqs. \eqref{eq321}, \eqref{EquPhiXab}, \eqref{YSDTm1} for $n=-1$, Eqs. \eqref{CurlcurlT0}, \eqref{YSDT0}, \eqref{SourcedW} for $n=0$ and Eqs. \eqref{eq321}, \eqref{YSDT} for $n>0$.

\subsection{Asymptotic behavior and antipodal matching for SDT harmonics}

We want to establish the asymptotic behavior of a general solution to Eq.  \eqref{generalSDTTensorEquations}. We will not explicitly write the asymptotic behaviors and antipodal matching conditions for tensors built from scalars and vectors since they can be straightforwardly obtained from the scalar and vector cases. 

First, let us compute the asymptotics of the $\ell=0,1$ part of a SDT tensor :
\begin{align}
    T^{\ell\le1}_{\tau\tau} &=8 C^{\pm}(x^A)e^{-3\vert\tau\vert} -4\nabla^A J^{\pm}_A(x^A) e^{-4\vert\tau\vert}+ o(e^{-4\vert\tau\vert}) ,\\
    T^{\ell\le1}_{\tau A} &= \pm J^{\pm}_A(x^A)e^{-2\vert\tau\vert} \mp 4 \nabla_A C^{\pm}(x^A) e^{-3\vert\tau\vert}+ o(e^{-3\vert\tau\vert}) ,\\
    T^{\ell\le1}_{AB} &= C^{\pm}(x^A) \gamma_{AB} e^{-\vert\tau\vert}-\frac{1}{2}\nabla^C J^{\pm}_C(x^A) \gamma_{AB} e^{-2\vert\tau\vert}+ o(e^{-2\vert\tau\vert})  ,
\end{align}
where $C^{\pm}(x^A)$ is a combination of  $\ell=0,1$ scalars on the sphere and $J^{\pm}_A(x^A)$ a $\ell=1$ vector on the sphere. 
Regarding the $\ell\ge2$ harmonics, we observe that a general $\ell\ge2$ SDT tensor can be written as follows:
\begin{align}\label{TSDTtautau}
    T^{(SDT)}_{\tau\tau} &=2 \nabla^C\nabla^D N^{(E)}_{CD}(\tau,x^A) \sech^2\tau ,\\     T^{(SDT)}_{\tau A} &= 2 \left(\partial_\tau+\tanh\tau \right) \nabla^C N^{(E)}_{AC}(\tau,x^A)+\nabla^C N^{(B)}_{AC}(\tau,x^A) \label{TSDTAtau} ,
        \end{align}
    \begin{align}
   T^{(SDT)}_{AB} &= - \nabla^2 N^{(E)}_{AB}(\tau,x^A)+ \Big(4 + 2 \cosh( 2 \tau)+2 \cosh^2\tau \nonumber \\ 
  &  \times (3 \tanh\tau \partial_\tau+\partial_\tau^2)\Big)N^{(E)}_{AB}(\tau,x^A)\nn\\
  &+\nabla^C\nabla^D N^{(E)}_{CD}(\tau,x^A)\gamma_{AB}+\cosh^2 \tau (2\tanh\tau+ \partial_\tau) N^{(B)}_{AB},\label{TSDTAB}
\end{align}
where $N_{AB}^{(E)}(\tau,x^A)$ and $N_{AB}^{(B)}(\tau,x^A)$ are traceless tensors of the form 
\begin{align}
    N_{AB}^{(E)}(\tau,x^A) = \nabla_{\langle A} \nabla_{B \rangle} f(\tau,x^A),\\
     N_{AB}^{(B)}(\tau,x^A) = \nabla_{(A}\epsilon_{B)C}\nabla^C \tilde{f}(\tau,x^A).
\end{align}

By imposing Eq.  \eqref{generalSDTTensorEquations}, the functions $f$ and $\tilde f$ must satisfy the scalar equation \eqref{generalscalarequation}. Hence, $N_{AB}^{(E)n}$ and $N_{AB}^{(B)n}$ have the following asymptotics:
\begin{align}\label{CABasympt0}
    &N_{AB}^{(E)n}(\tau,x^A) \nn\\&\quad= e^{n\vert\tau\vert}\sum_{r=0}^n N_{n,2}^{(r,0)}[\Gamma_{AB}^{(\mathbb L,E)n\pm}(x^A)]e^{-2r\vert\tau\vert}\nonumber \\
    &\quad+ e^{-(n+2)\vert\tau\vert}\left(\Gamma_{AB}^{(\mathbb S,E)n\pm}(x^A)+\vert\tau\vert N_{n,2}^{(n+1,1)}[\Gamma_{AB}^{(\mathbb L,E)n\pm}(x^A)]\right)\nonumber\\ 
    &\quad+o(e^{-(n+2)\vert\tau\vert}),
\\
    &N_{AB}^{(B)n}(\tau,x^A) \nn\\&\quad= e^{n\vert\tau\vert}\sum_{r=0}^n N_{n,2}^{(r,0)}[ \Gamma_{AB}^{(\mathbb L,B)n\pm}(x^A)]e^{-2r\vert\tau\vert}\nonumber  \\&\quad+ e^{-(n+2)\vert\tau\vert}\left( \Gamma_{AB}^{(\mathbb S,B)n\pm}(x^A)+\vert\tau\vert\, N_{n,2}^{(n+1,1)}[\Gamma_{AB}^{(\mathbb L,B)n\pm}(x^A)]\right) \nonumber\\ 
    &\quad+o(e^{-(n+2)\vert\tau\vert}),\label{CABasymptilde}
\end{align}
where the operators $N_{n,2}^{(r,0)}[\cdot]$ and $N_{n,2}^{(n+1,1)}[\cdot]$ are defined in Eqs. \eqref{NnsOp1}-\eqref{NnsOp2} with $s=2$.

The 8 traceless tensors $\Gamma_{AB}^{(\mathbb L,E)n\pm}(x^A),\Gamma_{AB}^{(\mathbb S,E)n\pm}(x^A)$ and $\Gamma_{AB}^{(\mathbb L,B)n\pm}(x^A),\Gamma_{AB}^{(\mathbb S,B)n\pm}(x^A)$ are related to 8 functions $\psi^{(\mathbb L)\pm}_n(x^A)$,$\psi^{(\mathbb S)\pm}_n(x^A)$ and $\tilde \psi^{(\mathbb L)\pm}_n(x^A)$,$\tilde \psi^{(\mathbb S)\pm}_n(x^A)$ as 
\begin{align}
    \Gamma_{AB}^{(\mathbb L,E)n\pm}(x^A)&:=\nabla_{\langle A }\nabla_{B\rangle}\psi^{(\mathbb L)\pm}_n(x^A),\\
    \Gamma_{AB}^{(\mathbb L,B)n\pm}(x^A)&:=\nabla_{(A} \epsilon_{B)C}\nabla^C \tilde \psi^{(\mathbb L)\pm}_n(x^A),\\
    \Gamma_{AB}^{(\mathbb S,E)n\pm}(x^A)&:=\nabla_{\langle A }\nabla_{B\rangle} \psi^{(\mathbb S)\pm}_n(x^A),\\
    \Gamma_{AB}^{(\mathbb S,B)n\pm}(x^A)&:=\nabla_{(A} \epsilon_{B)C}\nabla^C \tilde \psi^{(\mathbb S)\pm}_n(x^A),
\end{align}
which obey the antipodal matching conditions \eqref{Matchingscalarend}. 

Implementing Eqs. \eqref{CABasympt0}-\eqref{CABasymptilde} into \eqref{TSDTtautau}, we find that the $\tau\tau$ component of any $\ell\ge2$ SDT tensor satisfying Eq. \eqref{generalSDTTensorEquations} behaves as :
\begin{align}
&T^{(SDT)n}_{\tau\tau}(\tau\to\pm\infty,x^A) =8(1-\delta_{n,-1}) e^{(n-2)\vert\tau\vert} \nn\\&\times \sum_{r=0}^{n}\;\sum_{k=0}^{r} \frac{(-1)^{r}(n-k)!\;(r+2)! }{k!(k+2)!(r-k)!}\Delta_{k,0}\nabla^{A}\nabla^B\Gamma_{AB}^{(\mathbb L,E)n\pm}(x^A) e^{-2r\vert\tau\vert}  \nn
        \end{align}
    \begin{align}
&+8e^{-(n+4)\vert\tau\vert}\Big[-2\delta_{n,0}\nabla^{A}\nabla^B\Gamma_{AB}^{(\mathbb L,E)n\pm}(x^A)\nn \\
&+(1-\delta_{n,0})\frac{(-1)^{n+1}}{n!}\Delta_{2,n+1,0}\Big((4+n)+2 \nabla^2\Big)\nabla^{A}\nabla^B\Gamma_{AB}^{(\mathbb L,E)n\pm}(x^A)\nn \\&+\nabla^{A}\nabla^B\Gamma_{AB}^{(\mathbb S,E)n\pm}(x^A)+\vert\tau\vert\frac{2(-1)^{n+1}}{(n+1)!} \Delta_{n+1,0}\nabla^{A}\nabla^B\Gamma_{AB}^{(\mathbb L,E)n\pm}(x^A)\Big]\nonumber\\
& + o( e^{-(n+4)\vert\tau\vert}), 
\end{align}
where we used the identity 
\begin{align}
    &\sum_{k=0}^n\Big(\frac{(n+1)!}{k!(k+1)!}+\frac{(n+2)!}{k!(k+2)!}\Big)\Delta_{k,0} \nn\\&= \frac{1-\delta_{n,0}}{n!}\Delta_{2,n+1,0}
    \Big((4+n)+2 \nabla^2\Big)+ 2 \delta_{n,0}.
\end{align}
This operator annihilates all $2\le\ell\le n$ harmonics. 

Regarding the $\tau A$ component \eqref{TSDTAtau}, we get 
\begin{align}
    &T_{A\tau}^{(SDT)n(E)}(\tau\to\pm\infty,x^A)  = \pm 2(1-\delta_{n,-1})e^{n\vert\tau\vert}\Big[\sum_{r=0}^{n}e^{-2r\vert\tau\vert}\sum_{k=0}^r\nn\\
    &\big((n+1)(1+k)\!-\!2 k(r+1)\big) \frac{(-1)^{r}(n-k)!r!}{k!(k+1)!(r-k)!}\Delta_{k,1}\!\nabla^B\Gamma_{AB}^{(\mathbb L,E)n\pm}(x^A)\Big]\nn\\&\pm 2 e^{-(n+2)\vert\tau\vert}\Big[ -2\vert\tau\vert \frac{ (-1)^{n+1} }{n!}\Delta_{n+1,1}\nabla^B\Gamma_{AB}^{(\mathbb L,E)n\pm}(x^A)\nn
    \end{align}\begin{align}
   & -(n+1)\nabla^B\Gamma_{AB}^{(\mathbb S,E)n\pm}(x^A) +2\frac{(-1)^{n+1}}{(n+1)!} \Big( (n+1) \Delta_{1,n+1,1}\nn\\ 
&+\Delta_{0,n+1,1}\Big)\nabla^B\Gamma_{AB}^{(\mathbb L,E)n\pm}(x^A)\Big]+o(e^{-(n+2)\vert\tau\vert}),
\end{align}
and
\begin{align}
    &T_{A\tau}^{(SDT)n(B)}(\tau\to\pm\infty,x^A) = e^{n\vert\tau\vert}(1-\delta_{n,-1})\Big[\sum_{r=0}^{n}e^{-2r\vert\tau\vert}\nn\\ 
    &\times \sum_{k=0}^r \frac{(-1)^{r}(n-k)! r!}{(k!)^2(r-k)!}\Delta_{k,1} \nabla^B\Gamma_{AB}^{(\mathbb L,B)n\pm}(x^A)\Big]\nn\\&+
    e^{-(n+2)\vert\tau\vert}\Big[2\vert\tau\vert \frac{ (-1)^{n+1} }{(n+1)!} \Delta_{n+1,1} \nabla^B\Gamma_{AB}^{(\mathbb L,B)n\pm}(x^A)
    +\nabla^B\Gamma_{AB}^{(\mathbb S,B)n\pm}(x^A)\Big]\nn\\
    &+o(e^{-(n+2)\vert\tau\vert}). 
\end{align}
We used the identity \eqref{DeltaIdentity}. 
For the $AB$ components \eqref{TSDTAB}, the trace with respect to the two-sphere admits the following asymptotics 
\begin{widetext}
\begin{align}
     &T_{A}^{(SDT)n\,A}(\tau\to\pm\infty,x^A) = 2 e^{n\vert\tau\vert}(1-\delta_{n,-1})\Big[\sum_{r=0}^{n}e^{-2r\vert\tau\vert} \sum_{k=0}^r \frac{(-1)^{r}(n-k)! r!}{(k!)^2(r-k)!}\Delta_{k,0}\nabla^A\nabla^B \Gamma_{AB}^{(\mathbb L,E)n\pm}(x^A)\Big]+
   2 e^{-(n+2)\vert\tau\vert}\Big[\nn\\ 
   &  2\vert\tau\vert  \frac{ (-1)^{n+1}}{(n+1)!}\Delta_{n+1,0} \nabla^A\nabla^B\Gamma_{AB}^{(\mathbb L,E)n\pm}(x^A)  +\nabla^A\nabla^B \Gamma_{AB}^{(\mathbb S,E)n\pm}(x^A)\Big]+o(e^{-(n+2)\vert\tau\vert}),
\end{align}
and the traceless part behaves as
\begin{align}
    &T_{\langle AB\rangle}^{(SDT)n(E)}(\tau\to\pm\infty,x^A)  = \frac{1}{2} e^{(n+2)\vert\tau\vert}(1-\delta_{n,-1})\Big[(n+2)!\Gamma_{AB}^{(\mathbb L,E)n\pm}(x^A)-(1-\delta_{n,0})(n+1)!(\nabla^2-(n+4))\Gamma_{AB}^{(\mathbb L,E)n\pm}(x^A)e^{-2\vert\tau\vert}\nn \\
    &+(1-\delta_{n,0}-\delta_{n,1})\sum_{r=2}^{n}e^{-2r\vert\tau\vert}(-1)^{r}(r-2)!\Big(2(n-1)!\frac{r-1}{(r-2)!}\Delta_{2,2}\Gamma_{AB}^{(\mathbb L,E)n\pm}(x^A) +\sum_{k=2}^r \frac{(n-k)! }{(k!)^2(r-k)!}\Delta_{k,2}(2(r-1)(r-k)\Delta_{1,2}+\alpha^{n,r,k}) \nn\\
    &\times \Gamma_{AB}^{(\mathbb L,E)n\pm}(x^A)\Big)\Big] +
   \frac{1}{2}e^{-n\vert\tau\vert}\Big[2n\vert\tau\vert \frac{ (-1)^{n+1} }{n!}\Delta_{n+1,2} \Gamma_{AB}^{(\mathbb L,E)n\pm}(x^A)
   + n(n+1) \Gamma_{AB}^{(\mathbb S,E)n\pm}(x^A)-(2n+1)2  \frac{ (-1)^{n+1}}{(n+1)!}\Delta_{n+1,2} \Gamma_{AB}^{(\mathbb L,E)n\pm}(x^A)\nn\\
   &-(1-\delta_{n,-1})(-1)^{n+1}\sum_{k=0}^n \frac{(n(n+1)+k(n-1)-2\Delta_{1,2})n!}{(k!)^2}\Delta_{k,2} \Gamma_{AB}^{(\mathbb L,E)n\pm}(x^A)\Big]+o(e^{-n\vert\tau\vert});        \end{align}
    \begin{align} 
    &T_{\langle AB\rangle}^{(SDT)n(B)}(\tau\to\pm\infty,x^A) =\frac{(\pm1)}{4} e^{(n+2)\vert\tau\vert}(1-\delta_{n,-1})\Big[(n+2)n!\Gamma_{AB}^{(\mathbb L,B)n\pm}(x^A)\nn\\&-(1-\delta_{n,0})n!(\nabla^2-(n+4))\Gamma_{AB}^{(\mathbb L,B)n\pm}(x^A)e^{-2\vert\tau\vert}+(1-\delta_{n,0}-\delta_{n,1})\sum_{r=2}^{n}e^{-2r\vert\tau\vert}(n+2-2r)\sum_{k=2}^r \frac{(-1)^{r}(n-k)! (r-2)!}{k!(k-2)!(r-k)!}\Delta_{k,2}\Gamma_{AB}^{(\mathbb L,B)n\pm}(x^A)\Big]\nn\\&+
   \frac{(\pm1)}{4}e^{-n\vert\tau\vert}\Big[-2 n\vert\tau\vert \frac{(-1)^{n+1}}{(n+1)!} \Delta_{n+1,2}\Gamma_{AB}^{(\mathbb L,B)n\pm}(x^A)
    -n\Gamma_{AB}^{(\mathbb S,B)n\pm}(x^A)+2 \frac{ (-1)^{n+1}}{(n+1)!} \Delta_{n+1,2}\Gamma_{AB}^{(\mathbb L,B)n\pm}(x^A)\nn\\&+(1-\delta_{n,-1})(-1)^{n+1}\sum_{k=0}^n \frac{(n+k)n!}{(k!)^2}\Delta_{k,2} \Gamma_{AB}^{(\mathbb L,B)n\pm}(x^A)\Big]+o(e^{-n\vert\tau\vert})
\end{align}
where we defined $\alpha^{n,r,k}=k (n-2 r+2) (k (n-2 r+3)-n-1)$.
We can summarize the generic asymptotic behavior of a SDT tensor as follows. 
We define 
\begin{align}
    H_{AB}^{(\mathbb L)n\pm}(x^A):=&(n+1)\Gamma_{AB}^{(\mathbb L,E)n\pm}(x^A)\pm \frac{1}{2}\Gamma_{AB}^{(\mathbb L,B)n\pm}(x^A),\\
    H_{AB}^{(\mathbb S)n\pm}(x^A):=& n(n+1) \Gamma_{AB}^{(\mathbb S,E)n\pm}(x^A)\mp \frac{n}{2}  \Gamma_{AB}^{(\mathbb S,B)n\pm}(x^A)-(1-\delta_{n,-1}-\delta_{n,0})(-1)^{n+1}\Big((n+1)\Delta_{1,2}+n \Big) \Big((n+1)!\Gamma_{AB}^{(\mathbb L,E)n\pm}(x^A)\nn\\&\mp\frac{n!}{2}\Gamma_{AB}^{(\mathbb L,B)n\pm}(x^A) \Big)+2 \delta_{n,-1}\Gamma_{AB}^{(\mathbb L,E)-1\pm}(x^A)\mp \delta_{n,0}\Delta_{1,2}\Gamma_{AB}^{(\mathbb L,B)0\pm}(x^A),\\
    H_{AB}^{(\mathbb L,\tau)n\pm}(x^A):=&-2\frac{(-1)^{n+1}(1-\delta_{n,0})}{(n+1+\delta_{n,-1})(n+1)!}\big(\Delta_{n+1,2}\delta_A^C\delta_B^D -4 \Delta_{2,n+1,2}\nabla_{\langle A}\nabla_{B\rangle }\nabla^C\nabla^D \big)H_{CD}^{(\mathbb L)n\pm}(x^A),\label{relationKABn1}\\ \label{relationLA}
    H_A^{(\mathbb S)n\pm}(x^A):=& (n+1)\nabla^B\Gamma_{AB}^{(\mathbb S,E)n\pm}(x^A)-\frac{(\pm 1)}{2}\nabla^B\Gamma_{AB}^{(\mathbb S,B)n\pm}(x^A)+ (-1)^{n}2(n+1)!\nabla^B\Gamma_{AB}^{(\mathbb L,E)n\pm}(x^A) - \frac{1}{2}\delta_{n,0} J_A^\pm(x^A)\\
    H_{A}^{(\mathbb L,\tau)n\pm}(x^A):=&2  \frac{(-1)^{n+1}}{(n+1)!} \big(\Delta_{n+1,1} \delta_A^C -2\Delta_{1,n+1,1}\nabla_{A}\nabla^C\big)\nabla^B H_{BC}^{(\mathbb L)n\pm}(x^A),\\h^{(\mathbb S)n\pm}(x^A) :=& \nabla^A\nabla^B\Gamma_{AB}^{(\mathbb S,E)n\pm}(x^A)+\delta_{n,-1}C^\pm(x^A)-\frac{1}{2}\delta_{n,0} \nabla^A J_A^\pm.
\end{align}
\end{widetext}
These definitions imply the relationships 
\begin{align}\label{relationDA}
    H_{A}^{(\mathbb L,\tau)n\pm}(x^A) &= - (n+1+\delta_{n,-1})\nabla^BH_{AB}^{(\mathbb L,\tau)n\pm}(x^A), \qquad n\ne0;\\
    h^{(\mathbb S)n\pm}(x^A) &=\frac{1}{n(n+1)} \Big(\nabla^A\nabla^B H_{AB}^{(\mathbb S)n\pm}(x^A)+(-1)^{n+1}n!\nn\\& \times \big((n+1)\Delta_{1,0}+n\big)\nabla^A\nabla^B H_{AB}^{(\mathbb L)n\pm}(x^A)\Big),\;n\ge1.\label{relationf}
\end{align}
In addition, for any E parity tensor $T^{(E)}_{AB}$, we get
\begin{equation}
    \Delta_{2,2}T^{(E)}_{\langle AB\rangle} = 2 \nabla_{\langle A}\nabla_{B\rangle}\nabla^C\nabla^D T^{(E)}_{CD}
\end{equation}
and for any B parity tensor $T^{(B)}_{AB}$, 
\begin{equation}
    \nabla^C\nabla^D T^{(B)}_{CD} = 0. 
\end{equation}

This leads to the expressions 
\begin{widetext}
\begin{align}
&T^{(SDT)n}_{\langle AB\rangle}(\tau\to\pm\infty,x^A) = \frac{1}{2} e^{(n+2)\vert\tau\vert}(1-\delta_{n,-1})\Big[(n+2)n!H_{AB}^{(\mathbb L)n\pm}(x^A)-(1-\delta_{n,0})n!(\nabla^2-(n+4))H_{AB}^{(\mathbb L)n\pm}(x^A)e^{-2\vert\tau\vert}+(1-\delta_{n,0})\sum_{r=2}^{n}e^{-2r\vert\tau\vert}\nn\\ & \times \frac{2(-1)^{r}(r-1)!}{(n+1)}\Big(\Big\{ \frac{2(n-1)!}{(r-2)!} +\sum_{k=2}^r \frac{(n-k)! }{(k!)^2(r-k)!}\Delta_{2,k,2}(2(r-k)\Delta_{1,2}-2(2+n-2r)k^2) \Big\}\nabla_{\langle A}\nabla_{B\rangle}\nabla^C\nabla^D H_{CD}^{(\mathbb L)n\pm}(x^A)\nn\\
&+\frac{(n+2-2r)(n+1)}{2(r-1)}\sum_{k=2}^r \frac{(n-k)!}{k!(k-2)!(r-k)!}\Delta_{k,2}H_{AB}^{(\mathbb L)n\pm}(x^A)\Big)\Big]+
   \frac{1}{2}e^{-n\vert\tau\vert}\Big[\vert\tau\vert n(n+1+\delta_{n,-1}) H_{AB}^{(\mathbb L,\tau)n\pm}(x^A)+ 2\delta_{n,-1} H^{(\mathbb L)-1\pm}_{AB}(x^A)
   \nn\\&+H_{AB}^{(\mathbb S)n\pm}(x^A)+(1-\delta_{n,-1}-\delta_{n,0})\frac{4(-1)^{n+1}n! }{(n+1)}\nabla_{\langle A}\nabla_{B\rangle }\nabla^C\nabla^DH_{CD}^{(\mathbb L)n\pm}(x^A)  \nn\\&+2(1-\delta_{n,-1}-\delta_{n,0})\frac{ (-1)^{n+1}}{(n+1)!} (\Delta_{n+1,2}\delta_A^C\delta_B^D-\frac{2(3n+2)}{(n+1)}\Delta_{2,n+1,2}\nabla_{\langle A}\nabla_{B\rangle }\nabla^C\nabla^D)H_{CD}^{(\mathbb L)n\pm}(x^A)\nn\\&+(1-\delta_{n,-1}-\delta_{n,0})(-1)^{n+1}n!\sum_{k=2}^n \frac{(n+k)}{(k!)^2}(\Delta_{k,2}\delta_A^C\delta_B^D-4 \Delta_{2,k,2}\nabla_{\langle A}\nabla_{B\rangle }\nabla^C\nabla^D)H_{CD}^{(\mathbb L)n\pm}(x^A)\nn\\&+(1-\delta_{n,-1}-\delta_{n,0})4(-1)^{n+1}n!\sum_{k=2}^n \frac{(\Delta_{1,2}+k)}{(n+1)(k!)^2}\Delta_{2,k,2}\nabla_{\langle A}\nabla_{B\rangle}\nabla^C\nabla^D H_{CD}^{(\mathbb L)n\pm}(x^A)\Big]+o(e^{-n\vert\tau\vert}),\\
 &T_{A}^{(SDT)n \; A}(\tau\to\pm\infty,x^A)= \frac{2}{(n+1)} e^{n\vert\tau\vert}(1-\delta_{n,-1})\Big[\sum_{r=0}^{n}e^{-2r\vert\tau\vert}\sum_{k=0}^r \frac{(-1)^{r}(n-k)! r!}{(k!)^2(r-k)!}\Delta_{k,0}\nabla^A\nabla^B H_{AB}^{(\mathbb L)n\pm}(x^A)\Big]\nn\\&+
   2 e^{-(n+2)\vert\tau\vert}\Big[-\frac{1}{n+1+\delta_{n,-1}}\vert\tau\vert \nabla^AH_{A}^{(\mathbb L,\tau)n\pm}(x^A) 
    +\delta_{n,-1}\vert\tau\vert \nabla^A\nabla^B H^{(\mathbb S)-1\pm}_{AB}(x^A)+h^{(\mathbb S)n\pm}(x^A)\Big]+o(e^{-(n+2)\vert\tau\vert}),
    \end{align}\begin{align}
&T_{\tau\tau}^{(SDT)n}(\tau\to\pm\infty,x^A) =\frac{8}{(n+1)}(1-\delta_{n,-1}) e^{(n-2)\vert\tau\vert} \sum_{r=0}^{n}\;\sum_{k=0}^{r} \frac{(-1)^{r}(n-k)!\;(r+2)! }{k!(k+2)!(r-k)!}\Delta_{k,0}\nabla^{A}\nabla^BH_{AB}^{(\mathbb L)n\pm}(x^A) e^{-2r\vert\tau\vert}  \nn\\&+8e^{-(n+4)\vert\tau\vert}\Big[-2\delta_{n,0}\nabla^{A}\nabla^B H_{AB}^{(\mathbb L)n\pm}(x^A)+(1-\delta_{n,0}-\delta_{n,-1})\frac{(-1)^{n+1}}{(n+1)!}\Delta_{2,n+1,0}\Big((4+n)+2 \nabla^2\Big)\nabla^{A}\nabla^BH_{AB}^{(\mathbb L)n\pm}(x^A)\nn \\&+h^{(\mathbb S)n\pm}(x^A)-\frac{1}{n+1+\delta_{n,-1}}\vert\tau\vert\nabla^A H_{A}^{(\mathbb L,\tau)n\pm}(x^A)+\delta_{n,-1}\vert\tau\vert \nabla^A\nabla^B H^{(\mathbb S)-1\pm}_{AB}(x^A)\Big]+ o( e^{-(n+4)\vert\tau\vert}),
\end{align}
and 
\begin{align}
    &T_{A\tau}^{(SDT)n}(\tau\to\pm\infty,x^A)=\pm 2 (1-\delta_{n,-1})e^{n\vert\tau\vert}\Big[\sum_{r=0}^{n}e^{-2r\vert\tau\vert}\sum_{k=0}^r \frac{(-1)^{r}(n-k)!r!}{(k!)^2(r-k)!}\Big(\Delta_{k,1}\delta^B_A -\frac{2k(r+1)}{(k+1)(n+1)}\Delta_{1,k,1}\nabla_{A}\nabla^B\Big)\nabla^CH_{CB}^{(\mathbb L)n\pm}(x^A)\Big]\nn\\&\pm 2 e^{-(n+2)\vert\tau\vert}\Big[ \vert\tau\vert H_{A}^{(\mathbb L,\tau)n\pm}(x^A)
    -H_A^{(\mathbb S)n\pm}(x^A) +2\frac{(-1)^{n+1}}{(n+1)!(n+1+\delta_{n,-1})}  \Delta_{1,n+1,1}\nabla_A\nabla^C\nabla^B H_{CB}^{(\mathbb L)n\pm}(x^A)\nn\\&+2 (1-\delta_{n,-1}-\delta_{n,0})(-1)^{n+1}n! \sum_{k=1}^{n}\frac{1}{k!(k+1)!}\Delta_{1,k,1}\nabla_A\nabla^C\nabla^B H_{CB}^{(\mathbb L)n\pm}(x^A)\Big] +o(e^{-(n+2)\vert\tau\vert}),
\end{align}
\end{widetext} 
where we used again Eq. \eqref{DeltaIdentity} for the $A\tau$ component.

Using a similar procedure as in the vector case, we can find the antipodal matching conditions for the asymptotic data. 
The mode functions $\nabla^BH_{AB}^{(\mathbb L)n\pm}(x^A)$, $H_A^{(\mathbb S)n\pm}(x^A)$, $H_{A}^{(\mathbb L,\tau)n\pm}(x^A)$ and $h^{(\mathbb S)n\pm}(x^A)$ obey similar antipodal matching conditions as, respectively, $K_A^{(\mathbb L)n\pm}(x^A)$, $K_A^{(\mathbb S)n\pm}(x^A)$, $K_A^{(\mathbb L,\tau)n\pm}(x^A)$ and $k^{(\mathbb S)n\pm}(x^A)$ that appear in the vector case up to a global minus sign. The explicit antipodal relationships read as follows: 
\begin{align}\label{AntipodalmathcT1}
   &H_A^{(\mathbb S)n+}(x^A)- O^{(q)n}_V(\nabla^BH_{AB}^{(\mathbb L)n+}(x^A)) = (-1)^{n+1}\Upsilon^*\Big(H_A^{(\mathbb S)n-}(x^A) \nn\\
   &\quad -  O^{(q)n}_V(\nabla^BH_{AB}^{(\mathbb L)n-}(x^A))\Big), \quad n\ge0,\\
&
   \epsilon^{AB}\nabla_AH_B^{(\mathbb S)-1,+}(x^A)- \epsilon^{AB}\nabla_A O^{(q)n=-1}_V(\nabla^C H_{CB}^{(\mathbb L)-1,+}(x^A)) = \nn\\ 
   &\quad  \Upsilon^*\Big(\epsilon^{AB}\nabla_AH_B^{(\mathbb S)-1,-}(x^A)\nn\\
   &\quad-  \epsilon^{AB}\nabla_AO^{(q)n=-1}_V(\nabla^C H_{CB}^{(\mathbb L)-1,-}(x^A))\Big),\\
   &
   \nabla^AH_A^{(\mathbb S)-1,+}(x^A) =- \Upsilon^* \nabla^A H_A^{(\mathbb S)-1,-}(x^A),\\
   &\nabla^BH_{AB}^{(\mathbb L)n+}(x^A)- O^{(p)n}_{VE}(\mathcal H_A^{(\mathbb S)n+}(x^A)) + O^{(p)n}_{VB}(H_A^{(\mathbb S)n+}(x^A))  \nn\\&= (-1)^{n}\Upsilon^*\Big(\nabla^BH_{AB}^{(\mathbb L)n-}(x^A)- O^{(p)n}_{VE}(\mathcal H_A^{(\mathbb S)n-}(x^A))\nn\\&\quad + O^{(p)n}_{VB}(H_A^{(\mathbb S)n-}(x^A))\Big),  \\
&h^{(\mathbb S)n+}(x^A)-\frac{1}{n+1}O^{(q)n}(\nabla^A \nabla^BH_{AB}^{(\mathbb L)n+}(x^A))=(-1)^{n+1} \times  \nn \\ 
   &\quad\Upsilon^*\Big(h^{(\mathbb S)n-}(x^A) -\frac{1}{n+1}O^{(q)n}(\nabla^A \nabla^BH_{AB}^{(\mathbb L)n-}(x^A))\Big), \quad n\ge0, 
             \\
    &h^{(\mathbb S)-1,+}(x^A)+\frac{1}{2}O^{(q)n=-1}(\nabla^A H_A^{(\mathbb S)-1,+}(x^A))= \nn  \\
   &\quad\Upsilon^*\Big(h^{(\mathbb S)-1,-}(x^A) +\frac{1}{2}O^{(q)n=-1}(\nabla^A H_A^{(\mathbb S)-1,-}(x^A))\Big), 
                   \end{align}
    \begin{align}
   &\nabla^A \nabla^BH_{AB}^{(\mathbb L)n+}(x^A)-(n+1)O^{(p)n}(h^{(\mathbb S)n+}(x^A))= (-1)^{n}\times \nn \\
   &\quad \Upsilon^*\Big(\nabla^A \nabla^BH_{AB}^{(\mathbb L)n-}(x^A)-(n+1)O^{(p)n}(h^{(\mathbb S)n-}(x^A))\Big),\\&
   H_{A}^{(\mathbb L,\tau)n+}(x^A)  = (-1)^{n}\Upsilon^* H_{A}^{(\mathbb L,\tau)n-}(x^A).
\end{align}
where $\mathcal H_A^{(\mathbb S)n\pm}(x^A) =H_A^{(\mathbb S)n\pm}(x^A)+2(1-\delta_{n,-1})(-1)^{n+1}n!\nabla^BH_{AB}^{(\mathbb L)n\pm}(x^A)$.

Regarding $H_{AB}^{(\mathbb L)n\pm}(x^A)$, $H_{AB}^{(\mathbb S)n\pm}(x^A)$ and $H_{AB}^{(\mathbb L,\tau)n\pm}(x^A)$, we have the following antipodal matching conditions: \begin{align}
    &H_{AB}^{(\mathbb L)n+}(x^A)-O^{(p)n}_{TE}[\mathcal H^{(\mathbb S)n+}_{AB}(x^A)]+ O^{(p)n}_{TB}[\mathbb H^{(\mathbb S)n+}_{AB}(x^A)] \nn\\
    &\quad= (-1)^{n}\Upsilon^*\Big( H_{AB}^{(\mathbb L)n-}(x^A)-O^{(p)n}_{TE}[\mathcal H^{(\mathbb S)n-}_{AB}(x^A)]\nn\\
    &\quad\quad+ O^{(p)n}_{TB}[\mathbb H^{(\mathbb S)n-}_{AB}(x^A)]\Big),\\
    &H_{AB}^{(\mathbb S)n+}(x^A)-O^{(q)n}_T[H^{(\mathbb L)n+}_{AB}(x^A)] = (-1)^{n+1}\Upsilon^*\Big(H_{AB}^{(\mathbb S)n-}(x^A)\nn\\ 
    &\quad -O^{(q)n}_T[H^{(\mathbb L)n-}_{AB}(x^A)]\Big), \; n\ge1;\\
    &\epsilon^{BC}\nabla_C\nabla^A H_{AB}^{(\mathbb S)-1,+}(x^A)-\epsilon^{BC}\nabla_C\nabla^A O^{(q)-1}_T[H^{(\mathbb L)-1,+}_{AB}(x^A)] =\nn\\
    &\quad \Upsilon^*\Big(\epsilon^{BC}\nabla_C\nabla^A H_{AB}^{(\mathbb S)-1,-}(x^A)-\epsilon^{BC}\nabla_C\nabla^A O^{(q)-1}_T[H^{(\mathbb L)-1,-}_{AB}(x^A)]\Big);\\
    &\nabla^A \nabla^B H^{(\mathbb S)-1,+}_{AB}(x^A) = - \Upsilon^*\nabla^A \nabla^B H^{(\mathbb S)-1,-}_{AB}(x^A);\\
    &H_{AB}^{(\mathbb L,\tau)n+}(x^A)=(-1)^{n}\Upsilon^*H_{AB}^{(\mathbb L,\tau)n-}(x^A),\label{AntipodalmathcTend}
\end{align}
where $\mathcal H^{(\mathbb S)n\pm}_{AB}(x^A):=  H^{(\mathbb S)n\pm}_{AB}(x^A)+(-1)^{n+1}(1-\delta_{n,-1}-\delta_{n,0})n!((n+1)\Delta_{1,2}+n)H^{(\mathbb L)n\pm}_{AB}(x^A) $ and $\mathbb H^{(\mathbb S)n\pm}_{AB}(x^A):=  H^{(\mathbb S)n\pm}_{AB}(x^A)-(-1)^{n+1}(1-\delta_{n,-1}-\delta_{n,0})n!((n+1)\Delta_{1,2}+n)H^{(\mathbb L)n\pm}_{AB}(x^A) $. The operators $O^{(p)n}_{TE}$, $O^{(p)n}_{TB}$, $O^{(q)n}_T$ acting on a tensor $f_{AB}(x^A)$ are defined as follows  
\begin{align}
    &O^{(p)n}_{TE}[f_{AB}(x^A)] \nn\\&:= \sum_{\ell=2}^{+\infty}\sum_{m=-\ell}^{+\ell}(-1)^{\ell+n+1}\frac{(n+1)!}{2n\Gamma(n+\ell+2)\Gamma(n-\ell+1)} \nn\\ 
    &\hspace{-0.3cm}\times \mathring T^{(E)\ell m}_{AB}(x^A)\oint_{S^2}d\Omega' \overline{\mathring T^{(E)\ell m}_{CD}(x^{A'})}f^{CD}(x^{A'}),\quad n\ge2,\\
    &O^{(p)n}_{TB}[f_{AB}(x^A)]\nn\\ &:= \sum_{\ell=2}^{+\infty}\sum_{m=-\ell}^{+\ell}(-1)^{\ell+n+1}\frac{(n+1)!}{2n\Gamma(n+\ell+2)\Gamma(n-\ell+1)} \nn\\ &\hspace{-0.3cm}\times \mathring T^{(B)\ell m}_{AB}(x^A)\oint_{S^2}d\Omega' \overline{\mathring T^{(B)\ell m}_{CD}(x^{A'})}f^{CD}(x^{A'}),\quad n\ge2,
    \end{align}
    \begin{align}
    &O^{(p)n}_{TD}[f_{AB}(x^A)]:=0,\qquad D=E,B, \qquad n<2,\\
    &O^{(q)n}_T[f_{AB}(x^A)] \nn\\&:=\sum_{\ell=2}^{+\infty}\sum_{m=-\ell}^{+\ell} (-1)^{n+1}\Big[n\frac{(-\ell)_{n+1}(\ell+1)_{n+1}}{(n+1)!}(H_{n+1} \nn\\ 
    &- 2 H_{\ell} )-(1-\delta_{n,0}-\delta_{n,-1})\Big(-(n+1)\ell(\ell+1)+n\Big)n!\Big]  \nn\\&\times \mathring T^{(E)\ell m}_{AB}(x^A)\oint_{S^2}d\Omega' \overline{\mathring T^{(E)\ell m}_{CD}(x^{A'})}f^{CD}(x^{A'})\nn\\&-\sum_{\ell=2}^{+\infty}\sum_{m=-\ell}^{+\ell} (-1)^{n+1}\Big[n\frac{(-\ell)_{n+1}(\ell+1)_{n+1}}{(n+1)!}(H_{n+1} \nn\\ 
    &- 2 H_{\ell} )-(1-\delta_{n,0}-\delta_{n,-1})\Big(-(n+1)\ell(\ell+1)+n\Big)n!\Big]\nn\\&\times \mathring T^{(B)\ell m}_{AB}(x^A)\oint_{S^2}d\Omega' \overline{\mathring T^{(B)\ell m}_{CD}(x^{A'})}f^{CD}(x^{A'}).
\end{align}
The operators $O^{(p)n}_{TE}$, $O^{(p)n}_{TB}$ annihilate all $\ell\ge n+1$ harmonics. Similarly to the vector case, the operator $O^{(q)n}_T$ does not annihilate $\ell< n+1$ harmonics.  Antipodal matching conditions are generally non local for the asymptotic data $H_A^{(\mathbb S)n\pm}(x^A)$, $h^{(\mathbb S)n\pm}(x^A)$, $H_{AB}^{(\mathbb L)n\pm}(x^A)$ and $H_{AB}^{(\mathbb S)n\pm}(x^A)$ and always local for $H_{A}^{(\mathbb L,\tau)n\pm}(x^A)$ and $H_{AB}^{(\mathbb L,\tau)n\pm}(x^A)$. If one assumes $\ell\ge n+1$, then the leading asymptotic data $H_{AB}^{(\mathbb L)n\pm}(x^A)$ also admit local antipodal matching conditions.

For inhomogeneous solutions to Eqs. \eqref{generalInhomTensEqu1}-\eqref{generalInhomTensEqu3} for $n \geq -1$, the procedure to extract antipodal matching conditions, which results in the existence of conserved quantities between $\tau =\infty$ and $\tau=-\infty$, goes as follows. First, we construct the source $S_{ab}^{(SDT)n}$ using Eq. \eqref{Kab}. The source then allows to define the source scalars $\rho^E_n$ and $\rho^B_n$ from Eq. \eqref{kappaEB}. This allows to compute $A_{n\ell m}^{(S,D)}$ and $B_{n\ell m}^{(S,D)}$, $D=E,B$ using Eqs. \eqref{ASETensorbeg}-\eqref{BSETesnorend}, which can be substituted into the decomposition \eqref{SDTNHdecomp}. We can then subtract these contributions to obtain a ``subtracted'' tensor $\hat T^{n}_{ab}$ that behaves asymptotically as a homogeneous solution. Then, we can identify from the asymptotic behavior of $\hat T^{n}_{ab}$ the asymptotic data $\hat H_{AB}^{(\mathbb L)n\pm}(x^A)$, $\hat H_{AB}^{(\mathbb S)n\pm}(x^A)$, $\hat H_{AB}^{(\mathbb L,\tau)n\pm}(x^A)$, $\hat H_A^{(\mathbb S)n\pm}(x^A)$, $\hat H_{A}^{(\mathbb L,\tau)n\pm}(x^A)$ and $\hat h^{(\mathbb S)n\pm}(x^A)$ that satisfy the antipodal matching conditions \eqref{AntipodalmathcT1}-\eqref{AntipodalmathcTend}. These antipodal relationships are equivalent to the conservation of the following charges. 

For $n\ge1$, 
\begin{subequations}\label{Q2A}
\begin{align}
    \mathcal Q_{U}[\hat T^{(P)n}]& := \oint_{S^2}d\Omega \hat T^{(P)n+}_{AB} \overline{U^{AB}} \nonumber \\ 
    &=(-1)^{n+1}\oint_{S^2}d\Omega \hat T^{(P)n-}_{AB} \Upsilon^*\overline{U^{AB}},\\
     \mathcal Q_{U}[\hat T^{(Q)n}]&:= \oint_{S^2}d\Omega \hat T^{(Q)n+}_{AB} \overline{U^{AB}}\nonumber \\ 
     &=(-1)^{n}\oint_{S^2}d\Omega \hat T^{(Q)n-}_{AB} \Upsilon^*\overline{U^{AB}},
\end{align}    
\end{subequations}
where $U_{AB}(x^B)$ are arbitrary symmetric traceless tensors on the sphere and
\begin{align}
    \hat T^{(P)n\pm}_{AB}& :=\hat H_{AB}^{(\mathbb S)n\pm}(x^A)-O^{(q)n}_T[\hat H^{(\mathbb L)n\pm}_{AB}(x^A)], \\
    \hat T^{(Q)n\pm}_{AB}& := \hat H_{AB}^{(\mathbb L)n\pm}(x^A)-O^{(p)n}_{TE}[\hat{\mathcal H}^{(\mathbb S)n\pm}_{AB}(x^A)]\nn\\&\quad\;+ O^{(p)n}_{TB}[\hat{\mathbb H}^{(\mathbb S)n\pm}_{AB}(x^A)]. 
\end{align}

For $n=0$, a vector formulation is suitable to express the subleading conserved charges while the leading conserved charges are defined using tensors. They read as 
\begin{subequations}\label{Q2B}
\begin{align}
    \mathcal Q_{Y}[\hat T^{(P)0}]&:= \oint_{S^2}d\Omega \hat T^{(P)0+}_{A} \overline{Y^{A}}=-\oint_{S^2}d\Omega \hat T^{(P)0-}_{A} \Upsilon^*\overline{Y^{A}},\\
     \mathcal Q_{U}[\hat H^{(\mathbb L)0}]&:= \oint_{S^2}d\Omega \hat H_{AB}^{(\mathbb L)0+}(x^A) \overline{U^{AB}}\nn\\&=\oint_{S^2}d\Omega \hat H_{AB}^{(\mathbb L)0-}(x^A) \Upsilon^*\overline{U^{AB}},
\end{align}    
\end{subequations}
where $\hat H_{AB}^{(\mathbb L)0\pm}(x^A)$ is the leading asymptotic data of $\hat T_{\langle A B\rangle}$ and 
\begin{align}
    \hat T^{(P)0\pm}_{A}&:= \hat H_A^{(\mathbb S)0\pm}(x^A)- O^{(q)0}_V(\nabla^B\hat H_{AB}^{(\mathbb L)0\pm}(x^A)).
\end{align}

Finally, for $n=-1$, a scalar formulation is suitable to express the conserved charges as 
\begin{subequations}\label{Q2C}
\begin{align}
    \mathcal Q_{T}[\hat t^{(P)D}]&:= \oint_{S^2}d\Omega\, \hat t^{(P)D+} \overline{T}=\oint_{S^2}d\Omega\, \hat t^{(P)D-} \Upsilon^*\overline{T},\\
     \mathcal Q_{T}[\hat t^{(Q)D}]&:= \oint_{S^2}d\Omega\, \hat t^{(Q)D+} \overline{T}=-\oint_{S^2}d\Omega\, \hat t^{(Q)D-} \Upsilon^*\overline{T},
\end{align}    
\end{subequations}
with  $D=E,B$ and 
\begin{align}
    \hat t^{(P)E\pm} &:= \hat h^{(\mathbb S)-1,\pm}(x^A)+\frac{1}{2}O^{(q)-1}(\nabla^A \hat H_A^{(\mathbb S)-1,\pm}(x^A)), \\
    \hat t^{(P)B\pm} &:=   \epsilon^{BC}\nabla_C\nabla^A\hat H_{AB}^{(\mathbb S)-1,\pm}(x^A)\nn\\&\hspace{1.2cm}-\epsilon^{BC}\nabla_C\nabla^A O^{(q)-1}_T[\hat H^{(\mathbb L)-1,\pm}_{AB}(x^A)], \\
    \hat t^{(Q)E\pm} &:= \nabla^A\nabla^B\hat H_{AB}^{(\mathbb S)-1,\pm}(x^A), \\
    \hat t^{(Q)B\pm} &:=  \epsilon^{AB}\nabla_A \nabla^C\hat H_{CB}^{(\mathbb L)-1\pm}(x^A).
\end{align}

Other charges that could be built with other asymptotic data that appear in the relationships \eqref{AntipodalmathcT1}-\eqref{AntipodalmathcTend} are redundant  with the charges $ \mathcal Q_{U}[\hat T^{(P)n}]$, $ \mathcal Q_{U}[\hat T^{(Q)n}]$ built explicitly here. This can be seen from the relations \eqref{relationKABn1}, \eqref{relationLA},\eqref{relationDA}, \eqref{relationf}.

Let us discuss the conserved charges in terms of the relevant KG inner products. In the generic case $n\ge1$, the tensor KG inner product between $\hat T_{ab}$ and any SDT tensor on $dS_3$ that satisfies the homogeneous equation \eqref{generalSDTTensorEquations} will converge for $\tau\to\pm\infty$. Such an arbitrary SDT tensor is a linear combination of a $p$-parity tensor $T^{(SDT)n,P}_{ab}$ and a $q$-parity tensor  $T^{(SDT)n,Q}_{ab}$ which satisfy Eq. \eqref{generalSDTTensorEquations}. We can reproduce the charges $ \mathcal Q_{U}[\hat T^{(P)n}]$ and $ \mathcal Q_{U}[\hat T^{(Q)n}]$ upon identifying $U_{AB}$ to the asymptotic data of $T^{(SDT)n,Q}_{\langle A B\rangle}$, or, to the asymptotic data of $T^{(SDT)n,P}_{\langle A B\rangle}$, respectively.

In the special case  $n=0$, the vector KG inner product between, on the one hand, the vector $\xi^a_{\hat T}$ built from the tensor $\hat T_{ab}$ and, on the other hand, the $p$-parity vector $\xi^a_{T^{(SDT)n,P}}$ built from the $p$-parity tensor $T^{(SDT)n,P}_{ab}$ satisfying Eqs. \eqref{generalSDTTensorEquations} corresponds to $\mathcal Q_{U}[\hat H^{(\mathbb L)0}]$ where $U_{AB}$ has been suitably identified to the asymptotic data of $\xi^a_{T^{(SDT)n,P}}$. In addition, the vector KG inner product between, on the one hand, the vector $\xi^a_{\hat T}$ built from the tensor $\hat T_{ab}$ and, on the other hand, a transverse vector $V^{(T)n,Q}_a$ satisfying Eqs. \eqref{generalEquationvectorsHarms} will match $\mathcal Q_{Y}[\hat T^{(P)0}]$, with $Y_A$ corresponding to the asymptotic data of $V^{(T)n,Q}_a$. 

Finally, for the special case $n=-1$, the scalar KG inner products between, on the one hand, the scalars $\psi^E_\xi$ and $\psi^B_\xi$ built from the transverse vector $\xi^a_{\hat T}$ (see Eqs. \eqref{PsiEVect}-\eqref{PsiBVect}) and, on the other hand, the  scalars of $p$-parity $\psi^{P}_n(\tau,x^A)$ and $q$-parity $\psi^{Q}_n(\tau,x^A)$ which satisfy Eq. \eqref{generalscalarequation}, will match up to the charges \eqref{Q2C} once $T$ is identified with the asymptotic data of  $\psi_n^{Q}(\tau,x^A)$ or $\psi^{P}_n(\tau,x^A)$, as in the scalar case.

\begin{table}[!hbt]
    \centering
    \begin{tabular}{|l|l|c|}\hline
    {\bf Rank} & {\bf Class} & {\bf Eigenvalue} \\ \hline
     Scalar &    Scalar & $1-(n+1)^2$  \\ \hline
     Vector &   Transverse vector (E or B)& $2-(n+1)^2$ \\ 
    &    Longitudinal vector  & $3-(n+1)^2$ \\ \hline
    Tensor &   Pure trace tensor & $1-(n+1)^2$  \\
    &  Traceless tensor built from scalars  &  $7-(n+1)^2$  \\
    &   Traceless tensor built from vectors (E or B) & $6-(n+1)^2$ \\ 
    &   Divergence-free traceless tensor (E or B) & $3-(n+1)^2$ \\ 
       \hline
    \end{tabular}
    \caption{Eigenvalue of the D'Alembertian operator $\alpha$ for the distinct classes of harmonics of $dS_3$ of ranks 0 (scalar) ,1 (vector) and 2 (tensor). All tensors are symmetric. 
    }
    \label{alpha}
\end{table}

\section{Conclusion}

We provided a systematic construction of scalar, vector and tensor harmonics on $dS_3$. The eigenvalues of the D'Alembertian operator appearing in Eq. \eqref{wave} are provided for the various classes of harmonics on Table \ref{alpha}. Each harmonic $\Psi^{(c) n\ell m}_{i_1 \dots i_T}$ is labelled by $n$, the spherical harmonic mode $\ell m$ and the type $c=p$ or $q$. The $p$ type and $q$ type originate from the hyperboloidal nature of the wave equation. For scalars ($r=0$), transverse vectors ($r=1$) and SDT tensors ($r=2$) and each integer $n$, the $p$ type and $q$ type are distinguished by their parity under the antipodal map (i.e. combined time reversal and parity flip) on $dS_3$, which is respectively $(-1)^{n+r+1}$ and $(-1)^{n+r}$. For all cases, the asymptotic behavior of $q$ type harmonics in either limit $\tau \to \pm \infty$ is slower than, or equal to, the asymptotic behavior of $p$ type harmonics. In other words, the $q$ harmonics are more leading, or equally leading, as the $p$ harmonics. Since $q$ harmonics are defined with Legendre $Q$ functions, they display logarithmic branches for $\ell\ge n+1$, which manifest themselves as linear terms in $\vert\tau\vert$ in the asymptotic behavior. 

Since the initial value problem is defined from two sets of initial data over the sphere, one can completely define a homogeneous solution to the wave equation \eqref{wave} by its leading and (suitably defined) subleading asymptotic behavior around either boundary. We provided the explicit formulae that allow to extract the leading and subleading asymptotic data defined over the sphere. These asymptotic data allow to construct conserved charges at the two boundaries $\tau \to \pm \infty$, which are related by an antipodal relationship, which we derived for each case.

In the presence of arbitrary sources, the inhomogeneous solutions acquire new asymptotic branches that obscure the antipodal relationships. Nevertheless, we were able to provide a general procedure that allows to identify the asymptotic data even in the presence of arbitrary sources by constructing a ``subtracted field'' that admits the same asymptotic behavior as a homogeneous solution. We therefore provided the antipodal relationships and the construction of conserved charges for the inhomogeneous case as well in full generality. We expect that this analysis will be relevant for the study of interacting 4-dimensional theories. 

The resulting charges can be classified as follows. Each scalar on $dS_3$ defines two scalar conservation laws, one of $p$-type and one of $q$-type, that are both defined for an arbitrary function over the sphere, see Eqs. \eqref{Q0}. Similarly, each transverse vector with $n \geq 0$ on $dS_3$ defines a $p$-type and a $q$-type vector conservation law that are both defined for an arbitrary vector over the sphere, see Eqs. \eqref{Q1A}. In the special case $n =-1$, one can instead construct a pair of $p$-type and a pair of $q$-type scalar conservation laws, see Eqs. \eqref{Q1B}. Finally, each symmetric divergence-free traceless  (SDT) tensor on $dS_3$ with $n \geq 1$ defines a $p$-type and a $q$-type tensor conservation law that are both defined for an arbitrary traceless tensor over the sphere, see Eqs. \eqref{Q2A}. In the special case $n=0$, one can instead construct a $p$-type vector and a $q$-type tensor conservation law, see Eqs. \eqref{Q2B}. In the special case $n=-1$, one constructs a pair of $p$-type and $q$-type scalar conservation laws, see Eqs. \eqref{Q2C}.

These conservation laws are realized in many 4-dimensional asymptotically flat physical systems. They take many names, depending upon the physical theory considered and the asymptotic field within that theory considered in the expansion around spatial infinity. Typically, boundary conditions impose that one of the two types, either the $p$ type or $q$ type is absent. In that case, half of the conservation laws are trivial (i.e. of the form $0=0$). In several cases, the $q$ branches start to appear in the interacting theory beyond tree level, which brings logarithmic branches related to logarithmic soft theorems \cite{Laddha:2018myi,Laddha:2018vbn,Sahoo:2018lxl}. A non-exhaustive but illustrative list of examples of physical applications is given in Table \ref{table2}. Our results on scalar harmonics can be compared with \cite{Henneaux:2018mgn,Fuentealba:2024lll} where their $k$ is related to our $n$ as $n=k-1$. The antipodal relationships \eqref{Matchingscalarend0} are consistent with their analysis and make their analysis more explicit, while the antipodal relationships related to the subleading behavior of the harmonics \eqref{Matchingscalarend} are novel. Our results on vector harmonics extend the earlier work \cite{Campiglia:2018dyi,Compere:2025tzr,Fuentealba:2025ekj,Briceno:2025cdu,Henneaux_2018,Henneaux:2019yqq} by providing generic matching conditions at leading and subleading orders for all types of vector harmonics. The matching conditions obtained for tensor fields will be exploited in an upcoming work \cite{Compere:2026jaa} to derive the conservation law of the quadrupole in General Relativity. We note that for all cases considered, in a theory describing a massless helicity $s$ 4-dimensional field, a rank $r$ tensor in $dS_3$ describes a  
$\vert L \vert := n+1+r-s$ order component of the field. For example, the field $\sigma$ ($r=0$) in General Relativity ($s=2$) is leading order ($\vert L \vert=0$) while $V_{ab}$ is subleading order ($\vert L \vert=1$).

In our analysis, we noted that the conservation laws can be either local or non-local over the sphere. The local conservation laws relate an asymptotic field defined at a specific angle at future infinity with the corresponding asymptotic field defined at the antipodally related angle at past infinity. The non-local conservation laws are instead only globally defined over the sphere. While the asymptotic fields appearing in the leading logarithmic branches of the $q$ parity harmonics 
admit local antipodal relationships, the other asymptotic fields admit a non-local mapping when no assumption is made on the asymptotic data.  To make explicit these maps, we defined various non-local operators $O^{(p)n}$, $O^{(q)n}$ for each case. An important exception occurs for all scalars, vectors and tensors with harmonic content $ \ell \ge n+1$. In that case, all leading asymptotic fields admit local antipodal relationships. This case precisely occurs in the description at spatial infinity of several physical 4-dimensional asymptotically flat systems such as massless scalar at leading order or General Relativity at leading ($\vert L \vert=0$), subleading  ($\vert L \vert=1$) and subsubleading ($\vert L \vert=2$) orders \footnote{Massless scalar has $s=0$ and leading $r=0$ fields have $n=-1$ so that $n+1 \le \ell$ for any $\ell \geq 0$. For General Relativity, first order SDT tensors satisfy $n+1=0\le\ell$, second order SDT tensors $(n+1=1)$ can not have $\ell=0$ harmonics and third order SDT tensors $(n+1=2)$ can not have $\ell=0,1$ harmonics.}. This suggests an intriguing connection between locality, conservation laws and the discrete spectrum of fields in the theory which deserves more exploration, see e.g. \cite{Pasterski:2016qvg,Freidel:2022skz} for related work.

As part of our analysis, we proved several theorems on the decomposition of vector and tensors on $dS_3$. Most of the 9 lemmae described in Sections \ref{sec:TransverseVectors}, \ref{SDTsec},\ref{sec:PropertiesTensors} have not yet appeared in the literature to the best of our knowledge with the exception of Lemma \ref{lemma:T3} stated in \cite{Ashtekar:1978zz} and proven in \cite{Compere:2011db} as Lemma 1. These lemmae generalize the lemmae proven in \cite{Compere:2011db}.  

As a final perspective, we expect that our work can be extended to fermionic fields, higher rank tensors and to higher dimensions.

\begin{widetext}
\begin{table*}[!tbh]
    \centering
    \begin{tabular}{|l|c|c|c|c|c|c|c|}\hline
   {\bf Theory}&  {\bf Field name } &  {\bf Class} & $n$ & {\bf Type } &  {\bf Parity } & {\bf Conserved charge} & {\bf References}\\ \hline
  General Relativity & $\sigma$ &   Scalar  & 1  & $p$ & Even &  (Super-)momentum & \cite{Ashtekar:1978zz,Ashtekar:1990gc} \\ \hline
  & $k_{ab}$ &  Traceless tensor built from scalar   & 1 & $q$ & Odd  & Supertranslation frame \footnote{The charge is given as Eq. \eqref{defQ} and can be realized canonically as associated with logarithmic supertranslations \cite{Fuentealba:2022xsz}.} & \cite{Compere:2011ve,Troessaert:2017jcm,Henneaux:2018cst,Fuentealba:2022xsz,Compere:2023qoa}\\ \hline
     & $i_{ab}$ &  SDT tensor (E and B) & 0 & $q$ & Even & Logarithmic charge & \cite{compere_asymptotic_2011,Compere:2023qoa} \\   \hline
   & $V_{ab}$ &  SDT tensor (E) &0 & $p$ & Odd & (Super-)center-of-mass  & \cite{beig1984integration,compere_asymptotic_2011,Prabhu:2021cgk,Compere:2023qoa} \\   \hline
      & $V_{ab}$ &  SDT tensor (B) &0 & $p$ & Odd & (Super-)angular momentum & \cite{beig1984integration,compere_asymptotic_2011,Prabhu:2021cgk,Compere:2023qoa} \\   \hline
      & $V_{ab}$ &  SDT tensor (E and B) &0 & $q$ & Even & Tail charges & \cite{Compere:2023qoa,Compere:2026jmk} \\   \hline
Massless QED & $\overset{n+1}{F_{\rho a}}$ & Transverse vector & $n$ & $p$ & $(-1)^n$ & Sub-$n$ charges & \cite{Campiglia:2018dyi,AtulBhatkar:2020hqz,Compere:2025tzr} \\ \hline
Massless scalar & $\phi^{(k)}$ & Scalar & $k-1$  & $p/q$ & $+/-$ $(-1)^{k}$ & Scalar charges &  
\cite{Henneaux:2018mgn,Fuentealba:2024lll,Briceno:2025ivl}\\\hline 
    \end{tabular}
    \caption{Some examples of physical realizations of conservation laws of $p$-type and $q$-type in 4-dimensional asymptotically flat spacetimes. The parity under the antipodal map is $(-1)^{n+r+1}$ (resp. $(-1)^{n+r}$) for a $p$-type (resp. $q$-type) rank $r$ SDT tensor. 
    }
\label{table2}
\end{table*}    
\end{widetext}

\section*{Acknowledgments}

We thank Marc Henneaux for his useful feedback on our manuscript. We used the Riemannian Geometry $\&$ Tensor Calculus (RGTC) package for Mathematica, written by Sotirios Bonanos (\textgreek{Σωτήριος Μπονάνος}). GC is Research Director of the FNRS.



\begin{thebibliography}{74}%
\makeatletter
\providecommand \@ifxundefined [1]{%
 \@ifx{#1\undefined}
}%
\providecommand \@ifnum [1]{%
 \ifnum #1\expandafter \@firstoftwo
 \else \expandafter \@secondoftwo
 \fi
}%
\providecommand \@ifx [1]{%
 \ifx #1\expandafter \@firstoftwo
 \else \expandafter \@secondoftwo
 \fi
}%
\providecommand \natexlab [1]{#1}%
\providecommand \enquote  [1]{``#1''}%
\providecommand \bibnamefont  [1]{#1}%
\providecommand \bibfnamefont [1]{#1}%
\providecommand \citenamefont [1]{#1}%
\providecommand \href@noop [0]{\@secondoftwo}%
\providecommand \href [0]{\begingroup \@sanitize@url \@href}%
\providecommand \@href[1]{\@@startlink{#1}\@@href}%
\providecommand \@@href[1]{\endgroup#1\@@endlink}%
\providecommand \@sanitize@url [0]{\catcode `\\12\catcode `\$12\catcode
  `\&12\catcode `\#12\catcode `\^12\catcode `\_12\catcode `\%12\relax}%
\providecommand \@@startlink[1]{}%
\providecommand \@@endlink[0]{}%
\providecommand \url  [0]{\begingroup\@sanitize@url \@url }%
\providecommand \@url [1]{\endgroup\@href {#1}{\urlprefix }}%
\providecommand \urlprefix  [0]{URL }%
\providecommand \Eprint [0]{\href }%
\providecommand \doibase [0]{http://dx.doi.org/}%
\providecommand \selectlanguage [0]{\@gobble}%
\providecommand \bibinfo  [0]{\@secondoftwo}%
\providecommand \bibfield  [0]{\@secondoftwo}%
\providecommand \translation [1]{[#1]}%
\providecommand \BibitemOpen [0]{}%
\providecommand \bibitemStop [0]{}%
\providecommand \bibitemNoStop [0]{.\EOS\space}%
\providecommand \EOS [0]{\spacefactor3000\relax}%
\providecommand \BibitemShut  [1]{\csname bibitem#1\endcsname}%
\let\auto@bib@innerbib\@empty
\bibitem [{\citenamefont {Regge}\ and\ \citenamefont
  {Wheeler}(1957)}]{PhysRev.108.1063}%
  \BibitemOpen
  \bibfield  {author} {\bibinfo {author} {\bibfnamefont {T.}~\bibnamefont
  {Regge}}\ and\ \bibinfo {author} {\bibfnamefont {J.~A.}\ \bibnamefont
  {Wheeler}},\ }\href {\doibase 10.1103/PhysRev.108.1063} {\bibfield  {journal}
  {\bibinfo  {journal} {Phys. Rev.}\ }\textbf {\bibinfo {volume} {108}},\
  \bibinfo {pages} {1063} (\bibinfo {year} {1957})}\BibitemShut {NoStop}%
\bibitem [{\citenamefont {{Zerilli}}(1970)}]{1970PhRvD...2.2141Z}%
  \BibitemOpen
  \bibfield  {author} {\bibinfo {author} {\bibfnamefont {F.~J.}\ \bibnamefont
  {{Zerilli}}},\ }\href {\doibase 10.1103/PhysRevD.2.2141} {\bibfield
  {journal} {\bibinfo  {journal} {\prd}\ }\textbf {\bibinfo {volume} {2}},\
  \bibinfo {pages} {2141} (\bibinfo {year} {1970})}\BibitemShut {NoStop}%
\bibitem [{\citenamefont {Bardeen}(1980)}]{Bardeen:1980kt}%
  \BibitemOpen
  \bibfield  {author} {\bibinfo {author} {\bibfnamefont {J.~M.}\ \bibnamefont
  {Bardeen}},\ }\href {\doibase 10.1103/PhysRevD.22.1882} {\bibfield  {journal}
  {\bibinfo  {journal} {Phys. Rev. D}\ }\textbf {\bibinfo {volume} {22}},\
  \bibinfo {pages} {1882} (\bibinfo {year} {1980})}\BibitemShut {NoStop}%
\bibitem [{\citenamefont {{Kodama}}\ and\ \citenamefont
  {{Sasaki}}(1984)}]{1984PThPS..78....1K}%
  \BibitemOpen
  \bibfield  {author} {\bibinfo {author} {\bibfnamefont {H.}~\bibnamefont
  {{Kodama}}}\ and\ \bibinfo {author} {\bibfnamefont {M.}~\bibnamefont
  {{Sasaki}}},\ }\href {\doibase 10.1143/PTPS.78.1} {\bibfield  {journal}
  {\bibinfo  {journal} {Progress of Theoretical Physics Supplement}\ }\textbf
  {\bibinfo {volume} {78}},\ \bibinfo {pages} {1} (\bibinfo {year}
  {1984})}\BibitemShut {NoStop}%
\bibitem [{\citenamefont {Seljak}\ and\ \citenamefont
  {Zaldarriaga}(1997)}]{Seljak:1996gy}%
  \BibitemOpen
  \bibfield  {author} {\bibinfo {author} {\bibfnamefont {U.}~\bibnamefont
  {Seljak}}\ and\ \bibinfo {author} {\bibfnamefont {M.}~\bibnamefont
  {Zaldarriaga}},\ }\href {\doibase 10.1103/PhysRevLett.78.2054} {\bibfield
  {journal} {\bibinfo  {journal} {Phys. Rev. Lett.}\ }\textbf {\bibinfo
  {volume} {78}},\ \bibinfo {pages} {2054} (\bibinfo {year} {1997})},\ \Eprint
  {http://arxiv.org/abs/astro-ph/9609169} {arXiv:astro-ph/9609169} \BibitemShut
  {NoStop}%
\bibitem [{\citenamefont {Kamionkowski}\ \emph {et~al.}(1997)\citenamefont
  {Kamionkowski}, \citenamefont {Kosowsky},\ and\ \citenamefont
  {Stebbins}}]{Kamionkowski:1996zd}%
  \BibitemOpen
  \bibfield  {author} {\bibinfo {author} {\bibfnamefont {M.}~\bibnamefont
  {Kamionkowski}}, \bibinfo {author} {\bibfnamefont {A.}~\bibnamefont
  {Kosowsky}}, \ and\ \bibinfo {author} {\bibfnamefont {A.}~\bibnamefont
  {Stebbins}},\ }\href {\doibase 10.1103/PhysRevLett.78.2058} {\bibfield
  {journal} {\bibinfo  {journal} {Phys. Rev. Lett.}\ }\textbf {\bibinfo
  {volume} {78}},\ \bibinfo {pages} {2058} (\bibinfo {year} {1997})},\ \Eprint
  {http://arxiv.org/abs/astro-ph/9609132} {arXiv:astro-ph/9609132} \BibitemShut
  {NoStop}%
\bibitem [{\citenamefont {Lifshitz}(1946)}]{Lifshitz:1945du}%
  \BibitemOpen
  \bibfield  {author} {\bibinfo {author} {\bibfnamefont {E.}~\bibnamefont
  {Lifshitz}},\ }\href {\doibase 10.1007/s10714-016-2165-8} {\bibfield
  {journal} {\bibinfo  {journal} {J. Phys. (USSR)}\ }\textbf {\bibinfo {volume}
  {10}},\ \bibinfo {pages} {116} (\bibinfo {year} {1946})}\BibitemShut
  {NoStop}%
\bibitem [{\citenamefont {Lifshitz}\ and\ \citenamefont
  {Khalatnikov}(1963)}]{Lifshitz:1963ps}%
  \BibitemOpen
  \bibfield  {author} {\bibinfo {author} {\bibfnamefont {E.~M.}\ \bibnamefont
  {Lifshitz}}\ and\ \bibinfo {author} {\bibfnamefont {I.~M.}\ \bibnamefont
  {Khalatnikov}},\ }\href {\doibase 10.1080/00018736300101283} {\bibfield
  {journal} {\bibinfo  {journal} {Adv. Phys.}\ }\textbf {\bibinfo {volume}
  {12}},\ \bibinfo {pages} {185} (\bibinfo {year} {1963})}\BibitemShut
  {NoStop}%
\bibitem [{\citenamefont {Mashhoon}(1973)}]{PhysRevD.8.4297}%
  \BibitemOpen
  \bibfield  {author} {\bibinfo {author} {\bibfnamefont {B.}~\bibnamefont
  {Mashhoon}},\ }\href {\doibase 10.1103/PhysRevD.8.4297} {\bibfield  {journal}
  {\bibinfo  {journal} {Phys. Rev. D}\ }\textbf {\bibinfo {volume} {8}},\
  \bibinfo {pages} {4297} (\bibinfo {year} {1973})}\BibitemShut {NoStop}%
\bibitem [{\citenamefont {Hu}(1974)}]{10.1063/1.1666537}%
  \BibitemOpen
  \bibfield  {author} {\bibinfo {author} {\bibfnamefont {B.~L.}\ \bibnamefont
  {Hu}},\ }\href {\doibase 10.1063/1.1666537} {\bibfield  {journal} {\bibinfo
  {journal} {Journal of Mathematical Physics}\ }\textbf {\bibinfo {volume}
  {15}},\ \bibinfo {pages} {1748} (\bibinfo {year} {1974})}\BibitemShut
  {NoStop}%
\bibitem [{\citenamefont {Jantzen}(1978)}]{10.1063/1.523778}%
  \BibitemOpen
  \bibfield  {author} {\bibinfo {author} {\bibfnamefont {R.~T.}\ \bibnamefont
  {Jantzen}},\ }\href {\doibase 10.1063/1.523778} {\bibfield  {journal}
  {\bibinfo  {journal} {Journal of Mathematical Physics}\ }\textbf {\bibinfo
  {volume} {19}},\ \bibinfo {pages} {1163} (\bibinfo {year} {1978})},\ \Eprint
  {http://arxiv.org/abs/https://pubs.aip.org/aip/jmp/article-pdf/19/5/1163/19149335/1163\_1\_online.pdf}
  {https://pubs.aip.org/aip/jmp/article-pdf/19/5/1163/19149335/1163\_1\_online.pdf}
  \BibitemShut {NoStop}%
\bibitem [{\citenamefont {Gerlach}\ and\ \citenamefont
  {Sengupta}(1978)}]{Gerlach:1978gy}%
  \BibitemOpen
  \bibfield  {author} {\bibinfo {author} {\bibfnamefont {U.~H.}\ \bibnamefont
  {Gerlach}}\ and\ \bibinfo {author} {\bibfnamefont {U.~K.}\ \bibnamefont
  {Sengupta}},\ }\href {\doibase 10.1103/PhysRevD.18.1773} {\bibfield
  {journal} {\bibinfo  {journal} {Phys. Rev. D}\ }\textbf {\bibinfo {volume}
  {18}},\ \bibinfo {pages} {1773} (\bibinfo {year} {1978})}\BibitemShut
  {NoStop}%
\bibitem [{\citenamefont {Tomita}(1982)}]{10.1143/PTP.68.310}%
  \BibitemOpen
  \bibfield  {author} {\bibinfo {author} {\bibfnamefont {K.}~\bibnamefont
  {Tomita}},\ }\href {\doibase 10.1143/PTP.68.310} {\bibfield  {journal}
  {\bibinfo  {journal} {Progress of Theoretical Physics}\ }\textbf {\bibinfo
  {volume} {68}},\ \bibinfo {pages} {310} (\bibinfo {year} {1982})},\ \Eprint
  {http://arxiv.org/abs/https://academic.oup.com/ptp/article-pdf/68/1/310/5319939/68-1-310.pdf}
  {https://academic.oup.com/ptp/article-pdf/68/1/310/5319939/68-1-310.pdf}
  \BibitemShut {NoStop}%
\bibitem [{\citenamefont {Beig}\ and\ \citenamefont
  {Schmidt}(1982)}]{beig1982einstein}%
  \BibitemOpen
  \bibfield  {author} {\bibinfo {author} {\bibfnamefont {R.}~\bibnamefont
  {Beig}}\ and\ \bibinfo {author} {\bibfnamefont {B.~G.}\ \bibnamefont
  {Schmidt}},\ }\href@noop {} {\bibfield  {journal} {\bibinfo  {journal}
  {Communications in Mathematical Physics}\ }\textbf {\bibinfo {volume} {87}},\
  \bibinfo {pages} {65} (\bibinfo {year} {1982})}\BibitemShut {NoStop}%
\bibitem [{\citenamefont {Beig}(1984)}]{beig1984integration}%
  \BibitemOpen
  \bibfield  {author} {\bibinfo {author} {\bibfnamefont {R.}~\bibnamefont
  {Beig}},\ }\href@noop {} {\bibfield  {journal} {\bibinfo  {journal}
  {Proceedings of the Royal Society of London. A. Mathematical and Physical
  Sciences}\ }\textbf {\bibinfo {volume} {391}},\ \bibinfo {pages} {295}
  (\bibinfo {year} {1984})}\BibitemShut {NoStop}%
\bibitem [{\citenamefont {de~Boer}\ and\ \citenamefont
  {Solodukhin}(2003)}]{deBoer:2003vf}%
  \BibitemOpen
  \bibfield  {author} {\bibinfo {author} {\bibfnamefont {J.}~\bibnamefont
  {de~Boer}}\ and\ \bibinfo {author} {\bibfnamefont {S.~N.}\ \bibnamefont
  {Solodukhin}},\ }\href {\doibase 10.1016/S0550-3213(03)00494-2} {\bibfield
  {journal} {\bibinfo  {journal} {Nucl. Phys. B}\ }\textbf {\bibinfo {volume}
  {665}},\ \bibinfo {pages} {545} (\bibinfo {year} {2003})},\ \Eprint
  {http://arxiv.org/abs/hep-th/0303006} {arXiv:hep-th/0303006} \BibitemShut
  {NoStop}%
\bibitem [{\citenamefont {Dappiaggi}\ \emph {et~al.}(2006)\citenamefont
  {Dappiaggi}, \citenamefont {Moretti},\ and\ \citenamefont
  {Pinamonti}}]{Dappiaggi:2005ci}%
  \BibitemOpen
  \bibfield  {author} {\bibinfo {author} {\bibfnamefont {C.}~\bibnamefont
  {Dappiaggi}}, \bibinfo {author} {\bibfnamefont {V.}~\bibnamefont {Moretti}},
  \ and\ \bibinfo {author} {\bibfnamefont {N.}~\bibnamefont {Pinamonti}},\
  }\href {\doibase 10.1142/S0129055X0600270X} {\bibfield  {journal} {\bibinfo
  {journal} {Rev. Math. Phys.}\ }\textbf {\bibinfo {volume} {18}},\ \bibinfo
  {pages} {349} (\bibinfo {year} {2006})},\ \Eprint
  {http://arxiv.org/abs/gr-qc/0506069} {arXiv:gr-qc/0506069} \BibitemShut
  {NoStop}%
\bibitem [{\citenamefont {Mann}\ and\ \citenamefont
  {Marolf}(2006)}]{Mann:2005yr}%
  \BibitemOpen
  \bibfield  {author} {\bibinfo {author} {\bibfnamefont {R.~B.}\ \bibnamefont
  {Mann}}\ and\ \bibinfo {author} {\bibfnamefont {D.}~\bibnamefont {Marolf}},\
  }\href {\doibase 10.1088/0264-9381/23/9/010} {\bibfield  {journal} {\bibinfo
  {journal} {Class. Quant. Grav.}\ }\textbf {\bibinfo {volume} {23}},\ \bibinfo
  {pages} {2927} (\bibinfo {year} {2006})},\ \Eprint
  {http://arxiv.org/abs/hep-th/0511096} {arXiv:hep-th/0511096} \BibitemShut
  {NoStop}%
\bibitem [{\citenamefont {Barnich}\ and\ \citenamefont
  {Troessaert}(2010)}]{Barnich:2009se}%
  \BibitemOpen
  \bibfield  {author} {\bibinfo {author} {\bibfnamefont {G.}~\bibnamefont
  {Barnich}}\ and\ \bibinfo {author} {\bibfnamefont {C.}~\bibnamefont
  {Troessaert}},\ }\href {\doibase 10.1103/PhysRevLett.105.111103} {\bibfield
  {journal} {\bibinfo  {journal} {Phys. Rev. Lett.}\ }\textbf {\bibinfo
  {volume} {105}},\ \bibinfo {pages} {111103} (\bibinfo {year} {2010})},\
  \Eprint {http://arxiv.org/abs/0909.2617} {arXiv:0909.2617 [gr-qc]}
  \BibitemShut {NoStop}%
\bibitem [{\citenamefont {Bagchi}(2010)}]{Bagchi:2010zz}%
  \BibitemOpen
  \bibfield  {author} {\bibinfo {author} {\bibfnamefont {A.}~\bibnamefont
  {Bagchi}},\ }\href {\doibase 10.1103/PhysRevLett.105.171601} {\bibfield
  {journal} {\bibinfo  {journal} {Phys. Rev. Lett.}\ }\textbf {\bibinfo
  {volume} {105}},\ \bibinfo {pages} {171601} (\bibinfo {year} {2010})},\
  \Eprint {http://arxiv.org/abs/1006.3354} {arXiv:1006.3354 [hep-th]}
  \BibitemShut {NoStop}%
\bibitem [{\citenamefont {He}\ \emph {et~al.}(2015)\citenamefont {He},
  \citenamefont {Lysov}, \citenamefont {Mitra},\ and\ \citenamefont
  {Strominger}}]{He:2014laa}%
  \BibitemOpen
  \bibfield  {author} {\bibinfo {author} {\bibfnamefont {T.}~\bibnamefont
  {He}}, \bibinfo {author} {\bibfnamefont {V.}~\bibnamefont {Lysov}}, \bibinfo
  {author} {\bibfnamefont {P.}~\bibnamefont {Mitra}}, \ and\ \bibinfo {author}
  {\bibfnamefont {A.}~\bibnamefont {Strominger}},\ }\href {\doibase
  10.1007/JHEP05(2015)151} {\bibfield  {journal} {\bibinfo  {journal} {JHEP}\
  }\textbf {\bibinfo {volume} {05}},\ \bibinfo {pages} {151} (\bibinfo {year}
  {2015})},\ \Eprint {http://arxiv.org/abs/1401.7026} {arXiv:1401.7026
  [hep-th]} \BibitemShut {NoStop}%
\bibitem [{\citenamefont {Kapec}\ \emph {et~al.}(2017)\citenamefont {Kapec},
  \citenamefont {Mitra}, \citenamefont {Raclariu},\ and\ \citenamefont
  {Strominger}}]{Kapec:2016jld}%
  \BibitemOpen
  \bibfield  {author} {\bibinfo {author} {\bibfnamefont {D.}~\bibnamefont
  {Kapec}}, \bibinfo {author} {\bibfnamefont {P.}~\bibnamefont {Mitra}},
  \bibinfo {author} {\bibfnamefont {A.-M.}\ \bibnamefont {Raclariu}}, \ and\
  \bibinfo {author} {\bibfnamefont {A.}~\bibnamefont {Strominger}},\ }\href
  {\doibase 10.1103/PhysRevLett.119.121601} {\bibfield  {journal} {\bibinfo
  {journal} {Phys. Rev. Lett.}\ }\textbf {\bibinfo {volume} {119}},\ \bibinfo
  {pages} {121601} (\bibinfo {year} {2017})},\ \Eprint
  {http://arxiv.org/abs/1609.00282} {arXiv:1609.00282 [hep-th]} \BibitemShut
  {NoStop}%
\bibitem [{\citenamefont {Cheung}\ \emph {et~al.}(2017)\citenamefont {Cheung},
  \citenamefont {de~la Fuente},\ and\ \citenamefont
  {Sundrum}}]{Cheung:2016iub}%
  \BibitemOpen
  \bibfield  {author} {\bibinfo {author} {\bibfnamefont {C.}~\bibnamefont
  {Cheung}}, \bibinfo {author} {\bibfnamefont {A.}~\bibnamefont {de~la
  Fuente}}, \ and\ \bibinfo {author} {\bibfnamefont {R.}~\bibnamefont
  {Sundrum}},\ }\href {\doibase 10.1007/JHEP01(2017)112} {\bibfield  {journal}
  {\bibinfo  {journal} {JHEP}\ }\textbf {\bibinfo {volume} {01}},\ \bibinfo
  {pages} {112} (\bibinfo {year} {2017})},\ \Eprint
  {http://arxiv.org/abs/1609.00732} {arXiv:1609.00732 [hep-th]} \BibitemShut
  {NoStop}%
\bibitem [{\citenamefont {Pasterski}\ \emph {et~al.}(2017)\citenamefont
  {Pasterski}, \citenamefont {Shao},\ and\ \citenamefont
  {Strominger}}]{Pasterski:2016qvg}%
  \BibitemOpen
  \bibfield  {author} {\bibinfo {author} {\bibfnamefont {S.}~\bibnamefont
  {Pasterski}}, \bibinfo {author} {\bibfnamefont {S.-H.}\ \bibnamefont {Shao}},
  \ and\ \bibinfo {author} {\bibfnamefont {A.}~\bibnamefont {Strominger}},\
  }\href {\doibase 10.1103/PhysRevD.96.065026} {\bibfield  {journal} {\bibinfo
  {journal} {Phys. Rev. D}\ }\textbf {\bibinfo {volume} {96}},\ \bibinfo
  {pages} {065026} (\bibinfo {year} {2017})},\ \Eprint
  {http://arxiv.org/abs/1701.00049} {arXiv:1701.00049 [hep-th]} \BibitemShut
  {NoStop}%
\bibitem [{\citenamefont {Pasterski}\ and\ \citenamefont
  {Shao}(2017)}]{Pasterski:2017kqt}%
  \BibitemOpen
  \bibfield  {author} {\bibinfo {author} {\bibfnamefont {S.}~\bibnamefont
  {Pasterski}}\ and\ \bibinfo {author} {\bibfnamefont {S.-H.}\ \bibnamefont
  {Shao}},\ }\href {\doibase 10.1103/PhysRevD.96.065022} {\bibfield  {journal}
  {\bibinfo  {journal} {Phys. Rev. D}\ }\textbf {\bibinfo {volume} {96}},\
  \bibinfo {pages} {065022} (\bibinfo {year} {2017})},\ \Eprint
  {http://arxiv.org/abs/1705.01027} {arXiv:1705.01027 [hep-th]} \BibitemShut
  {NoStop}%
\bibitem [{\citenamefont {Donnay}\ \emph {et~al.}(2022)\citenamefont {Donnay},
  \citenamefont {Fiorucci}, \citenamefont {Herfray},\ and\ \citenamefont
  {Ruzziconi}}]{Donnay:2022wvx}%
  \BibitemOpen
  \bibfield  {author} {\bibinfo {author} {\bibfnamefont {L.}~\bibnamefont
  {Donnay}}, \bibinfo {author} {\bibfnamefont {A.}~\bibnamefont {Fiorucci}},
  \bibinfo {author} {\bibfnamefont {Y.}~\bibnamefont {Herfray}}, \ and\
  \bibinfo {author} {\bibfnamefont {R.}~\bibnamefont {Ruzziconi}},\ }\href@noop
  {} {\  (\bibinfo {year} {2022})},\ \Eprint {http://arxiv.org/abs/2212.12553}
  {arXiv:2212.12553 [hep-th]} \BibitemShut {NoStop}%
\bibitem [{\citenamefont {Higuchi}(1987)}]{Higuchi:1986wu}%
  \BibitemOpen
  \bibfield  {author} {\bibinfo {author} {\bibfnamefont {A.}~\bibnamefont
  {Higuchi}},\ }\href {\doibase 10.1063/1.527513} {\bibfield  {journal}
  {\bibinfo  {journal} {J. Math. Phys.}\ }\textbf {\bibinfo {volume} {28}},\
  \bibinfo {pages} {1553} (\bibinfo {year} {1987})},\ \bibinfo {note}
  {[Erratum: J.Math.Phys. 43, 6385 (2002)]}\BibitemShut {NoStop}%
\bibitem [{\citenamefont {{Chodos}}\ and\ \citenamefont
  {{Myers}}(1984)}]{1984AnPhy.156..412C}%
  \BibitemOpen
  \bibfield  {author} {\bibinfo {author} {\bibfnamefont {A.}~\bibnamefont
  {{Chodos}}}\ and\ \bibinfo {author} {\bibfnamefont {E.}~\bibnamefont
  {{Myers}}},\ }\href {\doibase 10.1016/0003-4916(84)90039-3} {\bibfield
  {journal} {\bibinfo  {journal} {Annals of Physics}\ }\textbf {\bibinfo
  {volume} {156}},\ \bibinfo {pages} {412} (\bibinfo {year}
  {1984})}\BibitemShut {NoStop}%
\bibitem [{\citenamefont {{Rubin}}\ and\ \citenamefont
  {{Ord{\'o}{\~n}ez}}(1985)}]{1985JMP....26...65R}%
  \BibitemOpen
  \bibfield  {author} {\bibinfo {author} {\bibfnamefont {M.~A.}\ \bibnamefont
  {{Rubin}}}\ and\ \bibinfo {author} {\bibfnamefont {C.~R.}\ \bibnamefont
  {{Ord{\'o}{\~n}ez}}},\ }\href {\doibase 10.1063/1.526749} {\bibfield
  {journal} {\bibinfo  {journal} {Journal of Mathematical Physics}\ }\textbf
  {\bibinfo {volume} {26}},\ \bibinfo {pages} {65} (\bibinfo {year}
  {1985})}\BibitemShut {NoStop}%
\bibitem [{\citenamefont {Marolf}\ and\ \citenamefont
  {Morrison}(2009{\natexlab{a}})}]{Marolf:2008hg}%
  \BibitemOpen
  \bibfield  {author} {\bibinfo {author} {\bibfnamefont {D.}~\bibnamefont
  {Marolf}}\ and\ \bibinfo {author} {\bibfnamefont {I.~A.}\ \bibnamefont
  {Morrison}},\ }\href {\doibase 10.1088/0264-9381/26/23/235003} {\bibfield
  {journal} {\bibinfo  {journal} {Class. Quant. Grav.}\ }\textbf {\bibinfo
  {volume} {26}},\ \bibinfo {pages} {235003} (\bibinfo {year}
  {2009}{\natexlab{a}})},\ \Eprint {http://arxiv.org/abs/0810.5163}
  {arXiv:0810.5163 [gr-qc]} \BibitemShut {NoStop}%
\bibitem [{\citenamefont {Ashtekar}\ and\ \citenamefont
  {Hansen}(1978)}]{Ashtekar:1978zz}%
  \BibitemOpen
  \bibfield  {author} {\bibinfo {author} {\bibfnamefont {A.}~\bibnamefont
  {Ashtekar}}\ and\ \bibinfo {author} {\bibfnamefont {R.~O.}\ \bibnamefont
  {Hansen}},\ }\href {\doibase 10.1063/1.523863} {\bibfield  {journal}
  {\bibinfo  {journal} {J. Math. Phys.}\ }\textbf {\bibinfo {volume} {19}},\
  \bibinfo {pages} {1542} (\bibinfo {year} {1978})}\BibitemShut {NoStop}%
\bibitem [{\citenamefont {Ashtekar}\ and\ \citenamefont
  {Romano}(1992)}]{Ashtekar:1991boundary}%
  \BibitemOpen
  \bibfield  {author} {\bibinfo {author} {\bibfnamefont {A.}~\bibnamefont
  {Ashtekar}}\ and\ \bibinfo {author} {\bibfnamefont {J.~D.}\ \bibnamefont
  {Romano}},\ }\href {\doibase 10.1088/0264-9381/9/4/019} {\bibfield  {journal}
  {\bibinfo  {journal} {Class. Quant. Grav.}\ }\textbf {\bibinfo {volume}
  {9}},\ \bibinfo {pages} {1069} (\bibinfo {year} {1992})}\BibitemShut
  {NoStop}%
\bibitem [{\citenamefont {Herberthson}\ and\ \citenamefont
  {Ludvigsen}(1992)}]{Herberthson:1992gcz}%
  \BibitemOpen
  \bibfield  {author} {\bibinfo {author} {\bibfnamefont {M.}~\bibnamefont
  {Herberthson}}\ and\ \bibinfo {author} {\bibfnamefont {M.}~\bibnamefont
  {Ludvigsen}},\ }\href {\doibase 10.1007/BF00756992} {\bibfield  {journal}
  {\bibinfo  {journal} {Gen. Rel. Grav.}\ }\textbf {\bibinfo {volume} {24}},\
  \bibinfo {pages} {1185} (\bibinfo {year} {1992})}\BibitemShut {NoStop}%
\bibitem [{\citenamefont {Perng}(1999)}]{Perng_1999}%
  \BibitemOpen
  \bibfield  {author} {\bibinfo {author} {\bibfnamefont {S.-M.}\ \bibnamefont
  {Perng}},\ }\href {\doibase 10.1063/1.532841} {\bibfield  {journal} {\bibinfo
   {journal} {Journal of Mathematical Physics}\ }\textbf {\bibinfo {volume}
  {40}},\ \bibinfo {pages} {1923–1950} (\bibinfo {year} {1999})}\BibitemShut
  {NoStop}%
\bibitem [{\citenamefont {Comp\`ere}\ \emph
  {et~al.}(2011{\natexlab{a}})\citenamefont {Comp\`ere}, \citenamefont
  {Dehouck},\ and\ \citenamefont {Virmani}}]{Compere:2011db}%
  \BibitemOpen
  \bibfield  {author} {\bibinfo {author} {\bibfnamefont {G.}~\bibnamefont
  {Comp\`ere}}, \bibinfo {author} {\bibfnamefont {F.}~\bibnamefont {Dehouck}},
  \ and\ \bibinfo {author} {\bibfnamefont {A.}~\bibnamefont {Virmani}},\ }\href
  {\doibase 10.1088/0264-9381/28/14/145007} {\bibfield  {journal} {\bibinfo
  {journal} {Class. Quant. Grav.}\ }\textbf {\bibinfo {volume} {28}},\ \bibinfo
  {pages} {145007} (\bibinfo {year} {2011}{\natexlab{a}})},\ \Eprint
  {http://arxiv.org/abs/1103.4078} {arXiv:1103.4078 [gr-qc]} \BibitemShut
  {NoStop}%
\bibitem [{\citenamefont {Troessaert}(2018)}]{Troessaert:2017jcm}%
  \BibitemOpen
  \bibfield  {author} {\bibinfo {author} {\bibfnamefont {C.}~\bibnamefont
  {Troessaert}},\ }\href {\doibase 10.1088/1361-6382/aaae22} {\bibfield
  {journal} {\bibinfo  {journal} {Class. Quant. Grav.}\ }\textbf {\bibinfo
  {volume} {35}},\ \bibinfo {pages} {074003} (\bibinfo {year} {2018})},\
  \Eprint {http://arxiv.org/abs/1704.06223} {arXiv:1704.06223 [hep-th]}
  \BibitemShut {NoStop}%
\bibitem [{\citenamefont
  {Strominger}(2014{\natexlab{a}})}]{Strominger:2013jfa}%
  \BibitemOpen
  \bibfield  {author} {\bibinfo {author} {\bibfnamefont {A.}~\bibnamefont
  {Strominger}},\ }\href {\doibase 10.1007/JHEP07(2014)152} {\bibfield
  {journal} {\bibinfo  {journal} {JHEP}\ }\textbf {\bibinfo {volume} {07}},\
  \bibinfo {pages} {152} (\bibinfo {year} {2014}{\natexlab{a}})},\ \Eprint
  {http://arxiv.org/abs/1312.2229} {arXiv:1312.2229 [hep-th]} \BibitemShut
  {NoStop}%
\bibitem [{\citenamefont
  {Strominger}(2014{\natexlab{b}})}]{Strominger:2013lka}%
  \BibitemOpen
  \bibfield  {author} {\bibinfo {author} {\bibfnamefont {A.}~\bibnamefont
  {Strominger}},\ }\href {\doibase 10.1007/JHEP07(2014)151} {\bibfield
  {journal} {\bibinfo  {journal} {JHEP}\ }\textbf {\bibinfo {volume} {07}},\
  \bibinfo {pages} {151} (\bibinfo {year} {2014}{\natexlab{b}})},\ \Eprint
  {http://arxiv.org/abs/1308.0589} {arXiv:1308.0589 [hep-th]} \BibitemShut
  {NoStop}%
\bibitem [{\citenamefont {Henneaux}\ and\ \citenamefont
  {Troessaert}(2019{\natexlab{a}})}]{Henneaux_2019}%
  \BibitemOpen
  \bibfield  {author} {\bibinfo {author} {\bibfnamefont {M.}~\bibnamefont
  {Henneaux}}\ and\ \bibinfo {author} {\bibfnamefont {C.}~\bibnamefont
  {Troessaert}},\ }\href {\doibase 10.1007/jhep05(2019)147} {\bibfield
  {journal} {\bibinfo  {journal} {Journal of High Energy Physics}\ }\textbf
  {\bibinfo {volume} {2019}} (\bibinfo {year} {2019}{\natexlab{a}}),\
  10.1007/jhep05(2019)147}\BibitemShut {NoStop}%
\bibitem [{\citenamefont {Henneaux}\ and\ \citenamefont
  {Troessaert}(2019{\natexlab{b}})}]{Henneaux:2019yqq}%
  \BibitemOpen
  \bibfield  {author} {\bibinfo {author} {\bibfnamefont {M.}~\bibnamefont
  {Henneaux}}\ and\ \bibinfo {author} {\bibfnamefont {C.}~\bibnamefont
  {Troessaert}},\ }\href {\doibase 10.1103/PhysRevD.99.125006} {\bibfield
  {journal} {\bibinfo  {journal} {Phys. Rev. D}\ }\textbf {\bibinfo {volume}
  {99}},\ \bibinfo {pages} {125006} (\bibinfo {year} {2019}{\natexlab{b}})},\
  \Eprint {http://arxiv.org/abs/1903.04437} {arXiv:1903.04437 [hep-th]}
  \BibitemShut {NoStop}%
\bibitem [{\citenamefont {Fuentealba}\ and\ \citenamefont
  {Henneaux}(2025{\natexlab{a}})}]{Fuentealba:2024lll}%
  \BibitemOpen
  \bibfield  {author} {\bibinfo {author} {\bibfnamefont {O.}~\bibnamefont
  {Fuentealba}}\ and\ \bibinfo {author} {\bibfnamefont {M.}~\bibnamefont
  {Henneaux}},\ }\href {\doibase 10.1007/JHEP03(2025)081} {\bibfield  {journal}
  {\bibinfo  {journal} {JHEP}\ }\textbf {\bibinfo {volume} {03}},\ \bibinfo
  {pages} {081} (\bibinfo {year} {2025}{\natexlab{a}})},\ \Eprint
  {http://arxiv.org/abs/2412.05088} {arXiv:2412.05088 [gr-qc]} \BibitemShut
  {NoStop}%
\bibitem [{\citenamefont {Fuentealba}\ and\ \citenamefont
  {Henneaux}(2025{\natexlab{b}})}]{Fuentealba:2025ekj}%
  \BibitemOpen
  \bibfield  {author} {\bibinfo {author} {\bibfnamefont {O.}~\bibnamefont
  {Fuentealba}}\ and\ \bibinfo {author} {\bibfnamefont {M.}~\bibnamefont
  {Henneaux}},\ }\href {\doibase 10.1007/JHEP07(2025)112} {\bibfield  {journal}
  {\bibinfo  {journal} {JHEP}\ }\textbf {\bibinfo {volume} {07}},\ \bibinfo
  {pages} {112} (\bibinfo {year} {2025}{\natexlab{b}})},\ \Eprint
  {http://arxiv.org/abs/2504.05385} {arXiv:2504.05385 [hep-th]} \BibitemShut
  {NoStop}%
\bibitem [{\citenamefont {Brice{\~n}o}\ \emph
  {et~al.}(2025{\natexlab{a}})\citenamefont {Brice{\~n}o}, \citenamefont
  {Gonz{\'a}lez},\ and\ \citenamefont {P{\'e}rez}}]{Briceno:2025ivl}%
  \BibitemOpen
  \bibfield  {author} {\bibinfo {author} {\bibfnamefont {M.}~\bibnamefont
  {Brice{\~n}o}}, \bibinfo {author} {\bibfnamefont {H.~A.}\ \bibnamefont
  {Gonz{\'a}lez}}, \ and\ \bibinfo {author} {\bibfnamefont {A.}~\bibnamefont
  {P{\'e}rez}},\ }\href@noop {} {\  (\bibinfo {year} {2025}{\natexlab{a}})},\
  \Eprint {http://arxiv.org/abs/2504.08612} {arXiv:2504.08612 [hep-th]}
  \BibitemShut {NoStop}%
\bibitem [{\citenamefont {Henneaux}\ and\ \citenamefont
  {Troessaert}(2019{\natexlab{c}})}]{Henneaux:2018mgn}%
  \BibitemOpen
  \bibfield  {author} {\bibinfo {author} {\bibfnamefont {M.}~\bibnamefont
  {Henneaux}}\ and\ \bibinfo {author} {\bibfnamefont {C.}~\bibnamefont
  {Troessaert}},\ }\href {\doibase 10.1007/JHEP05(2019)147} {\bibfield
  {journal} {\bibinfo  {journal} {JHEP}\ }\textbf {\bibinfo {volume} {05}},\
  \bibinfo {pages} {147} (\bibinfo {year} {2019}{\natexlab{c}})},\ \Eprint
  {http://arxiv.org/abs/1812.07445} {arXiv:1812.07445 [hep-th]} \BibitemShut
  {NoStop}%
\bibitem [{\citenamefont {Comp\`ere}\ and\ \citenamefont
  {Dehouck}(2011)}]{Compere:2011ve}%
  \BibitemOpen
  \bibfield  {author} {\bibinfo {author} {\bibfnamefont {G.}~\bibnamefont
  {Comp\`ere}}\ and\ \bibinfo {author} {\bibfnamefont {F.}~\bibnamefont
  {Dehouck}},\ }\href {\doibase 10.1088/0264-9381/28/24/245016} {\bibfield
  {journal} {\bibinfo  {journal} {Class. Quant. Grav.}\ }\textbf {\bibinfo
  {volume} {28}},\ \bibinfo {pages} {245016} (\bibinfo {year} {2011})},\
  \bibinfo {note} {[Erratum: Class.Quant.Grav. 30, 039501 (2013)]},\ \Eprint
  {http://arxiv.org/abs/1106.4045} {arXiv:1106.4045 [hep-th]} \BibitemShut
  {NoStop}%
\bibitem [{\citenamefont
  {Wald}(1995{\natexlab{a}})}]{wald1995quantumfieldtheorycurved}%
  \BibitemOpen
  \bibfield  {author} {\bibinfo {author} {\bibfnamefont {R.~M.}\ \bibnamefont
  {Wald}},\ }\href {https://arxiv.org/abs/gr-qc/9509057} {\enquote {\bibinfo
  {title} {Quantum field theory in curved spacetime},}\ } (\bibinfo {year}
  {1995}{\natexlab{a}}),\ \Eprint {http://arxiv.org/abs/gr-qc/9509057}
  {arXiv:gr-qc/9509057 [gr-qc]} \BibitemShut {NoStop}%
\bibitem [{\citenamefont {Bousso}\ \emph {et~al.}(2002)\citenamefont {Bousso},
  \citenamefont {Maloney},\ and\ \citenamefont {Strominger}}]{Bousso_2002}%
  \BibitemOpen
  \bibfield  {author} {\bibinfo {author} {\bibfnamefont {R.}~\bibnamefont
  {Bousso}}, \bibinfo {author} {\bibfnamefont {A.}~\bibnamefont {Maloney}}, \
  and\ \bibinfo {author} {\bibfnamefont {A.}~\bibnamefont {Strominger}},\
  }\href {\doibase 10.1103/physrevd.65.104039} {\bibfield  {journal} {\bibinfo
  {journal} {Physical Review D}\ }\textbf {\bibinfo {volume} {65}} (\bibinfo
  {year} {2002}),\ 10.1103/physrevd.65.104039}\BibitemShut {NoStop}%
\bibitem [{\citenamefont {Wald}(1995{\natexlab{b}})}]{Wald:1995yp}%
  \BibitemOpen
  \bibfield  {author} {\bibinfo {author} {\bibfnamefont {R.~M.}\ \bibnamefont
  {Wald}},\ }\href@noop {} {\emph {\bibinfo {title} {{Quantum Field Theory in
  Curved Space-Time and Black Hole Thermodynamics}}}},\ Chicago Lectures in
  Physics\ (\bibinfo  {publisher} {University of Chicago Press},\ \bibinfo
  {address} {Chicago, IL},\ \bibinfo {year} {1995})\BibitemShut {NoStop}%
\bibitem [{Note1()}]{Note1}%
  \BibitemOpen
  \bibinfo {note} {More precisely, since a primitive is defined only up to an
  additive constant, one may choose it such that $B^{(S)}_{n \ell m}(-\tau ) =
  -B^{(S)}_{n \ell m}(\tau ) $ whenever $s(\tau ,x^A)$ has $p$
  parity.}\BibitemShut {Stop}%
\bibitem [{Mat(2025)}]{Mathematica}%
  \BibitemOpen
  \href {https://functions.wolfram.com/} {\enquote {\bibinfo {title} {The
  mathematical functions site},}\ } (\bibinfo {year} {1998-2025})\BibitemShut
  {NoStop}%
\bibitem [{Note2()}]{Note2}%
  \BibitemOpen
  \bibinfo {note} {For $n=-1$, these relationships are equivalent only for
  $\ell \ge 1$ harmonics since $O^{(q)n=-1}$ vanishes when applied on a
  constant.}\BibitemShut {Stop}%
\bibitem [{\citenamefont {Marolf}\ and\ \citenamefont
  {Morrison}(2009{\natexlab{b}})}]{Marolf_2009}%
  \BibitemOpen
  \bibfield  {author} {\bibinfo {author} {\bibfnamefont {D.}~\bibnamefont
  {Marolf}}\ and\ \bibinfo {author} {\bibfnamefont {I.~A.}\ \bibnamefont
  {Morrison}},\ }\href {\doibase 10.1088/0264-9381/26/23/235003} {\bibfield
  {journal} {\bibinfo  {journal} {Classical and Quantum Gravity}\ }\textbf
  {\bibinfo {volume} {26}},\ \bibinfo {pages} {235003} (\bibinfo {year}
  {2009}{\natexlab{b}})}\BibitemShut {NoStop}%
\bibitem [{\citenamefont {Higuchi}\ and\ \citenamefont
  {Weeks}(2003)}]{Higuchi_2003}%
  \BibitemOpen
  \bibfield  {author} {\bibinfo {author} {\bibfnamefont {A.}~\bibnamefont
  {Higuchi}}\ and\ \bibinfo {author} {\bibfnamefont {R.~H.}\ \bibnamefont
  {Weeks}},\ }\href {\doibase 10.1088/0264-9381/20/14/303} {\bibfield
  {journal} {\bibinfo  {journal} {Classical and Quantum Gravity}\ }\textbf
  {\bibinfo {volume} {20}},\ \bibinfo {pages} {3005–3021} (\bibinfo {year}
  {2003})}\BibitemShut {NoStop}%
\bibitem [{\citenamefont {Fuentealba}\ \emph {et~al.}(2023)\citenamefont
  {Fuentealba}, \citenamefont {Henneaux},\ and\ \citenamefont
  {Troessaert}}]{Fuentealba:2022xsz}%
  \BibitemOpen
  \bibfield  {author} {\bibinfo {author} {\bibfnamefont {O.}~\bibnamefont
  {Fuentealba}}, \bibinfo {author} {\bibfnamefont {M.}~\bibnamefont
  {Henneaux}}, \ and\ \bibinfo {author} {\bibfnamefont {C.}~\bibnamefont
  {Troessaert}},\ }\href {\doibase 10.1007/JHEP02(2023)248} {\bibfield
  {journal} {\bibinfo  {journal} {JHEP}\ }\textbf {\bibinfo {volume} {02}},\
  \bibinfo {pages} {248} (\bibinfo {year} {2023})},\ \Eprint
  {http://arxiv.org/abs/2211.10941} {arXiv:2211.10941 [hep-th]} \BibitemShut
  {NoStop}%
\bibitem [{\citenamefont {Comp\`ere}\ \emph {et~al.}(2023)\citenamefont
  {Comp\`ere}, \citenamefont {Gralla},\ and\ \citenamefont
  {Wei}}]{Compere:2023qoa}%
  \BibitemOpen
  \bibfield  {author} {\bibinfo {author} {\bibfnamefont {G.}~\bibnamefont
  {Comp\`ere}}, \bibinfo {author} {\bibfnamefont {S.~E.}\ \bibnamefont
  {Gralla}}, \ and\ \bibinfo {author} {\bibfnamefont {H.}~\bibnamefont {Wei}},\
  }\href {\doibase 10.1088/1361-6382/acf5c1} {\bibfield  {journal} {\bibinfo
  {journal} {Class. Quant. Grav.}\ }\textbf {\bibinfo {volume} {40}},\ \bibinfo
  {pages} {205018} (\bibinfo {year} {2023})},\ \Eprint
  {http://arxiv.org/abs/2303.17124} {arXiv:2303.17124 [gr-qc]} \BibitemShut
  {NoStop}%
\bibitem [{Note3()}]{Note3}%
  \BibitemOpen
  \bibinfo {note} {More precisely, the conformal Killing equation is equivalent
  to $ \protect \mathcal {D}_a\protect \mathcal {D}_bH+q_{ab}H =c q_{ab}$,
  where $c$ is a constant. The general solution is $H=c+\psi _1^{(q)}$ with
  $\psi _1^{(q)}$ a linear combination of $n=1$, $\ell =0,1$ $q$-parity scalar
  harmonics. Since $H$ is defined up to an additive constant through the
  relation $\protect \hat \chi ^a=\protect \mathcal {D}^aH$, we may set
  $c=0$.}\BibitemShut {Stop}%
\bibitem [{\citenamefont {Deser}(1967)}]{Deser:1967zzb}%
  \BibitemOpen
  \bibfield  {author} {\bibinfo {author} {\bibfnamefont {S.}~\bibnamefont
  {Deser}},\ }\href@noop {} {\bibfield  {journal} {\bibinfo  {journal} {Ann.
  Inst. H. Poincare Phys. Theor. A}\ }\textbf {\bibinfo {volume} {7}},\
  \bibinfo {pages} {149} (\bibinfo {year} {1967})}\BibitemShut {NoStop}%
\bibitem [{Note4()}]{Note4}%
  \BibitemOpen
  \bibinfo {note} {Technically, the equation is $(\Box +3)\phi
  ^{(n)}=S^{(n)\phi }+ C$, where $C$ is a constant. However, this constant is
  not relevant since $\protect \mathcal {D}_a\protect \mathcal {D}_b C
  -\protect \frac {1}{3}\Box \protect \tmspace +\thickmuskip {.2777em}C
  q_{ab}=0$.}\BibitemShut {Stop}%
\bibitem [{\citenamefont {Laddha}\ and\ \citenamefont
  {Sen}(2018)}]{Laddha:2018myi}%
  \BibitemOpen
  \bibfield  {author} {\bibinfo {author} {\bibfnamefont {A.}~\bibnamefont
  {Laddha}}\ and\ \bibinfo {author} {\bibfnamefont {A.}~\bibnamefont {Sen}},\
  }\href {\doibase 10.1007/JHEP10(2018)056} {\bibfield  {journal} {\bibinfo
  {journal} {JHEP}\ }\textbf {\bibinfo {volume} {10}},\ \bibinfo {pages} {056}
  (\bibinfo {year} {2018})},\ \Eprint {http://arxiv.org/abs/1804.09193}
  {arXiv:1804.09193 [hep-th]} \BibitemShut {NoStop}%
\bibitem [{\citenamefont {Laddha}\ and\ \citenamefont
  {Sen}(2019)}]{Laddha:2018vbn}%
  \BibitemOpen
  \bibfield  {author} {\bibinfo {author} {\bibfnamefont {A.}~\bibnamefont
  {Laddha}}\ and\ \bibinfo {author} {\bibfnamefont {A.}~\bibnamefont {Sen}},\
  }\href {\doibase 10.1103/PhysRevD.100.024009} {\bibfield  {journal} {\bibinfo
   {journal} {Phys. Rev. D}\ }\textbf {\bibinfo {volume} {100}},\ \bibinfo
  {pages} {024009} (\bibinfo {year} {2019})},\ \Eprint
  {http://arxiv.org/abs/1806.01872} {arXiv:1806.01872 [hep-th]} \BibitemShut
  {NoStop}%
\bibitem [{\citenamefont {Sahoo}\ and\ \citenamefont
  {Sen}(2019)}]{Sahoo:2018lxl}%
  \BibitemOpen
  \bibfield  {author} {\bibinfo {author} {\bibfnamefont {B.}~\bibnamefont
  {Sahoo}}\ and\ \bibinfo {author} {\bibfnamefont {A.}~\bibnamefont {Sen}},\
  }\href {\doibase 10.1007/JHEP02(2019)086} {\bibfield  {journal} {\bibinfo
  {journal} {JHEP}\ }\textbf {\bibinfo {volume} {02}},\ \bibinfo {pages} {086}
  (\bibinfo {year} {2019})},\ \Eprint {http://arxiv.org/abs/1808.03288}
  {arXiv:1808.03288 [hep-th]} \BibitemShut {NoStop}%
\bibitem [{\citenamefont {Campiglia}\ and\ \citenamefont
  {Laddha}(2019)}]{Campiglia:2018dyi}%
  \BibitemOpen
  \bibfield  {author} {\bibinfo {author} {\bibfnamefont {M.}~\bibnamefont
  {Campiglia}}\ and\ \bibinfo {author} {\bibfnamefont {A.}~\bibnamefont
  {Laddha}},\ }\href {\doibase 10.1007/JHEP05(2019)207} {\bibfield  {journal}
  {\bibinfo  {journal} {JHEP}\ }\textbf {\bibinfo {volume} {05}},\ \bibinfo
  {pages} {207} (\bibinfo {year} {2019})},\ \Eprint
  {http://arxiv.org/abs/1810.04619} {arXiv:1810.04619 [hep-th]} \BibitemShut
  {NoStop}%
\bibitem [{\citenamefont {Comp{\`e}re}\ \emph {et~al.}(2025)\citenamefont
  {Comp{\`e}re}, \citenamefont {Fontaine},\ and\ \citenamefont
  {Nguyen}}]{Compere:2025tzr}%
  \BibitemOpen
  \bibfield  {author} {\bibinfo {author} {\bibfnamefont {G.}~\bibnamefont
  {Comp{\`e}re}}, \bibinfo {author} {\bibfnamefont {D.}~\bibnamefont
  {Fontaine}}, \ and\ \bibinfo {author} {\bibfnamefont {K.}~\bibnamefont
  {Nguyen}},\ }\href {\doibase 10.21468/SciPostPhysCore.8.4.066} {\bibfield
  {journal} {\bibinfo  {journal} {SciPost Phys. Core}\ }\textbf {\bibinfo
  {volume} {8}},\ \bibinfo {pages} {066} (\bibinfo {year} {2025})},\ \Eprint
  {http://arxiv.org/abs/2503.23937} {arXiv:2503.23937 [hep-th]} \BibitemShut
  {NoStop}%
\bibitem [{\citenamefont {Brice{\~n}o}\ \emph
  {et~al.}(2025{\natexlab{b}})\citenamefont {Brice{\~n}o}, \citenamefont
  {Gonz{\'a}lez}, \citenamefont {Henneaux},\ and\ \citenamefont
  {P{\'e}rez}}]{Briceno:2025cdu}%
  \BibitemOpen
  \bibfield  {author} {\bibinfo {author} {\bibfnamefont {M.}~\bibnamefont
  {Brice{\~n}o}}, \bibinfo {author} {\bibfnamefont {H.~A.}\ \bibnamefont
  {Gonz{\'a}lez}}, \bibinfo {author} {\bibfnamefont {M.}~\bibnamefont
  {Henneaux}}, \ and\ \bibinfo {author} {\bibfnamefont {A.}~\bibnamefont
  {P{\'e}rez}},\ }\href@noop {} {\  (\bibinfo {year} {2025}{\natexlab{b}})},\
  \Eprint {http://arxiv.org/abs/2510.21072} {arXiv:2510.21072 [hep-th]}
  \BibitemShut {NoStop}%
\bibitem [{\citenamefont {Henneaux}\ and\ \citenamefont
  {Troessaert}(2018{\natexlab{a}})}]{Henneaux_2018}%
  \BibitemOpen
  \bibfield  {author} {\bibinfo {author} {\bibfnamefont {M.}~\bibnamefont
  {Henneaux}}\ and\ \bibinfo {author} {\bibfnamefont {C.}~\bibnamefont
  {Troessaert}},\ }\href {\doibase 10.1007/jhep05(2018)137} {\bibfield
  {journal} {\bibinfo  {journal} {Journal of High Energy Physics}\ }\textbf
  {\bibinfo {volume} {2018}} (\bibinfo {year} {2018}{\natexlab{a}}),\
  10.1007/jhep05(2018)137}\BibitemShut {NoStop}%
\bibitem [{\citenamefont {Comp\`ere}\ and\ \citenamefont
  {Robert}()}]{Compere:2026jaa}%
  \BibitemOpen
  \bibfield  {author} {\bibinfo {author} {\bibfnamefont {G.}~\bibnamefont
  {Comp\`ere}}\ and\ \bibinfo {author} {\bibfnamefont {S.}~\bibnamefont
  {Robert}},\ }\href@noop {} {\enquote {\bibinfo {title} {{To be published}},}\
  }\BibitemShut {NoStop}%
\bibitem [{Note5()}]{Note5}%
  \BibitemOpen
  \bibinfo {note} {Massless scalar has $s=0$ and leading $r=0$ fields have
  $n=-1$ so that $n+1 \le \ell $ for any $\ell \geq 0$. For General Relativity,
  first order SDT tensors satisfy $n+1=0\le \ell $, second order SDT tensors
  $(n+1=1)$ can not have $\ell =0$ harmonics and third order SDT tensors
  $(n+1=2)$ can not have $\ell =0,1$ harmonics.}\BibitemShut {Stop}%
\bibitem [{\citenamefont {Freidel}\ \emph {et~al.}(2024)\citenamefont
  {Freidel}, \citenamefont {Pranzetti},\ and\ \citenamefont
  {Raclariu}}]{Freidel:2022skz}%
  \BibitemOpen
  \bibfield  {author} {\bibinfo {author} {\bibfnamefont {L.}~\bibnamefont
  {Freidel}}, \bibinfo {author} {\bibfnamefont {D.}~\bibnamefont {Pranzetti}},
  \ and\ \bibinfo {author} {\bibfnamefont {A.-M.}\ \bibnamefont {Raclariu}},\
  }\href {\doibase 10.1007/JHEP02(2024)176} {\bibfield  {journal} {\bibinfo
  {journal} {JHEP}\ }\textbf {\bibinfo {volume} {02}},\ \bibinfo {pages} {176}
  (\bibinfo {year} {2024})},\ \Eprint {http://arxiv.org/abs/2212.12469}
  {arXiv:2212.12469 [hep-th]} \BibitemShut {NoStop}%
\bibitem [{\citenamefont {Ashtekar}\ \emph {et~al.}(1990)\citenamefont
  {Ashtekar}, \citenamefont {Bombelli},\ and\ \citenamefont
  {Reula}}]{Ashtekar:1990gc}%
  \BibitemOpen
  \bibfield  {author} {\bibinfo {author} {\bibfnamefont {A.}~\bibnamefont
  {Ashtekar}}, \bibinfo {author} {\bibfnamefont {L.}~\bibnamefont {Bombelli}},
  \ and\ \bibinfo {author} {\bibfnamefont {O.}~\bibnamefont {Reula}},\
  }\href@noop {} {\  (\bibinfo {year} {1990})}\BibitemShut {NoStop}%
\bibitem [{\citenamefont {Henneaux}\ and\ \citenamefont
  {Troessaert}(2018{\natexlab{b}})}]{Henneaux:2018cst}%
  \BibitemOpen
  \bibfield  {author} {\bibinfo {author} {\bibfnamefont {M.}~\bibnamefont
  {Henneaux}}\ and\ \bibinfo {author} {\bibfnamefont {C.}~\bibnamefont
  {Troessaert}},\ }\href {\doibase 10.1007/JHEP03(2018)147} {\bibfield
  {journal} {\bibinfo  {journal} {JHEP}\ }\textbf {\bibinfo {volume} {03}},\
  \bibinfo {pages} {147} (\bibinfo {year} {2018}{\natexlab{b}})},\ \Eprint
  {http://arxiv.org/abs/1801.03718} {arXiv:1801.03718 [gr-qc]} \BibitemShut
  {NoStop}%
\bibitem [{\citenamefont {Comp\`ere}\ \emph
  {et~al.}(2011{\natexlab{b}})\citenamefont {Comp\`ere}, \citenamefont
  {Dehouck},\ and\ \citenamefont {Virmani}}]{compere_asymptotic_2011}%
  \BibitemOpen
  \bibfield  {author} {\bibinfo {author} {\bibfnamefont {G.}~\bibnamefont
  {Comp\`ere}}, \bibinfo {author} {\bibfnamefont {F.}~\bibnamefont {Dehouck}},
  \ and\ \bibinfo {author} {\bibfnamefont {A.}~\bibnamefont {Virmani}},\ }\href
  {\doibase 10.1088/0264-9381/28/14/145007} {\bibfield  {journal} {\bibinfo
  {journal} {Class. Quant. Grav.}\ }\textbf {\bibinfo {volume} {28}},\ \bibinfo
  {pages} {145007} (\bibinfo {year} {2011}{\natexlab{b}})},\ \Eprint
  {http://arxiv.org/abs/1103.4078} {arXiv:1103.4078 [gr-qc]} \BibitemShut
  {NoStop}%
\bibitem [{\citenamefont {Prabhu}\ and\ \citenamefont
  {Shehzad}(2022)}]{Prabhu:2021cgk}%
  \BibitemOpen
  \bibfield  {author} {\bibinfo {author} {\bibfnamefont {K.}~\bibnamefont
  {Prabhu}}\ and\ \bibinfo {author} {\bibfnamefont {I.}~\bibnamefont
  {Shehzad}},\ }\href {\doibase 10.1007/JHEP08(2022)029} {\bibfield  {journal}
  {\bibinfo  {journal} {JHEP}\ }\textbf {\bibinfo {volume} {08}},\ \bibinfo
  {pages} {029} (\bibinfo {year} {2022})},\ \Eprint
  {http://arxiv.org/abs/2110.04900} {arXiv:2110.04900 [gr-qc]} \BibitemShut
  {NoStop}%
\bibitem [{\citenamefont {Comp{\`e}re}\ and\ \citenamefont
  {Robert}(2026)}]{Compere:2026jmk}%
  \BibitemOpen
  \bibfield  {author} {\bibinfo {author} {\bibfnamefont {G.}~\bibnamefont
  {Comp{\`e}re}}\ and\ \bibinfo {author} {\bibfnamefont {S.}~\bibnamefont
  {Robert}},\ }\href {\doibase 10.1103/hlx7-frmx} {\bibfield  {journal}
  {\bibinfo  {journal} {Phys. Rev. Lett.}\ }\textbf {\bibinfo {volume} {137}},\
  \bibinfo {pages} {131401} (\bibinfo {year} {2026})},\ \Eprint
  {http://arxiv.org/abs/2603.08705} {arXiv:2603.08705 [hep-th]} \BibitemShut
  {NoStop}%
\bibitem [{\citenamefont {Atul~Bhatkar}(2021)}]{AtulBhatkar:2020hqz}%
  \BibitemOpen
  \bibfield  {author} {\bibinfo {author} {\bibfnamefont {S.}~\bibnamefont
  {Atul~Bhatkar}},\ }\href {\doibase 10.1007/JHEP02(2021)082} {\bibfield
  {journal} {\bibinfo  {journal} {JHEP}\ }\textbf {\bibinfo {volume} {02}},\
  \bibinfo {pages} {082} (\bibinfo {year} {2021})},\ \Eprint
  {http://arxiv.org/abs/2007.03627} {arXiv:2007.03627 [hep-th]} \BibitemShut
  {NoStop}%
\end{thebibliography}

%

\end{document}